\documentstyle[12pt,psfig,rotate]{article}

\newcommand{\nc}{\newcommand}
\nc{\degr}{\hbox{$^\circ$}}
\nc{\eg}{e.g.\ }
\nc{\etc}{etc.}
\nc{\cf}{cf.\ }
\nc{\ie}{i.e.\ }
\nc{\EE}[2]{$#1\times 10^{#2}$}
\nc{\Pdot}{$\dot{P}$}
\nc{\Msun}{M$_\odot$}
\nc{\Rsun}{{\rm R_\odot}}
\nc{\Lsun}{{\rm L_\odot}}

\begin{document}
\title{\bf A Trustworthy and Simple Method to Determine Pulsar Distances 
and the Electron Distribution in the Galaxy} 
\author{Oktay H.
Guseinov$\sp{1,2}$ \thanks{e-mail:huseyin@pascal.sci.akdeniz.edu.tr}
, Efe Yazgan$\sp2$
\thanks{email:yazgan@astroa.physics.metu.edu.tr}
, Sevin\c{c} Tagieva$\sp3$
\thanks{email:so$\_$tagieva@mail.ru} \\ \\
Ayb\"{u}ke K\"{u}pc\"{u} Yolda\c{s}$\sp2$
\thanks{email:aybuke@astroa.physics.metu.edu.tr} \\ \\
{$\sp1$Akdeniz University, Department of Physics,} \\
{Antalya, Turkey} \\
{$\sp2$Middle East Technical University, Department of Physics,} \\
{Ankara 06531, Turkey} \\
{$\sp3$Academy of Science, Physics Institute,} \\
{Baku 370143, Azerbaijan Republic} \\ \\ \\ \\ \\ \\  
}

\maketitle
\newpage
\begin{abstract}
\noindent
We present the model for determination of pulsar distances or 
average electron distribution using a method similar to the widely used 
dependence of A$_V$ on distances in different directions.
To have reliable pulsar distances, we 
have used several natural requirements and distances of pulsar-calibrators.
We have constructed  dispersion measure-distance relations for pulsars in 48 
different directions.

KEY WORDS: PULSAR DISTANCES, ELECTRON DISTRIBUTION, GALAXY
\end{abstract}

\clearpage
\parindent=0.2in

\section{Introduction}
In this paper, our goal is to give dependence between the dispersion 
measure (DM) and distances (d) for radio pulsars (PSRs) in each  
different 
direction, similar to the dependence between A$_V$ and distance. 
Naturally, degree of irregularity in electron distribution in 
the Galaxy is very small when compared with dust distribution. 
Most of the times, as a rule, a mathematical model for electron 
distribution is used to find distances of PSRs in the Galaxy. 
Using such models, PSR distances can be found with much 
error. Moreover, some of the PSR distances do not agree with the models.
For them, individual distances are given.
This is natural and this will not change even if perfectly complete 
models are constructed and used. We determined the distances to PSRs 
by using natural requirements (which will be explained) and  
distances which are well known by independent methods. This provided 
us with the knowledge to construct DM-distance relation for PSRs and the 
Galactic electron distribution very easily and with much less error.   

\section{Some Necessary General Information Which We Must Take into 
Consideration in the Construction of d-DM Relations} 
\begin{enumerate}
\item As it is known boundary mass of the progenitors of white dwarfs (WD) 
      and neutron stars (NS) ranges from about 5 \Msun (Lorimer et al. 
      1993; Strom 1994) 
      up to 6-11 \Msun (Weidemann 1990).
      We consider 7-8 \Msun (Nomoto 1984; Ayd\i n et al. 1996) as a better 
      value for the boundary mass which 
      corresponds to main sequence stars with spectral class B3V. Error in 
      this mass value (in spectral class) must be greater for NS which are 
      born in 
      close binary systems. 
\item According to Garmany \& Stencel (Garmany \& Stencel 1992) 75\% of O 
      stars and 58\% of stars with
      spectral classes B2V and earlier, are born in OB associations. In 
      addition, many young open clusters (they contain O or early type B 
      stars and massive supergiants) are located far from OB associations.
      However, progenitors of PSRs may be located in regions where
      there are no nearby open clusters or OB associations. For example, 
      near the Sun, within a
      region of radius 110 pc there are 6 stars with masses about 9 \Msun 
      where there are no young open clusters (Zombeck 1990). Similarly we 
      know that 
      there
      are no O stars or young open clusters and also no HII regions, near 
      the Crab 
      PSR (Lynga 1987; Barbon \& Hassan 1996). In these cases, where a 
      PSR is located far away from young 
      open clusters or OB associations, the progenitor may have a smaller 
      mass of about 6 \Msun or it might be a runaway star.

\item Sizes and masses of molecular clouds (MC) change in very large 
      intervals (Patel \& Pudritz 1994). MCs are generally located in 
galactic arms, and
      their distribution is highly inhomogeneous. Therefore stars
      may be born in small groups (open clusters-OC) and big clusters (OB
      association). One big
      association may consist of several open clusters with ages up to 
      (5-6) 10$^{7}$
      years. This means, the duration of star formation in one large OB
      association might continue for such long times (Ayd\i n et al. 
      1997). As the 
      average life times of most of the single PSRs do not exceed
      this value, in such a short time positions of arms, big star
      formation regions (SFRs, which contain several OB associations and 
      many OCs) and OB associations do not change in the Galaxy.

\item Dependence of the number of main sequence stars on their masses has an
      exponential character. For different star formation regions in the  
      Galaxy
      (also in Magellanic Clouds) the degree of this exponent may change from
      1.5 up to 2.5 (Salpeter 1955; Blaha \& Hamphreys 1989; Parker \& 
      Garmany 1993; Garmany et al. 1982). 
      Error in the value of the degree of the exponent is higher 
      for massive stars. 
      Naturally, these
      uncertainties lead to uncertainties in number-mass distribution of PSR
      progenitors. We may accept that the size of the birth place of PSRs not 
      to be limited with the size of the SFR. 
      This is because the distribution of young
      PSRs and SNRs in the neighborhood of Sun (closer than 3 kpc)
      indicate that these objects may be born far from SFR and OB  
      associations.\\
\item As well known, the distances for radio PSRs and all young objects 
      such as HII and HI regions and
      molecular clouds in average cannot be determined better than 30\%.
      Therefore the location of galactic arms and SFR, for which we have not 
      detected enough
      number of massive stars, are defined only roughly. Despite that, their 
      directions are very well known. 
\item It is well known that the luminosity of single PSRs 
      L=F$_\nu$d$^{2}$ mJy kpc at 400 MHz practically does not depend on 
      spin period (P) and period derivative (\.{P}), i.e. 
      the positions of PSRs on the P-\.{P} diagram. For this reason radio 
      luminosity of 
	single PSRs depend very weakly on the magnetic field strength of NSs 
        (B=3.3 10$^{19}$(P\.{P})$^{1/2}$) and 
	their 
	rate of rotational energy loss, \.{E}=4$\pi^2$I\.{P}/P$^3$ 
        (Allakhverdiev et al. 2002; Guseinov et al. 2002a). 
	Therefore we may accept that radio luminosity of single PSRs practically 
        does not depend on 
	their ages, the masses and duplicity of their progenitors (but close 
	binaries may be an exception).
\item	The average velocity of a single PSR is about 250 km/s 
	(Allakhverdiev et al. 1997; Hansen \& Phinney 1997). Therefore PSRs with 
        ages up to 5$\times$10$^{5}$ 
	years can not runaway from their 
	birth places to distances more than 200 pc (which corresponds to 160 pc 
	if we project this velocity to the Galactic plane). 
	But even for PSRs whose 
	velocities are very high (a very small number of them), it is practically 
	impossible for them
	to pass from one arm to another in a time shorter than 10$^{6}$ 
	years. Sizes of huge SFRs which are located in the 
	arms may approach to 1 kpc. The distance between the arms is about 1 kpc or 
	more. Therefore PSRs with ages up to 5$\times$10$^{5}$ years stay at  
	birth places.\\
\end{enumerate}
\section{Deviation of Star Formation Regions (Part of the Arms) From The 
Plane of Symmetry of the Galaxy}
Only old population stars (stars with ages about 10$^{10}$ years) can 
come to dynamical stability. 
The scale height and number density distributions in the Galactic center 
direction of old and young population stars are different.
This shows that old and young populations stars are not in 
equilibrium with each other, because the scale height and density 
distributions of the two populations are different. The total mass of gas 
and young stars 
located in the Galactic arms is only about 1-3\% of the mass of the whole 
Galaxy and the parameters of these arms change with a characteristic time 
$\sim$10$^9$ years. 
SFRs in the arms are yet more unstable having ages about an order of 
magnitude less 
than the characteristic ages of the arms of the Galaxy. Here, the arms 
are neither in dynamical equilibrium with the Galaxy, nor with SFRs. 
Therefore not in all regions on the geometrical plane, arms must be 
coincident with the geometrical plane of the Galaxy.
For some regions SFRs may stay higher or lower from the 
plane of the Galaxy. As shown by optical observations of cepheids with 
high luminosities (with long pulsation periods) and red supergiants, 
which are located at distances about 5-10 kpc from the Sun, in the 
direction l=220$\degr$-330$\degr$, SFRs lie about 300 pc below the 
plane of the Galaxy. At the same distance, in the direction  
l=70$\degr$-100$\degr$ it is found that SFRs lie about 400 pc above the 
plane of the Galaxy. The 
young objects closer than 3-5 kpc in the direction l=270$\degr$-320$\degr$ 
are located about 150 pc down the geometric plane of the Galaxy 
(Berdnikov 1987). Figure 1 displays l-z distribution of 108 
galactic 
PSRs with characteristic ages $\tau$$<$5$\times$10$^5$ years and 
distances $>$ 5 kpc. As seen from the figure between 250$\degr$ $<$ l $<$ 
300$\degr$ these young PSRs are below the galactic plane and between 
50$\degr$ $<$ l $<$ 80$\degr$ they are above the plane. These 
data 
must be taken into account in estimating distances of PSRs and in the 
construction of the electron distribution model of the Galaxy.\\

\section{Distribution Of Electrons In The Galaxy}
Irregularities have been observed in the distribution of dust, molecular
clouds and neutral Hydrogen (HI) in the Galaxy.
It is also natural to expect irregularities in the electron distribution
where the degree of irregularity is (naturally) considerably small.
Considerable variations in opacity and interstellar polarization can be 
observed for stars with the same distance in a very small region of sky
($\sim$1\degr\ square) close to the Galactic plane  
(Nickel \& Klare 1980). This is due to the highly inhomogeneous 
distribution of dust clouds.
For determination of hydrogen column density along the line of sight, there 
are two
surveys that included large number of stars; one with 554 stars
(Diplas \& Savage 1994) and the other with 594 stars (Fruscione et al. 1994).
Both of these surveys show that HI distribution and its degree of 
irregularity 
are quite different from the distribution (and its the degree of 
irregularity) of dust and MCs (Ankay \& Guseinov 1998).
The dispersion measure (DM), which is related with the electron
distribution, is different for PSRs with similar distances, but for
different locations on the sky. These irregularities in electron 
distribution are due to the contribution
of both HII regions and SNRs along the line of sight, and also the 
gravitational potential, the gas temperature and the cosmic ray distribution 
in the Galaxy. Even, in the central part (halo) of the Galaxy, novae and 
planetary nebulae 
have very small contribution to the electron density in large scale and 
to the background radio radiation. But irregularities in electron 
distribution is considerably smaller
than the ones in other components of interstellar medium that we have
mentioned above.
Even though these irregularities are small, there is no simple model
for Galactic electron distribution to calculate each pulsar's
distance.
Moreover, constructing a complex mathematical model which requires a lot of 
data for different components of the interstellar medium and for PSRs (e.g. 
Taylor \& Cordes 1993) cannot avoid large errors in the distances of 
individual pulsars.\\ Number density distribution of 
electrons in the Galaxy depends on the galactic longitude$-$l, latittude$-$b 
and distance from the Sun$-$d. However, in some directions the number 
density of electrons change considerably under small changes of l, b and 
d. This has been taken into account for a long time when estimating the PSR 
distances (Guseinov et al. 1978; Taylor \& Cordes 1993). \\
Dispersion measure for PSRs is defined through DM=$\int$ n$_e$dl; where l is 
the distance to the PSR and n$_e$ is the electron density in the line of sight.
DM naturally depend on 
the number of HII regions and SNRs on line of sight and their electron 
capacities (Lyne \& Graham-Smith 1998). Of course, for PSRs with small 
distance from the Sun the value 
of DM strongly depend on the number of HII regions and SNRs on the line of 
sight mainly in the anticenter direction. The best example is 
the Vela PSR. Is this statement valid as a rule for distant PSRs (with 
distances more than 3-4 kpc)? If it is valid, the value of DM must 
strongly depend on the number of  HII regions (OB 
associations) in the line 
of sight, and their electron capacities. If it is true, then contribution of 
electrons between the arms 
must be small in the value of DM. In other words, when we consider a 
direction between the arms we should expect 
small contributions to the value of DM.\\
As well known, the value of DM for PSRs are obtained very precisely by 
comparing their distances and distances for all objects (including also 
normal stars) which determines 
the location of the arms. We take this into account and show in 
Figure 2 
the dependence of the values of DM for PSRs with characteristic ages 
$\tau \le$ 5$\times$10$^{5}$ yrs on galactic longitude l. 
As seen from Figure 2 there are clusterings in the value of DM around 
some values of l (in the line of sight). The clusterings are easily seen 
for directions l=300$^o$ and l=360$^o$. If there are empty regions (where 
there is no SFR) in the line of sight, then there are clusterings.
Figure 3 represents the galactic distributions of PSRs with 
$\tau \le$ 5$\times$10$^{5}$ yrs and shows a similar effect. We compare 
these figures 
with the picture of galactic arms (see Figure 4) given in Georgelin and 
Georgelin (1976) which was constructed by utilizing the space distribution 
of giant HII regions. Note that the distance between the Sun and the 
galactic center is 
given as 10 kpc (correct value is 8.5 kpc) in this figure. 
Clustering of 
young PSRs may be an evidence that dominant number of PSRs are born inside or
close to OB associations and SFRs. 
As seen from Figures 2, 3 and 4, in the directions where the distance 
between 
the SFRs are large, young PSRs and giant HII regions are practically 
absent or very 
rare, value of DM continues rapidly to increase. 
How can we interpret this fact? SFRs and 
OB associations spatially
occupy only a small part of arms and their cross-section, where  
active PSRs are born. Of course, 
electron density and total number of electrons considerably increase in HII 
regions and SNRs. But 
the value of DM, as seen in Figure 2, changes 
smoothly in the plane of Galaxy. Therefore the value of electron density is 
also large for regions between the arms. Naturally the number density of 
electrons increase in galactic center direction rapidly. But this increase 
depends on and also form the ability SFR today in different directions in 
the Galaxy. In Figure 2, it is seen 
considerable difference between the DM values for PSR with approximately 
equal distance from the Sun (d=5.5$-$6.5 kpc) in symmetrical longitude 
directions l about 
300$\degr$ and l about 60$\degr$. As it is known these directions have 
been scanned carefully in search of PSRs with almost the same accuracy.\\  

\section{How we estimate the distances of PSRs}

In order to investigate the arm structure and the locations of SFRs around 
the Sun within a 
distance of 4--5 kpc, usually OB associations, OCs and red 
supergiants are studied.
For these objects the relative errors in estimating their distances  
can be 30\% (Humphreys 1978; Efremov 1989; Ayd\i n et al. 1996; Ahumada 
\& Lapasset 1995). There is no single good method to estimate the distance 
of all extended objects belonging to the arms (MCs, neutral
Hydrogen clouds, SNRs and HII regions).
In determining the distances to these objects using HI 21 cm line and
Galaxy rotation models, the error exceeds 30\%. It is known that it is 
impossible to calculate an object's distance
using HI line at 21 cm if the object's radial velocity component of
Galactic rotational velocity is small.
In addition to this, in certain directions (in the
vicinity of longitude l=0$\degr$) and distances the 
hydrogen lines gives two different 
distance value. However, it is the most widely used model. \\

It is a well known fact that no relation has been found in PSR
parameters to estimate their distance.
For ordinary distant stars, however, one can use the relations either
between luminosity and spectral class, or luminosity and pulsation
period to estimate their distance.
In the early days of PSR astronomy between 1970 and 1980, since 
the origin of pulsars, 
the mass of their progenitors and their birth rates were not well known 
(even 
worse than today). Therefore, a homogeneous electron density distribution in 
Galaxy had been assumed.
In doing this, one should also know some of the pulsar distances
independent of their DM. Thus, PSRs with distances 
estimated using HI line in general are used as an
extra distance calibrator (Manchester \& Taylor 1981). However, 
later on PSR distances have been estimated according to
the rough model of Galactic electron distribution and some natural
requirements (Guseinov \& Kasumov 1981; Johnston et al. 1992; Taylor \& 
Cordes 1993; G\"{o}k et al. 1996). Nowadays, calibrators are chosen from 
members of 
globular clusters (GCs) or Magellanic clouds (MC), PSRs connected to
Supernova remnants (SNRs) with well known distances and from PSRs whose 
distances are known from other available data (Lyne \& Graham-Smith 1998). 
In determining the calibrators for PSRs,
distance estimates calculated using the 21 cm line are not accepted as
a rule.\\

In estimating pulsar distances, the model of Galactic electron
distribution constructed by Taylor and Cordes 
 (Taylor \& Cordes 1993) has been widely used in recent years.
However, in estimating the pulsar distances, the approach of
G\"ok et al. (1996) gave smaller distances than the ones   
calculated using the model of Taylor and Cordes (1993) for PSRs
farther than 4 kpc and also for PSRs with Galactic latitudes greater than 
about 10\degr  at distances d$>$1 kpc.
To form a new model electron distribution, G\'omez et al. 
(2001) have published a huge PSR list which could be used as 
calibrators.
We have decided to revise their distance values to use them as
calibrators.
In Table 1 we present 39 pulsars for which errors in
distances should not be higher than 30\%.
Since the distances of PSRs in the same GC are the same, only one
pulsar from each GC has been included in the table.
Number of pulsars which we used as calibrators, is considerably smaller
than the one in the calibrator list of G\'omez et al. (2001).
In this table, one of the most problematic calibrators is PSRJ0835-4510
(in Vela SNR). Distance of this PSR has been discussed in detail by 
Guseinov et al. (2002b). This discussion shows how much information is 
necessary to analyze in order to obtain trustworthy distance values. \\

Radio luminosity of PSRs should not depend on their birth place and
should not considerably exceed luminosities of the strongest pulsars
(\eg Crab with a very well known distance and being the strongest  
pulsar in Magellanic clouds).
Luminosity of Crab is \EE{2.6}{3} mJy kpc$^2$, 56 mJy kpc$^2$ at 400
MHz and 1400 MHz, respectively.
Luminosity of the strongest PSR in the Large Magellanic Cloud (PSR
J0529-6652) is \EE{6}{3} mJy kpc$^2$ at 400 MHz and \EE{7.5}{2} mJy 
kpc$^2$ at 1400 MHz. PSRJ0045-7319 which is in the Small Magellanic Cloud 
has the highest 
luminosity at 400 MHz is \EE{9.5}{3} mJy kpc$^2$ whereas its luminosity at 
1400 MHz is \EE{1.3}{3} mJy kpc$^2$.
Therefore the upper limit for luminosities of PSRs might be close to
the value of \EE{1.5}{4} mJy kpc$^2$ and \EE{3.5}{3} mJy kpc$^2$ for
400 MHz and 1400 MHz, respectively (spectral indices of radiation in radio 
band of PSRs have been also taken into account).
In our list of PSRs the strongest one among the galactic PSRs is PSR 
J1644-4559 
with luminosities of \EE{7.6}{3} mJy kpc$^2$ and \EE{6.3}{3} mJy kpc$^2$
at 400 and 1400 MHz, respectively (Guseinov et al. 2002a and 
references 
therein). These PSRs have extraordinarily flat spectra (Taylor et al. 
1996).\\

Since PSRs were born in the Galactic plane and
surveys have scanned the Galactic plane many times, most of the PSRs,   
especially the farthest ones, have small Galactic latitudes
($|b| < 5\degr$) [Manchester et al. 2002; Morris et al. 2002, Lyne 
et al. 1998; D'Amico et al. 1998; Manchester et al. 1996). As can be 
seen in our calibrator table (Table 1), there are 14 PSRs with $|b|>30\degr$,
10 PSRs with $30\degr>|b|>7\degr$,
6 PSRs with $7\degr>|b|>3\degr$, and only
14 PSRs with $|b|<3\degr$.
Thus, the calibrators in Magellanic Clouds, GCs and calibrators with known
trigonometric parallax's become insignificant for PSRs with small $|b|$.
Only 3 from our calibrator list belong to $|b|<3$\degr and have
distances greater than 5 kpc.
Therefore, for the PSRs with large distances and low $|b|$ values there
are almost no calibrators.
In addition to this, for such distances it is quite difficult to judge
the value of the electron density.\\

Considering the reasons given above in adopting distances for PSRs, 
the following criteria become very important:
\begin{enumerate}
\item In the direction of 40\degr$<l<$320\degr we see the most luminous
      PSRs throughout the Galaxy.
\item For all Galactic longitudes ($l$), pulsars with equal
      characteristic times ($\tau$) must have, on the average, similar
      $|z|$ values except PSRs with $\tau\le$\EE{5}{6} years in
      the regions where SFRs are considerably above or below the
      Galactic plane.
\item PSRs with $\tau<$\EE{5}{5} years must still be closer to their
      birth places, \ie in the SFRs.
\item The PSR luminosity does not depend on $l$ and d, and it
      should not considerably exceed the luminosity of known strongest 
      pulsars at 400 and 1400 MHz. The value of luminosities at 400 and 
      1400 MHz must at most be 1.5$\times$10$^4$ mJy kpc$^2$ 
      and 3$\times$10$^3$ mJy kpc$^2$ respectively.
\item Electron density in the Galaxy must be correlated with the
      number density of HII regions and OB associations but not strongly 
      at large distances. Electron density must increase as one 
      approaches to the 
      galactic center and galactic plane similar to background radiation 
in radio band. 
      \item PSR distances must be arranged in such a way that their 
      value
      should correspond to a suitable distance value of PSRs in Table 1
      (value of DM and the direction of the PSR have to
      be taken into account).
\end{enumerate}

\section{PSR Distances and Electron Distribution in the Galaxy}
DM of PSRs (or electron colomn density) depend on number 
density of electrons and PSR distances. Therefore, if we know the PSR 
distances, we may have correct information about the electron distribution. 
And vice versa if we know electron distribution in the Galaxy then we also 
know the PSR distances. Therefore, there is no difference between 
electron or 
PSR distribution if our goal is to estimate the distances of PSRs. Of 
course if it was possible to have all information about electron 
distribution in the Galaxy and construct a mathematical model, it would be 
very good. However, it is impossible to have such information and 
construct a model which would be easy for practical use. At the same time, 
it is 
hard to use a model for PSRs having large $|b|$ values. There exists some 
PSRs even at galactic plane, whose distances do not fit the models.  
Hence, we have chosen 
a more practical and correct way. We continued the work of Guseinov et al. 
(2002a) and adapted distances for each PSR in such a 
way that satisfies our natural requirements taking into consideration the 
independently known distances (Table 1). 
We arranged pulsar distances 
in different directions for different longitude (l) and lattitude (b) 
intervals.
We have divided all PSRs in the Galaxy into 14 longitude intervals, and for 
each 
longitude interval we divide PSRs into different lattitude 
intervals. For example Figures 1a$-$1d represent distance (d) versus 
dispersion measure (DM) for PSRs with -10$\degr <$l$<10\degr$ and 
-3$\degr <$b$<$3$\degr$; -7$\degr$$<$b$<$-3$\degr$ and 
3$\degr$$<$b$<$7$\degr$; -15$\degr$$<$b$<$-7$\degr$ and 
7$\degr$$<$b$<$16$\degr$; -25$\degr$$<$b$<$-15$\degr$ and 
15$\degr$$<$b$<$25$\degr$. The total number of such figures, where 
electron distribution is different, is 48. As ssen from figures (from the 
inclination of d$-$DM dependence in each figure) the electron number 
density depends also on distances from the Sun in each interval of l and b. 
PSRs with independent distances listed in Table 1 are also shown in 
figures.
Our adapted distances for all 1328 PSRs, which we have used in the 48 
figures in this paper 
are included in the list of Guseinov et al. (2002b).
 What must we do to estimate the distances of newly observed PSRs? We must 
take the figure which corresponds to the direction of the newly observed 
PSR. 
Then choose the distance that corresponds to the value of DM with taking 
into consideration the value of lattitude. For example, Figure 1b 
corresponds to -10$\degr <$l$<10\degr$ and 3$\degr$$<|b|<$7$\degr$. If PSR 
has a $|b|$ value close to 7$\degr$ we must choose the value of the distance 
close to maximum among the distances which correspond to the value of DM. 
Of course, if $|b|$ is close to 3$\degr$, then we must choose the lower 
distance value.\\  

\section{Discussion and Conclusion}
To compare the distances of PSRs (new distances) that is obtained by our 
model and the distances obtained by the model of Taylor \& Cordes 
(1993), 
in Figure 5 we have plotted out our new distances vs. Taylor \& Cordes 
distances of PSRs. For most of these PSRs (from Table 1) almost the same 
independent distances are agreed on, but for some of them we have taken 
considerably 
different independent distances from that of Taylor \& Cordes (1993) 
model gives. The names of these PSRs, as shown in Figure 5, are as 
follows; PSR 
J1302-6350, J1748-2021, J1804-0735 and J1910-59. As seen in Figure 5, the 
new distances and the distances given by Taylor \& Cordes (1993) model 
differ 
up to more than twice. Most of these PSRs are located in south semi-sphere 
and were discovered during the 1400 MHz survey (with l between 
300$^o$-360$^o$). Among the PSRs which are close to the Sun, the greatest 
difference in the distance given by our model and the model of Taylor \& 
Cordes is owned by PSR J1939+2134 (see the distance value calculated 
from Taylor \& Cordes model in the book by Lyne \& Graham-Smith 
(1998).\\  

To decide which model give the true distances, we need to test. In the 
previous section, in finding the PSR distance we have several 
criteria. One of these criteria is that the PSRs at the same age should 
be at almost the same distances from the Galactic plane. As we said 
before, the surveys done at 1400 MHz scanned the Galactic plane with 
$|b|<$5$\degr$. Hence, they have discovered a 
lot of young PSRs. It is necessary to mention that the ratio of flux at 1400 
MHz 
to flux at 400 MHz of young PSRs is several times more than this ratio for 
old PSRs (Guseinov et al. 2002a). This makes the discovery of young 
PSRs 
easier. In Figure 6, latitude (b)-longitude (l) distribution of PSRs 
whose 
l is between 300$^o$-360$^o$ and characteristic ages below 10$^6$ years, is 
given. As seen from this figure, dominant number of these PSRs have 
$|b|<$2$\degr$. In Figures 7 and 8, the distance from the Galactic plane (z) 
vs. 
the distance from the Sun (d) for PSRs with age$<$10$^6$ years and in the 
longitude interval l=300$^o$-360$^o$ is given according to the new model and 
old model of electron distribution, respectively. The cone corresponds to 2 
degrees since most of the young PSRs have $|b|<$2$\degr$ as can be seen 
from Figure 6. It gives the limits of z at each distance d. As seen from 
these figures there is not an important difference neither in z nor in d. 
For only four of these PSRs Taylor \& Cordes (1993) model has given very 
large distance values. This indicates that for small b values 
($|b|<$2$\degr$) there 
is no significant difference in electron distributions of the two models 
for l between 300$^o$-360$^o$. This is in general also true for other 
longitude directions. \\

For PSRs whose characteristic ages are between 10$^{6}-$10$^7$ years, 
b-l distribution with l between 300$\degr$ and 360$\degr$, is shown in 
Figure 9. As seen from 
the figure dominant number of these PSRs have $|b|<$5$\degr$. In Figures 10 
and 11, the distance from Galactic plane z vs. the
distance from the Sun d is given according to the new model and old
model electron distribution, respectively. PSRs with high b values are 
nearer PSRs. As seen from Figures 10 and 11, for PSRs farther than $\sim$3 
kpc, the new distances (d,z) are considerably smaller than the distances 
(d,z) given by the old model. It is seen from Figure 10 that for all 
distances number density of PSRs are higher near the Galactic plane, and 
for PSRs farther than 6 kpc the distances from the Galactic plane is in 
average the same. The distance distribution according to the old model 
does not agree with this criterion. Thus according to the Taylor \& 
Cordes model, as the distances of PSRs increase, average 
distances from Galactic plane increases. It indicates that the space 
velocities of farther PSRs are larger, however there is no reason for 
this. \\

$Acknowledgments$
We thank T\"{U}B\.{I}TAK, the Scientific and Technical Research
Council of Turkey, for support through TBAG-\c{C}G4. \\ \\

\begin{table}
\scriptsize
\caption{Distance calibrators.}
\begin{tabular}{llllllll}\hline\hline
\multicolumn{1}{l}{\ Name} &
\multicolumn{1}{l}{\ l} &
\multicolumn{1}{l}{\ b} &
\multicolumn{1}{l}{\ d} &
\multicolumn{1}{l}{\ DM} &
\multicolumn{1}{l}{\ n$_e$} &
\multicolumn{1}{l}{\ Location} &
\multicolumn{1}{l}{\ Ref} \\
\hline
0024-7204W&305.9&  -44.9&  4.5&    24.3&   0.005&  GC NGC 104 (47 Tuc)&
[1,2] \\
0045-7319&303.5&  -43.8&  57&     105.4&  0.002&  SMC&    [3]\\
0113-7220&300.6&  -44.7&  57&     125&    0.002&  SMC&    [4]\\
0205+6449&130.7&  3.1&    3.2&    140.7&  0.044&  SNR G130.7+3.1&
[5]\\
0455-6951&281.2&  -35.2&  50&     94.9&   0.002&  LMC&    [4]\\
0529-6652&277.0&  -32.8&  50&     103.2&  0.002&  LMC&    [4]\\
0534+2200&184.6&  -5.8&   2&      56.79&  0.028&  SNR G184.6-5.8 
(Crab)&
[6]\\
0535-6935&280.1&  -31.9&  50&     89.4&   0.002&  LMC&    [4]\\
0540-6919&279.7&  -31.5&  50&     146&    0.003&  LMC&    [7]\\
0826+2637&196.9&  31.7&   0.4&    19.48&  0.049&  Parallax& [8]\\
0835-4510&263.6&  -2.8&   0.45&   68.2&   0.152&  SNR G263.9-3.3 
(Vela)&
[9,10,11]\\
0922+0638&225.4&  36.4&   1.21&   27.31&  0.020&  Parallax&
[12,13]\\
0953+0755&228.9&  43.7&   0.28&   2.97&   0.011&  Parallax&
[8,14]\\
1119-6127&292.2&  -0.54&  7.5&    707.4&  0.101&  SNR G292.2-0.5&
[15,11,16]\\
1124-5916&292.0&  1.8&    6&      330&    0.066&  SNR G292.0+1.8&   
[17]\\
1302-6350&304.2&  -0.9&   1.3&    146.72& 0.113&  Sp Binary, Be&
[18]\\
1312+1810&332.9&  79.8&   18.9&   24&     0.001&  GC NGC 5024 (M53)&
[1,19]\\ 
1341-6220&308.7&  -0.4&   8&      730&    0.097&  SNR G308.8-0.1&
[20,11]\\
1456-6843&313.9&  -8.5&   0.45&   8.6&    0.019&  Parallax& [21]\\
1513-5908&320.3&  -1.2&   4.2&    253.2&  0.060&  SNR G320.4-1.2&
[7,11,16]\\
1518+0204B&3.9&    46.8&   7&      30.5&   0.004&  GC NGC 5904 (M5)&
[1,22,23]\\
1623-2631&350.9&  15.9&   1.8&    62.86&  0.035&  GC NGC 6121 (M4)&
[1,24]\\
1641+3627B&59.8&   40.9&   7.7&    29.5&   0.004&  GC NGC 6205 (M13)&
[1,25]\\
1701-30&353.6&  7.3&    5&      114.4&  0.023&  GC NGC 6266 (M62)&
[1,26,27]\\
1721-1936&4.9&    9.7&    9&      71&     0.008&  GC NGC 6342&   
[1,28]\\
1740-53&338.2&  -11.9&  2.3&    71.8&   0.031&  GC NGC 6397&
[1,26,29]\\ 
1748-2021&7.7&    3.8&    6.6&    220&    0.033&  GC NGC 6440&
[1,30]\\
1748-2445B&3.85&   1.7&    7&      205&    0.029&  Ter 5&
[1,31]\\
1801-2451&5.27&   -0.9&   4.5&    289&    0.064&  SNR G5.27-0.9&
[32]\\
1803-2137&8.4&    0.1&    3.5&    233.9&  0.067&  SNR G8.7-0.1&  
[33,34,35,11,16]\\
1804-0735&20.8&   6.8&    7&      186.38& 0.027&  GC NGC 6539&
[1,36]\\
1807-2459&5.8&    -2.2&   2.6&    134&    0.052&  GC NGC 6544&
[1,26,37]\\
1823-3021B&2.8&    -7.9&   8&      87&     0.011&  GC NGC 6624&
[1,38]\\
1824-2452&7.8&    -5.6&   5.7&    119.83& 0.021&  GC NGC 6626 (M28)&
[1,39]\\
1856+0113&34.6&   -0.5&   2.8&    96.7&   0.035&  SNR G34.7-0.4& 
[11,16]\\
1910-59&336.5&  -25.6&  4&      34&     0.009&  GC NGC 6752&
[1,26,40]\\
1910+0004&35.2&   -4.2&   6.5&    201.5&  0.031&  GC NGC 6760&
[1,28]\\
1930+1852&54.1&   0.27&   5&      308&    0.062&  SNR G54.1+0.3&
[41]\\
1932+1059&47.4&   -3.9&   0.17&   3.18&   0.019&  Parallax&
[42,43,44,45]\\
1952+3252&68.8&   2.8&    2&      44.98&  0.022&  SNR G69.0+2.7& 
[7,34,11,16]\\
2022+5154&87.9&   8.4&    1.1&    22.58&  0.021&  Parallax& [46]\\
2129+1209H&65.1&   -27.3&  10&     67.15&  0.007&  GC NGC 7078 (M15)&
[7]\\
2337+6151&114.3&  0.2&    2.8&    58.38&  0.021&  SNR G114.3+0.3&
[47,11,16]\\
\hline
\end{tabular}
[1] Harris 1996; [2] Hesser et al. 1987; [3] Feast \& Walker 1987; [4] 
Crawford et al. 2001a; [5] Camilo et al. 2002b; [6] Trimble \& Woltjer 
1971; [7] Taylor et l. 1995; [8] Gwinn et al. 1986; [9] Cha et al. 1999; 
[10] Legge 2000; [11] Guseinov \& Ankay 2002; [12] Chatterjee et al. 
2001; [13] Fomalont et al. 1999; [14] Brisken et al. 2000; [15] Crawford 
et al. 2001b; [16] Kaspi \& Helfand 2002; [17] Camilo et al. 2002a; [18] 
Johnston et al. 1994; [19] Rey et al. 1998; [20] Caswell wt al. 1992; 
[21] Bailes et al. 1990; [22] Brocato et al. 1996a; [23] Sandquist et al. 
1996; [24] Cudworth \& Rees 1990; [25] Paltrinieri et al. 1998; [26] 
D'Amico et al. 2001; [27] Brocato et al. 1996b; [28] Heitsch \& Richtler 
1999; [29] Alcaino et al. 1987; [30] Ortolani et al. 1994; [31] Ortolani 
et al. 1996; [32] Thorsett et al. 2002; [33] Frail et al. 1994; [34] 
Allakhverdiev et al. 1997; [35] Finley \& \"{O}gelman 1994; [36] 
Armandroff 1988; [37] Kaspi et al. 1994; [38] arejedini \& Norris 1994; 
[39] Rees \& Cudworth 1991; [40] Buonanno et al. 1986; [41] Camilo et al. 
2002c; [42] Weisberg et al. 1980; [43] Salter et al. 1979; [44] Backer \& 
Sramek 1982; [45] Campbell 1995; [46] Campbell et al. 1996; [47] Furst et 
al. 1993
 \end{table}

\begin{figure*}
\centerline{\psfig{file=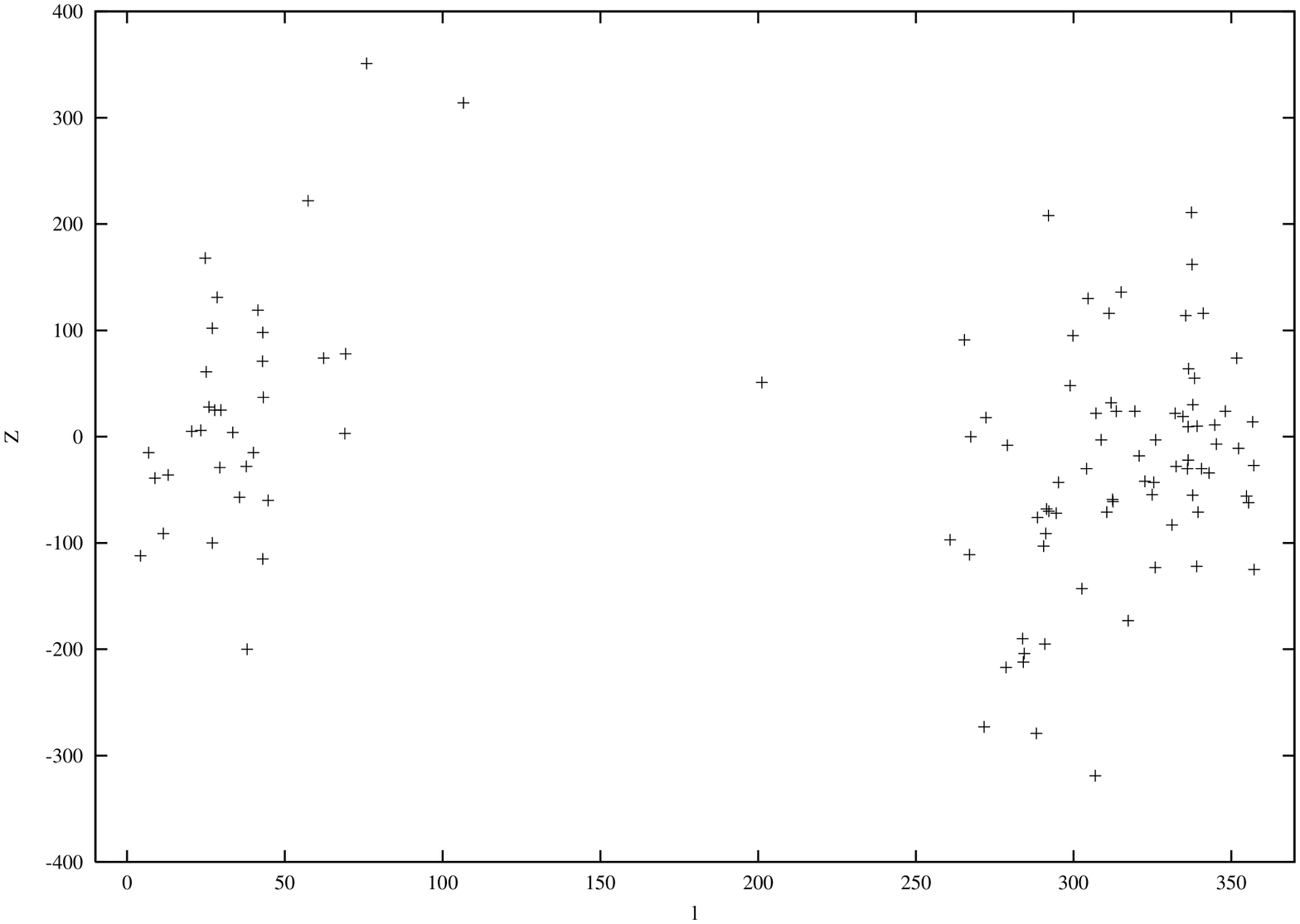,width=16cm,height=16cm,angle=-90}}
{Figure 1. Galactic height (z) vs. Galactic longitude (l) for 108 pulsars 
with age$<$5 10$^{5}$ years and d$>$5 kpc.}
\end{figure*}

\begin{figure*}
\centerline{\psfig{file=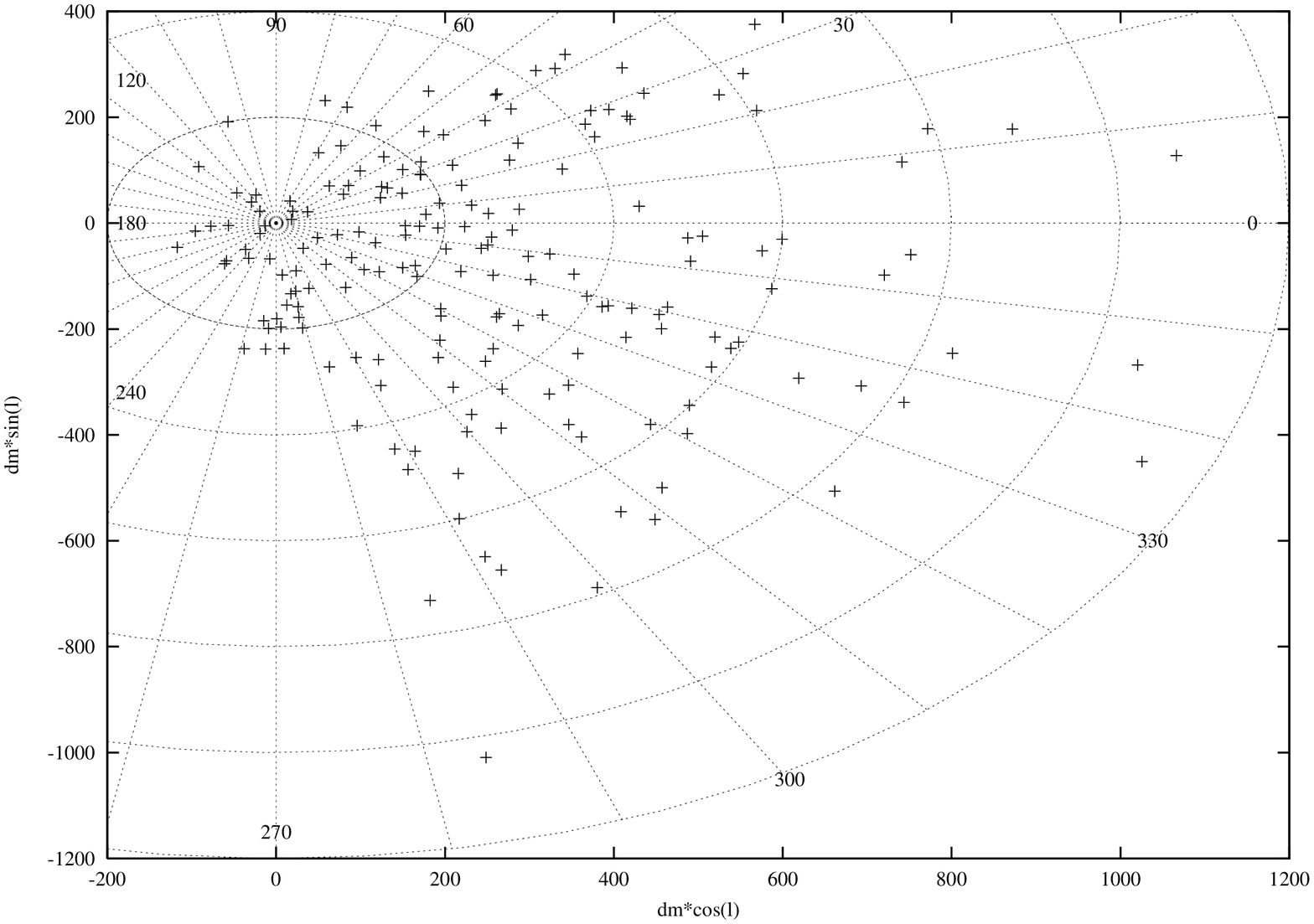,width=15cm,height=16.7cm,angle=-90}}
{Figure 2. Dispersion measure as a function of Galactic longitude (l) for 
189 pulsars with age$<$5 10$^5$ years} 
\end{figure*}

\begin{figure*}
\centerline{\psfig{file=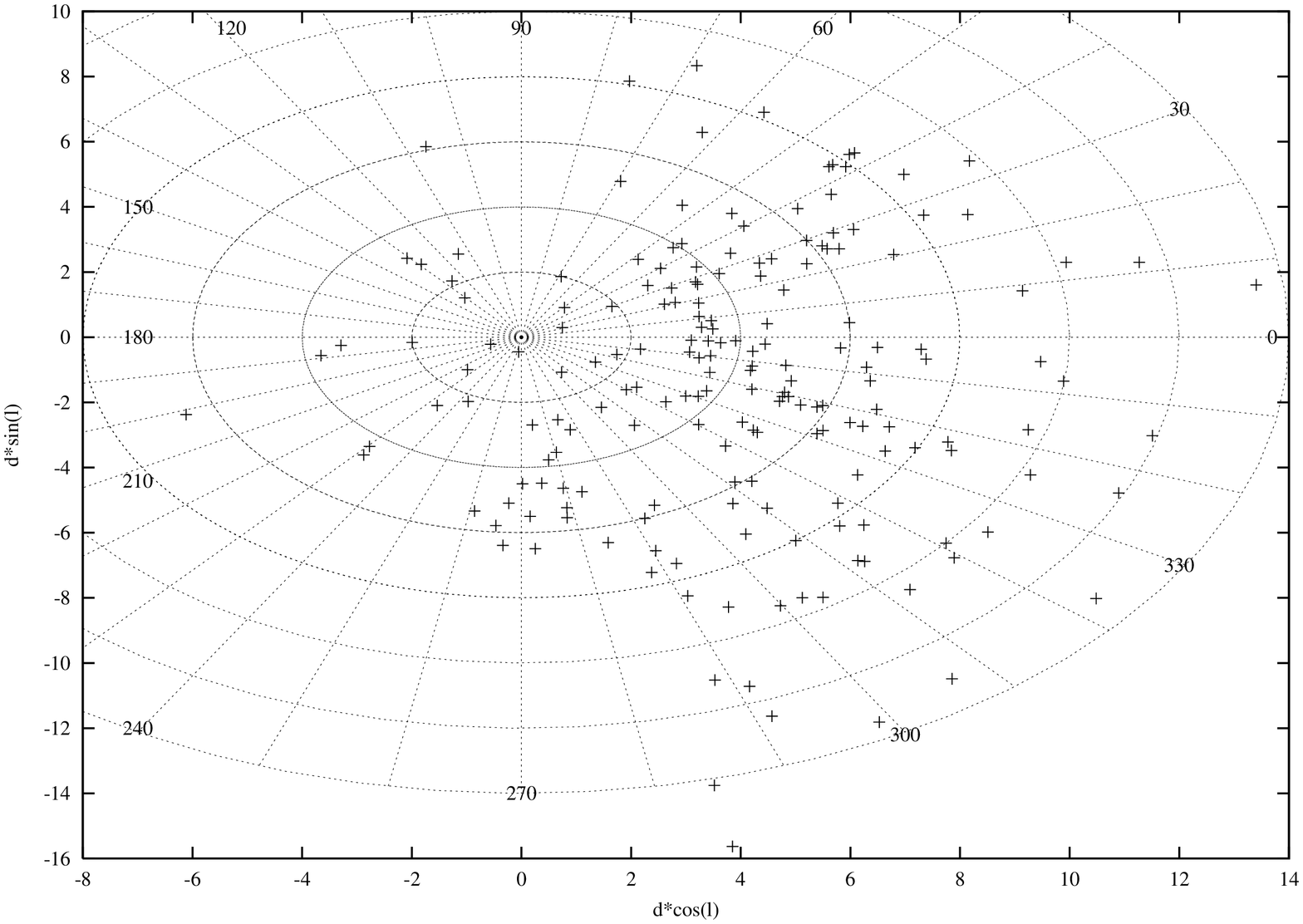,width=15cm,height=16.7cm,angle=-90}}
{Figure 3. Distance (d) as a function of Galactic longitude (l) for 189 
pulsars with age$<$5 10$^5$ years} 
\end{figure*}

\begin{figure*}
\centerline{\psfig{file=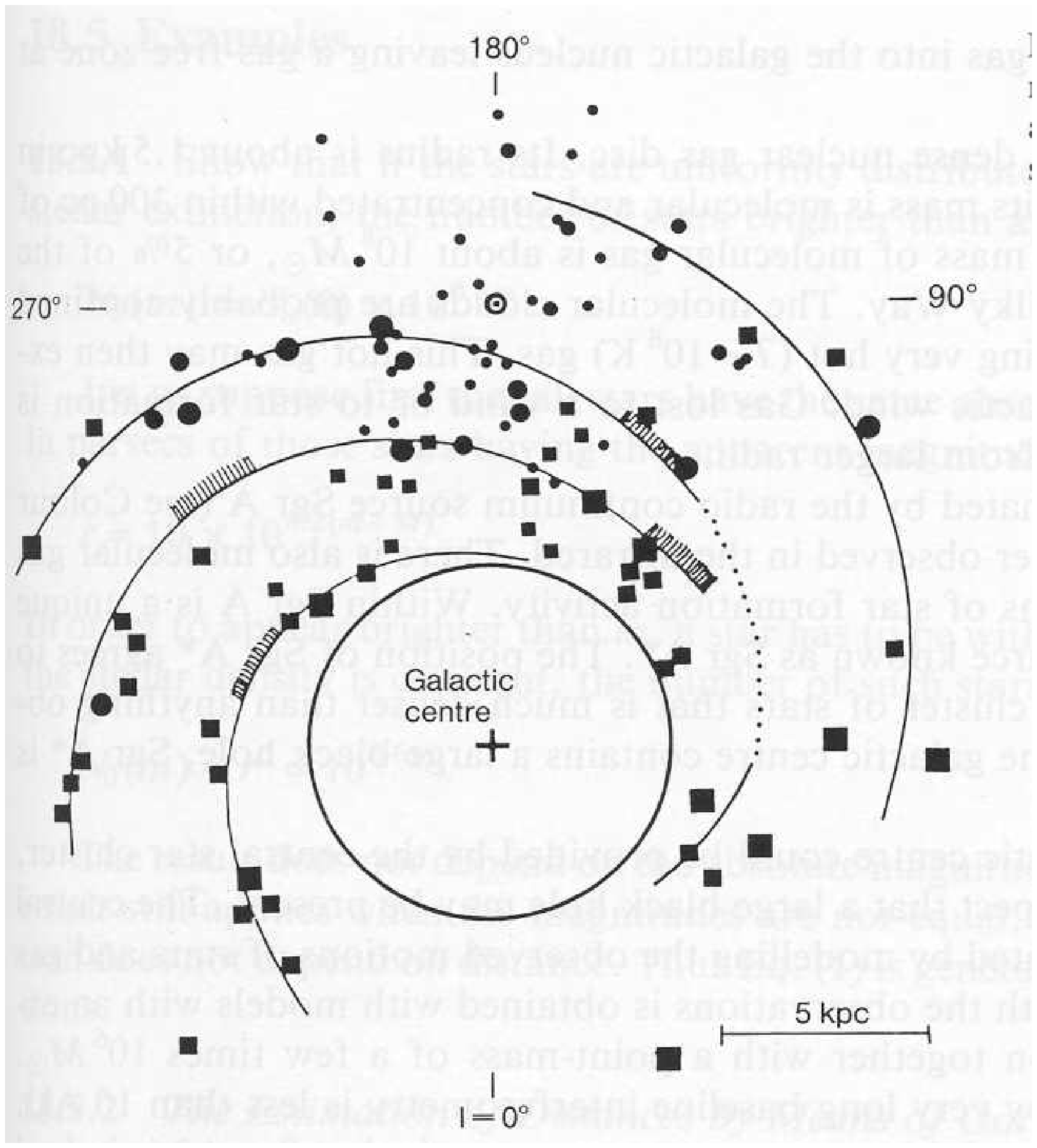,width=15cm,height=16.7cm}}
{Figure 4. Galactic arm structure constructed from the distribution of 
giant HII regions (Georgelin \& Georgelin 1976).} 
\end{figure*}

\begin{figure*}
\centerline{\psfig{file=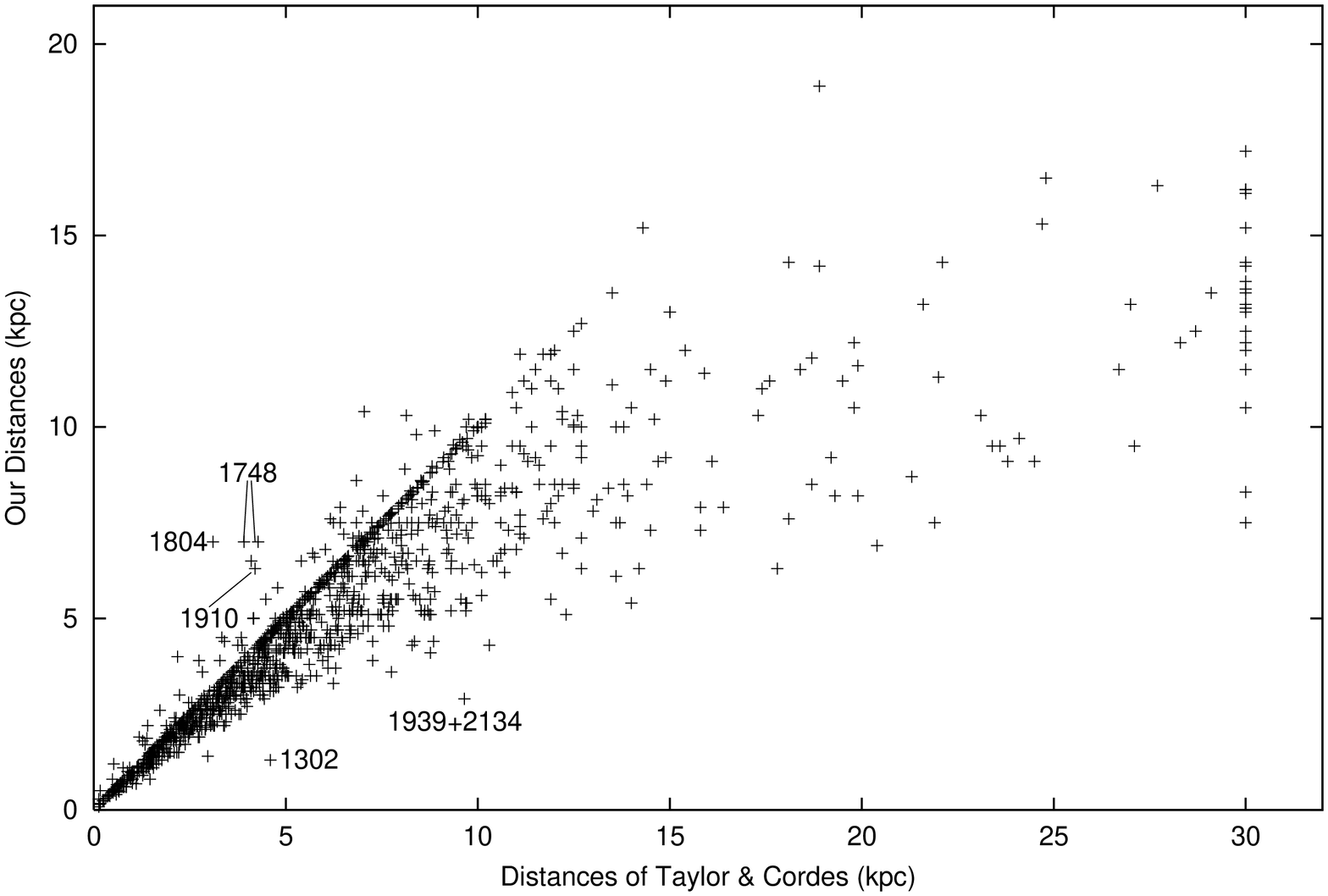,width=16cm,height=16cm,angle=-90}}
{Figure 5. Our distances vs. distances predicted by Taylor \& Cordes 
model for 1318 pulsars.} 
\end{figure*}

\begin{figure*}
\centerline{\psfig{file=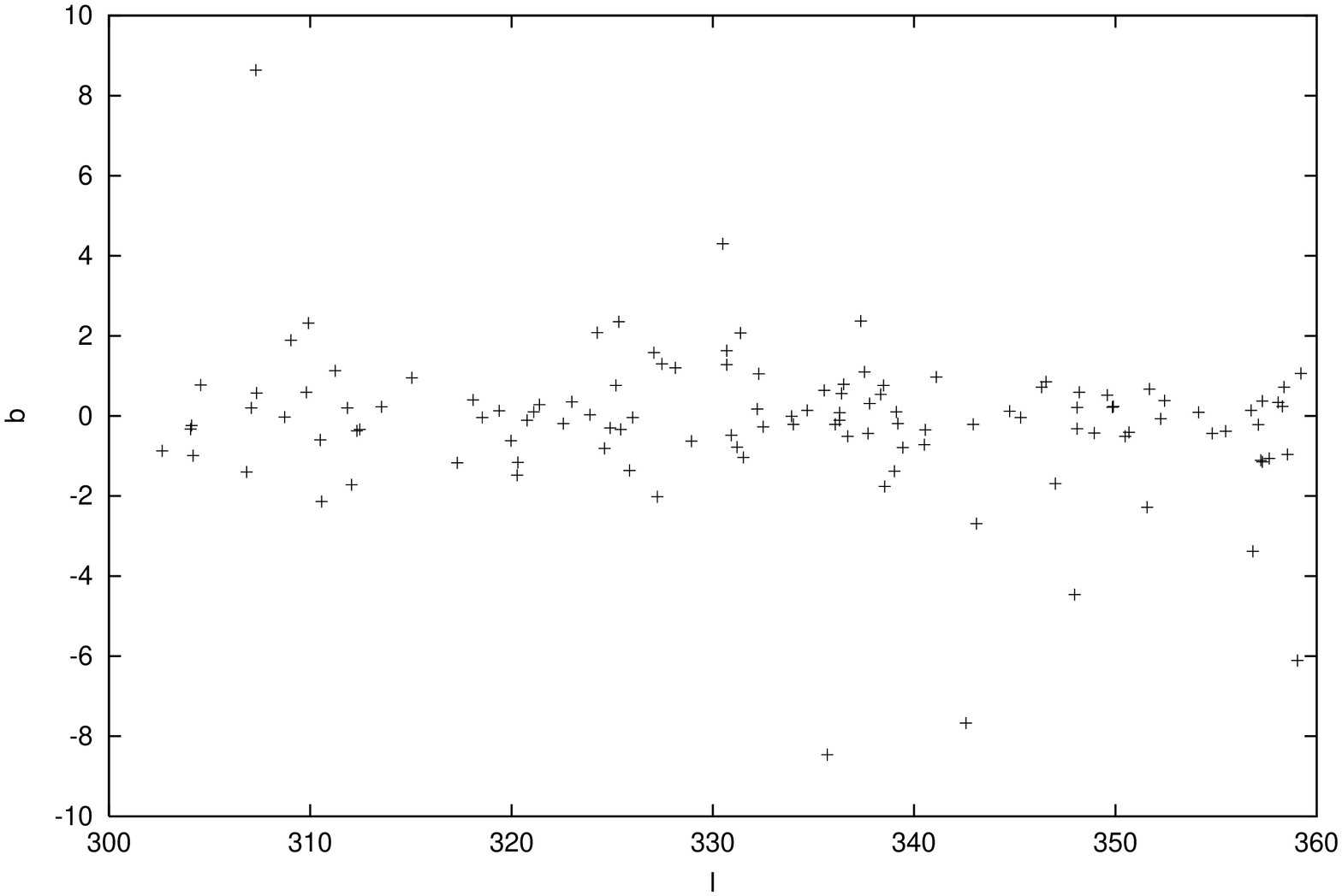,width=16cm,height=16cm,angle=-90}}
{Figure 6. Galactic longitude (l) vs. latitude (b) for pulsars within 
l=300$^o$-360$^o$ and younger than 10$^6$ years.} \end{figure*}

\begin{figure*}
\centerline{\psfig{file=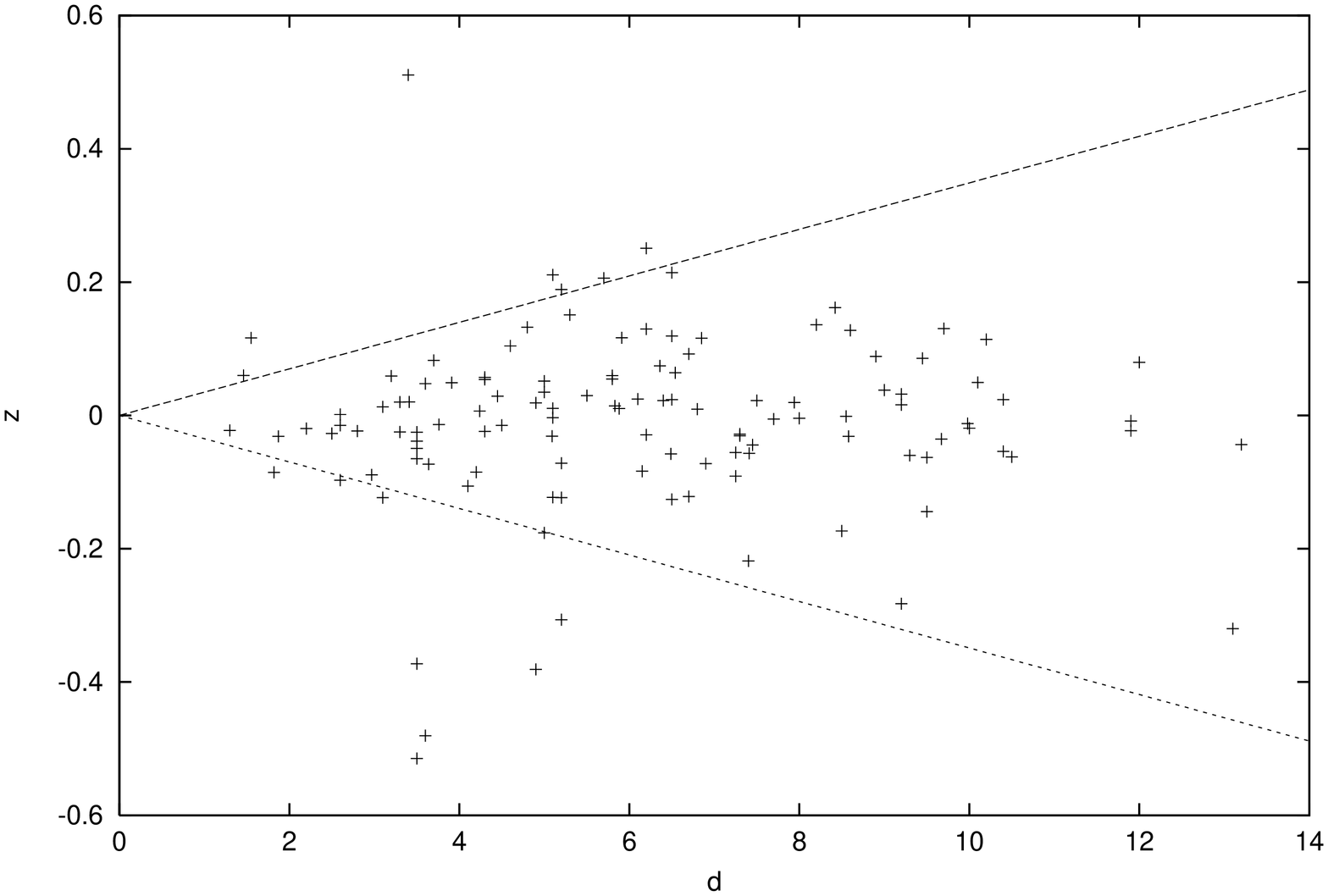,width=16cm,height=16cm,angle=-90}}
{Figure 7. Our distance values vs. Galactic height (z) for pulsars within 
l=300$^o$-360$^o$ and younger than 10$^6$ years.} 
\end{figure*}

\begin{figure*}
\centerline{\psfig{file=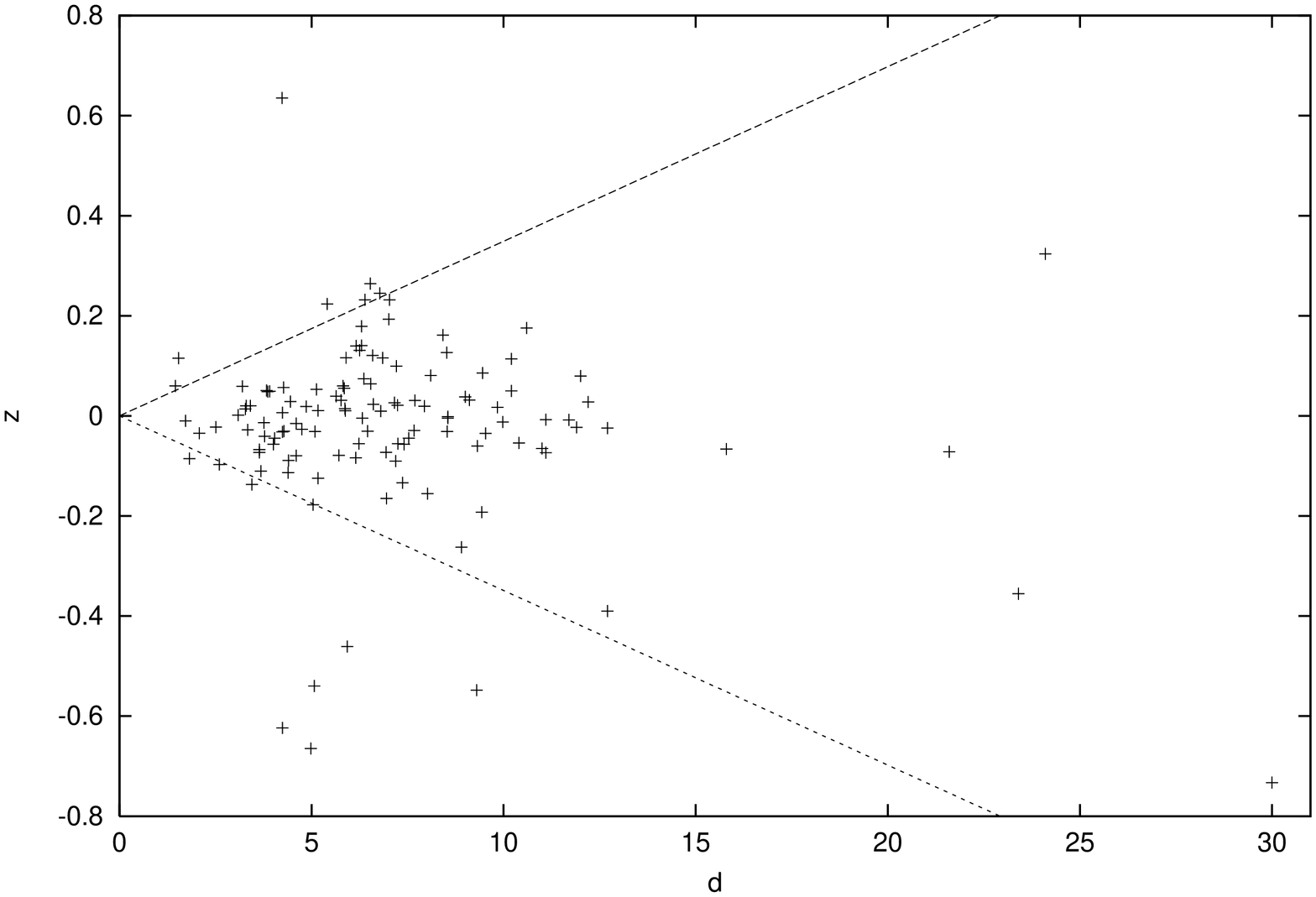,width=16cm,height=16cm,angle=-90}}
{Figure 8. Distances calculated from Taylor \& Cordes model vs. Galactic 
height (z) for pulsars within l=300$^o$-360$^o$ and younger than 10$^6$ 
years.} \end{figure*}

\begin{figure*}
\centerline{\psfig{file=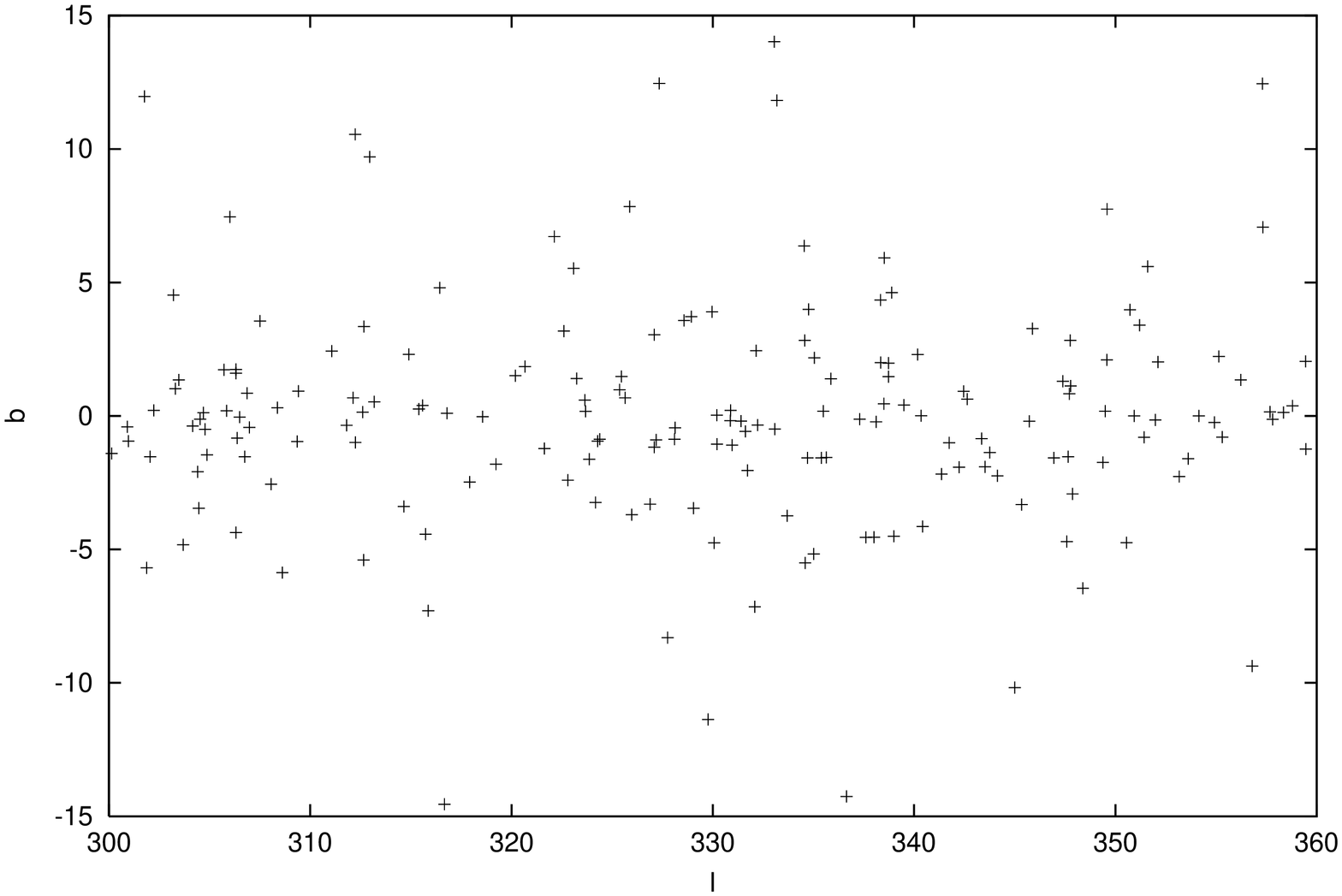,width=16cm,height=16cm,angle=-90}}
{Figure 9. Galactic distribution of pulsars with l=300$^o$-360$^o$ and 
$|$b$|<$15$^o$ and with 10$^6<$10$^7$ years.} 
\end{figure*}

\begin{figure*}
\centerline{\psfig{file=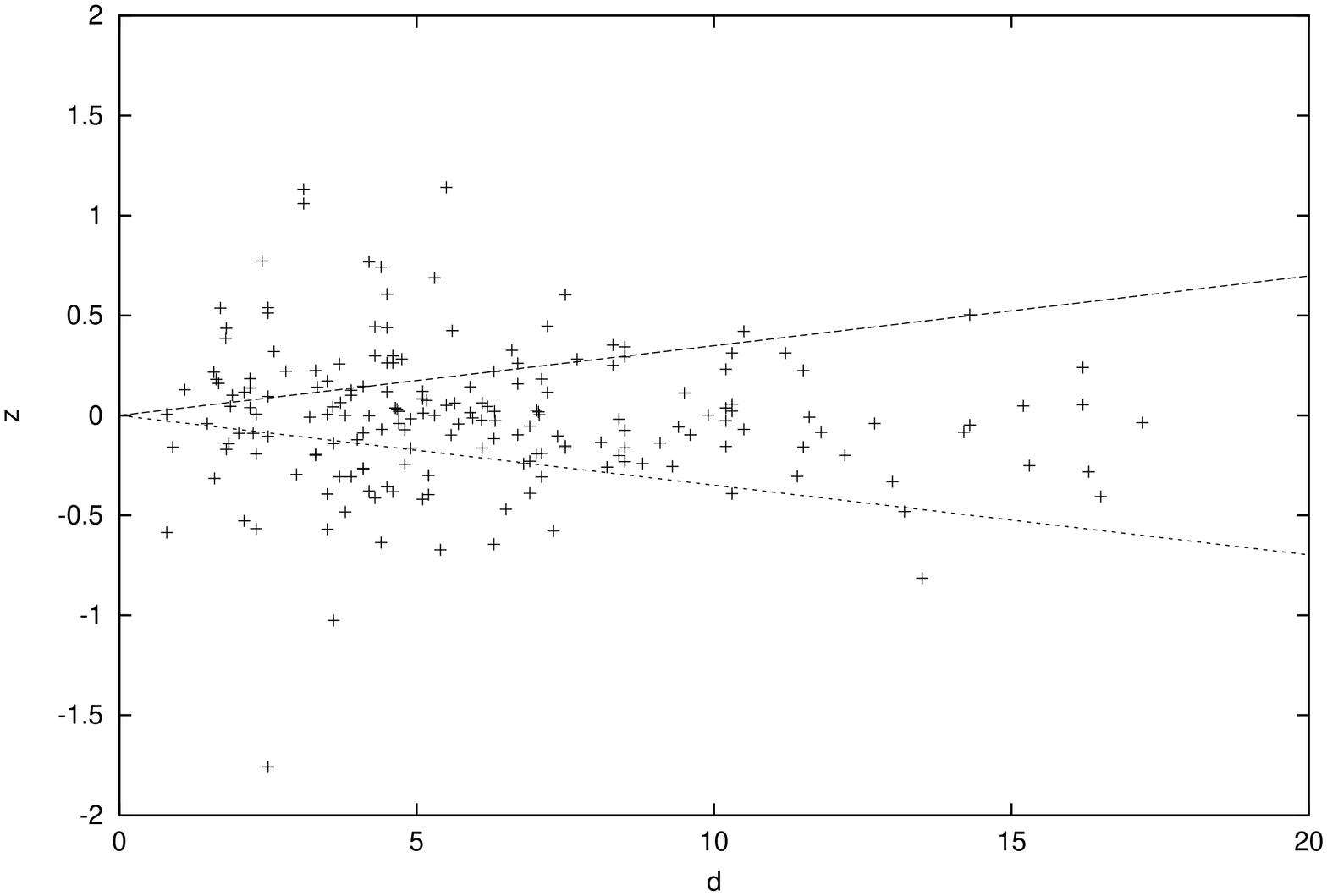,width=16cm,height=16cm,angle=-90}}
{Figure 10. Our distance values vs. Galactic height (z) for pulsars with 
l=300$^o$-360$^o$ which have ages of 10$^6$-10$^7$ years.} \end{figure*}

\begin{figure*}
\centerline{\psfig{file=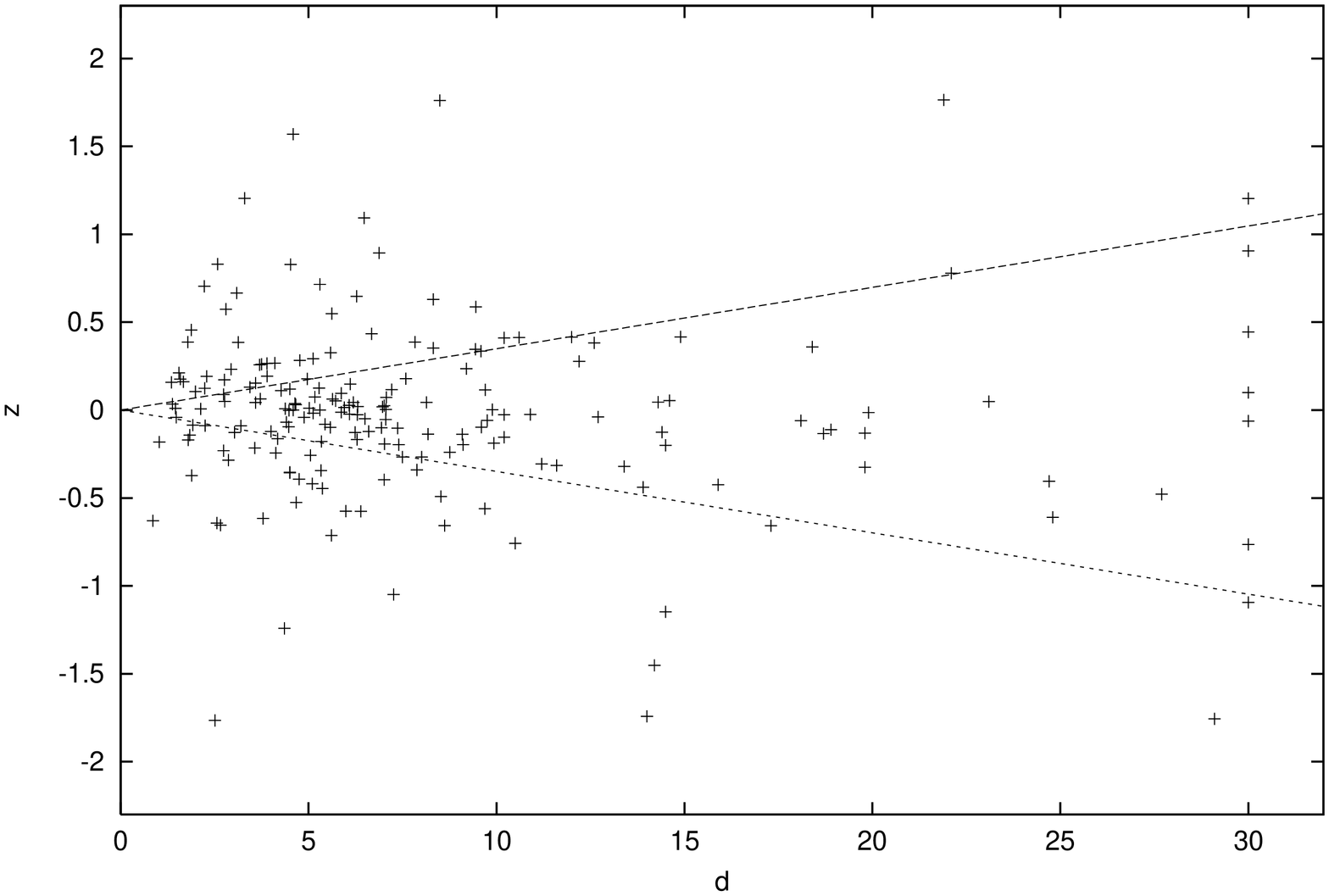,width=16cm,height=16cm,angle=-90}}
{Figure 11. Distance values from Taylor \& Cordes model for pulsars with 
l=300$^o$-360$^o$ which have ages of 10$^6$-10$^7$ years.} 
\end{figure*}

\begin{figure*}
\begin{tabular}{cc}
\psfig{file=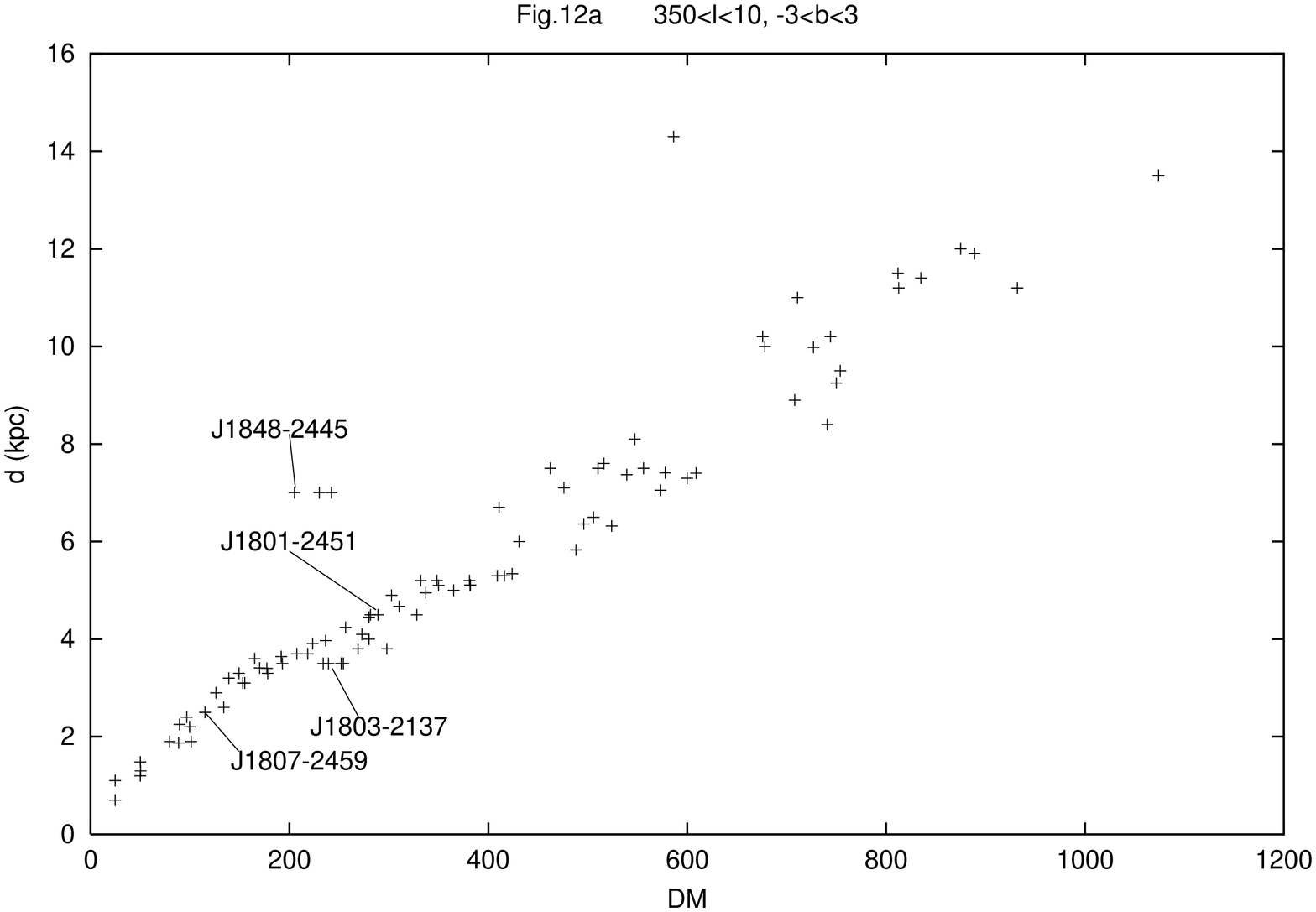,width=8cm,height=8cm,angle=-90} &
\psfig{file=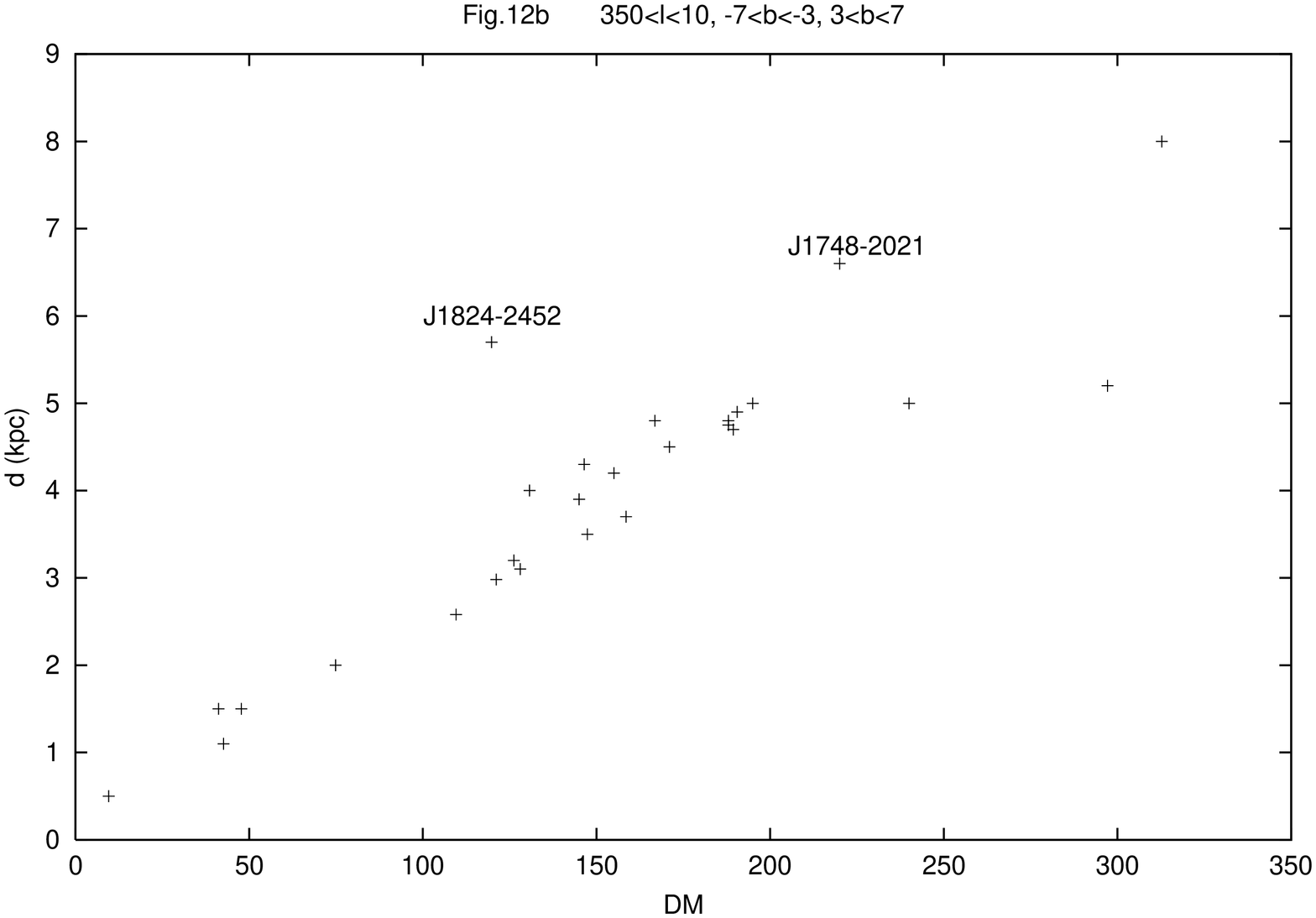,width=8cm,height=8cm,angle=-90} \\
\psfig{file=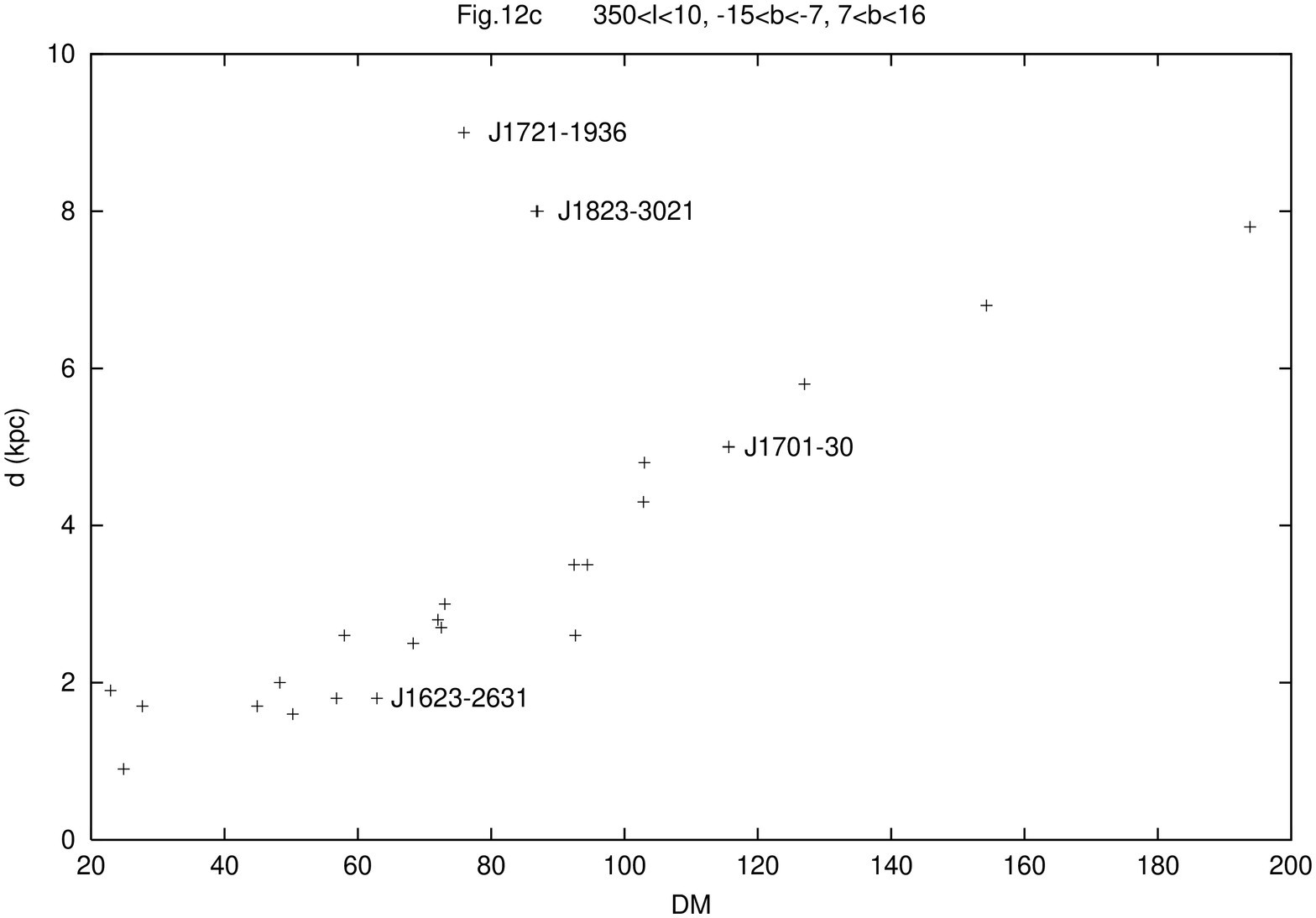,width=8cm,height=8cm,angle=-90} &
\psfig{file=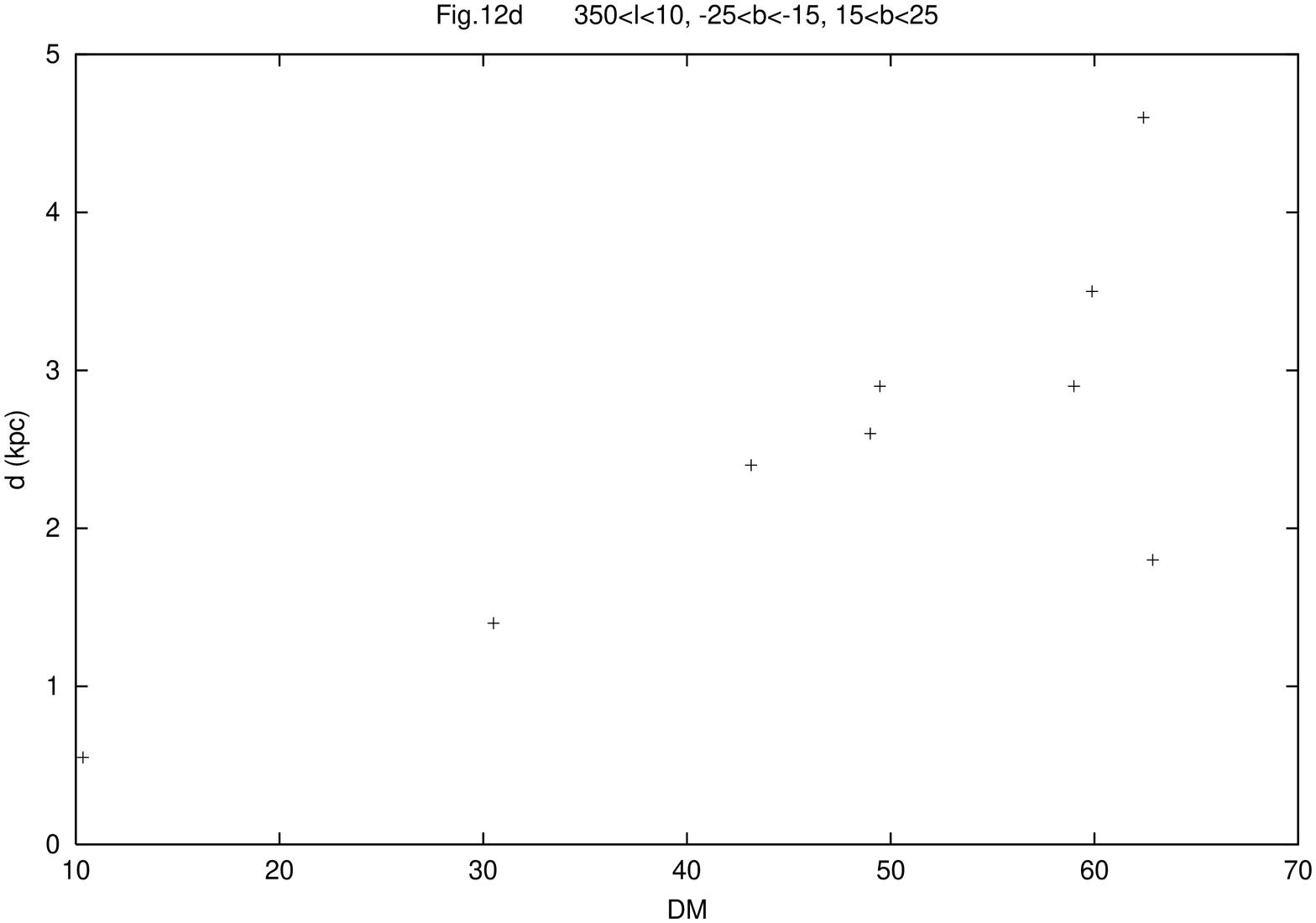,width=8cm,height=8cm,angle=-90} \\
\end{tabular}
\end{figure*}

\begin{figure*}
\begin{tabular}{cc}
\psfig{file=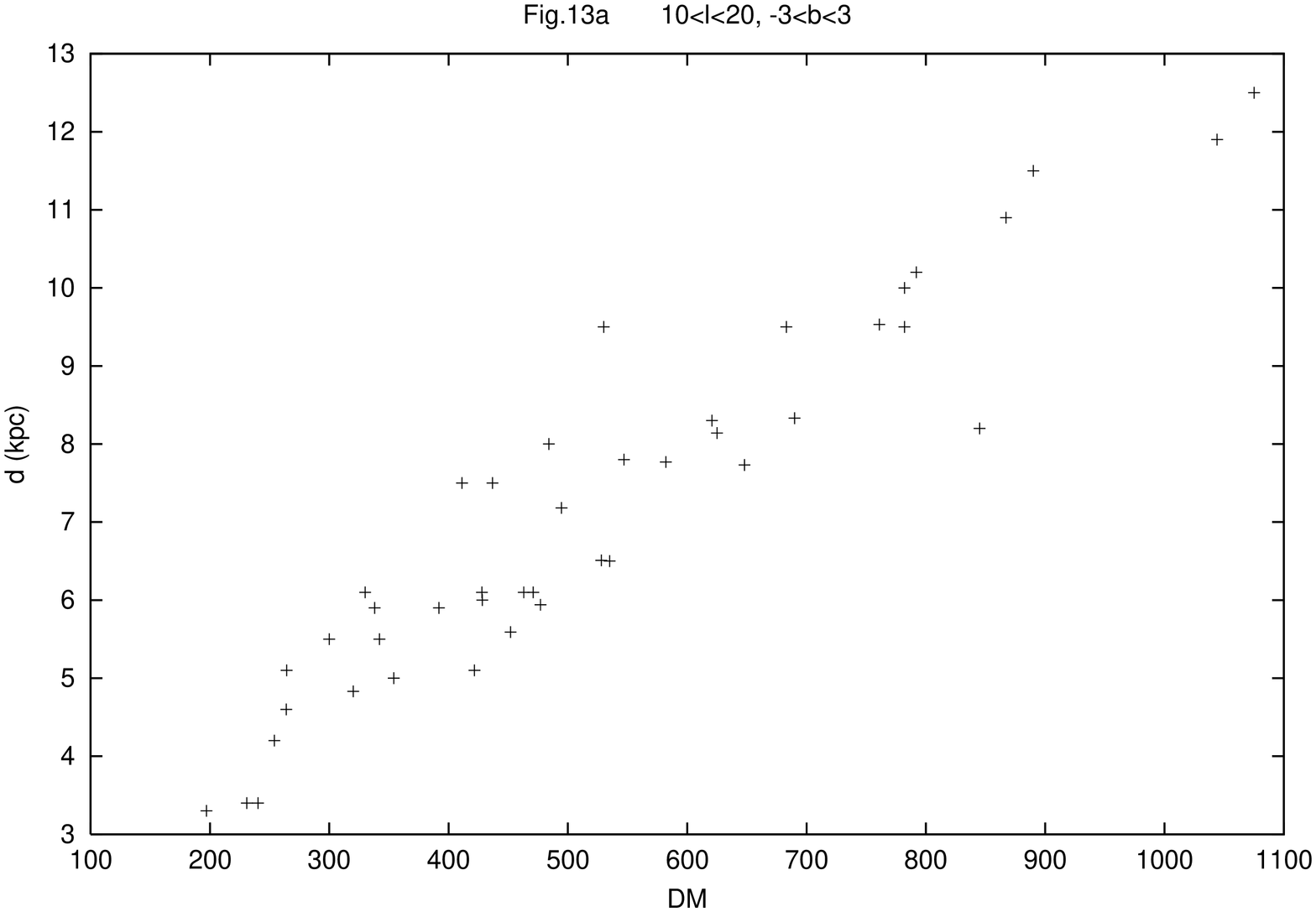,width=8cm,height=8cm,angle=-90} &
\psfig{file=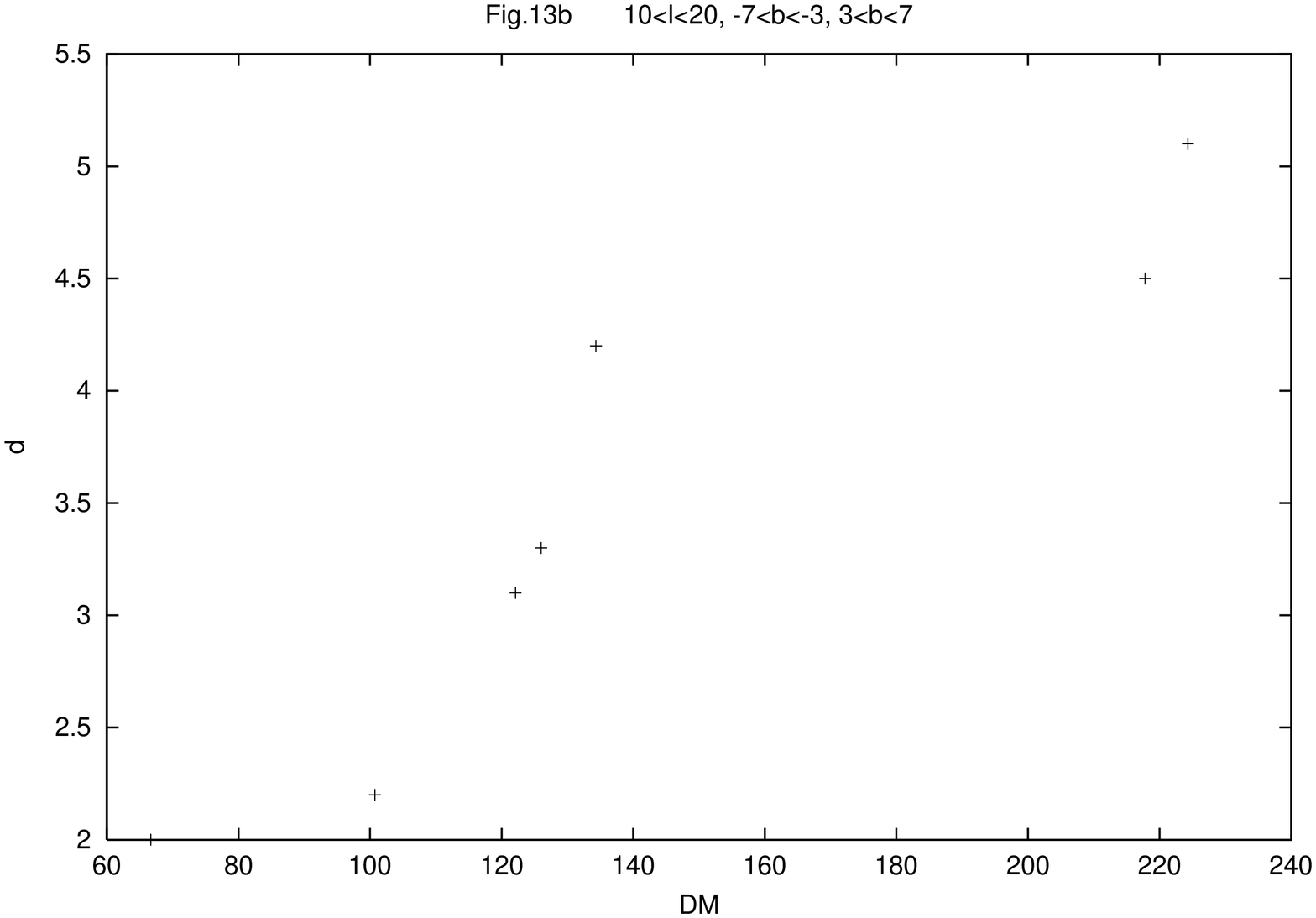,width=8cm,height=8cm,angle=-90} \\
\psfig{file=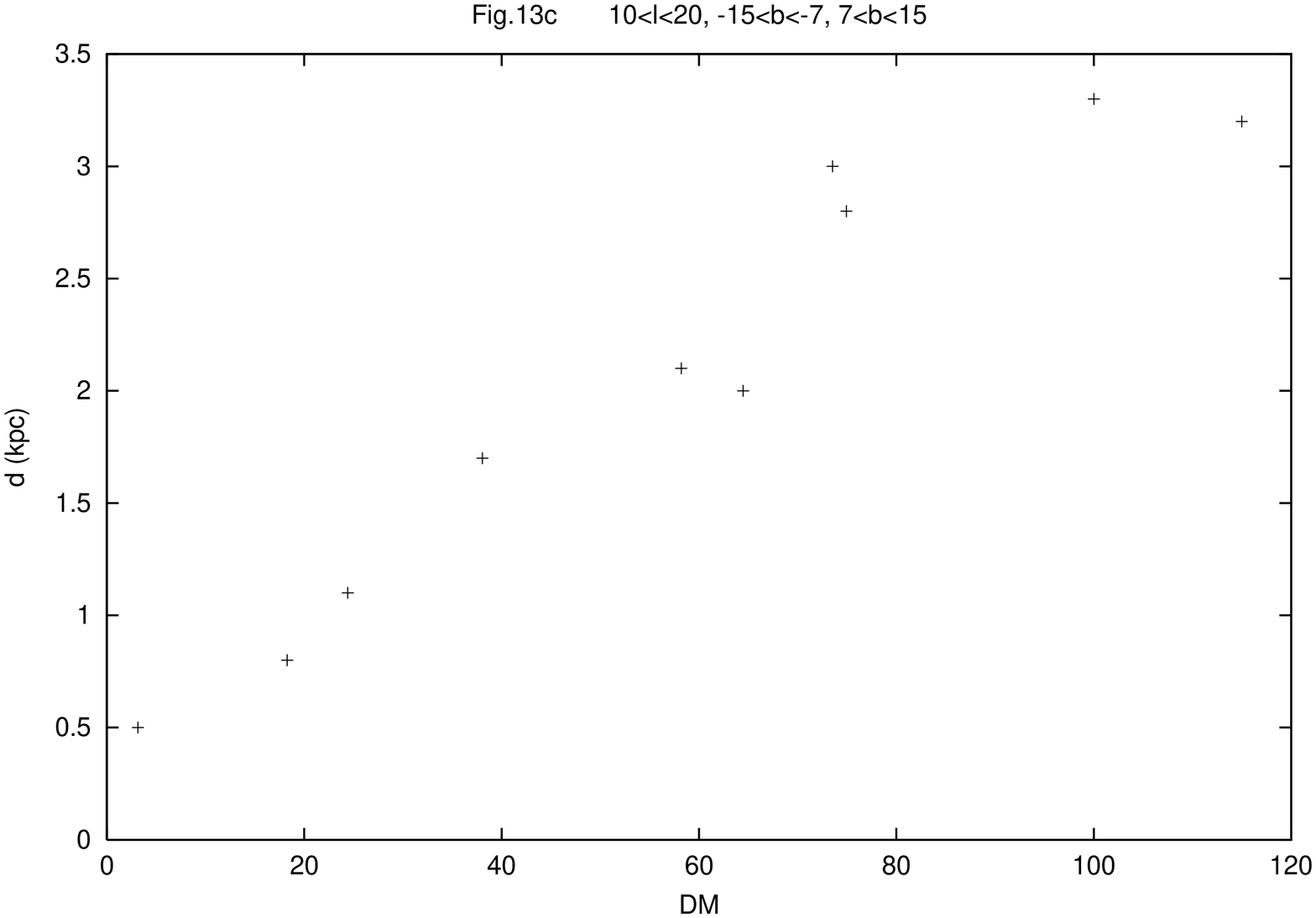,width=8cm,height=8cm,angle=-90} &
\psfig{file=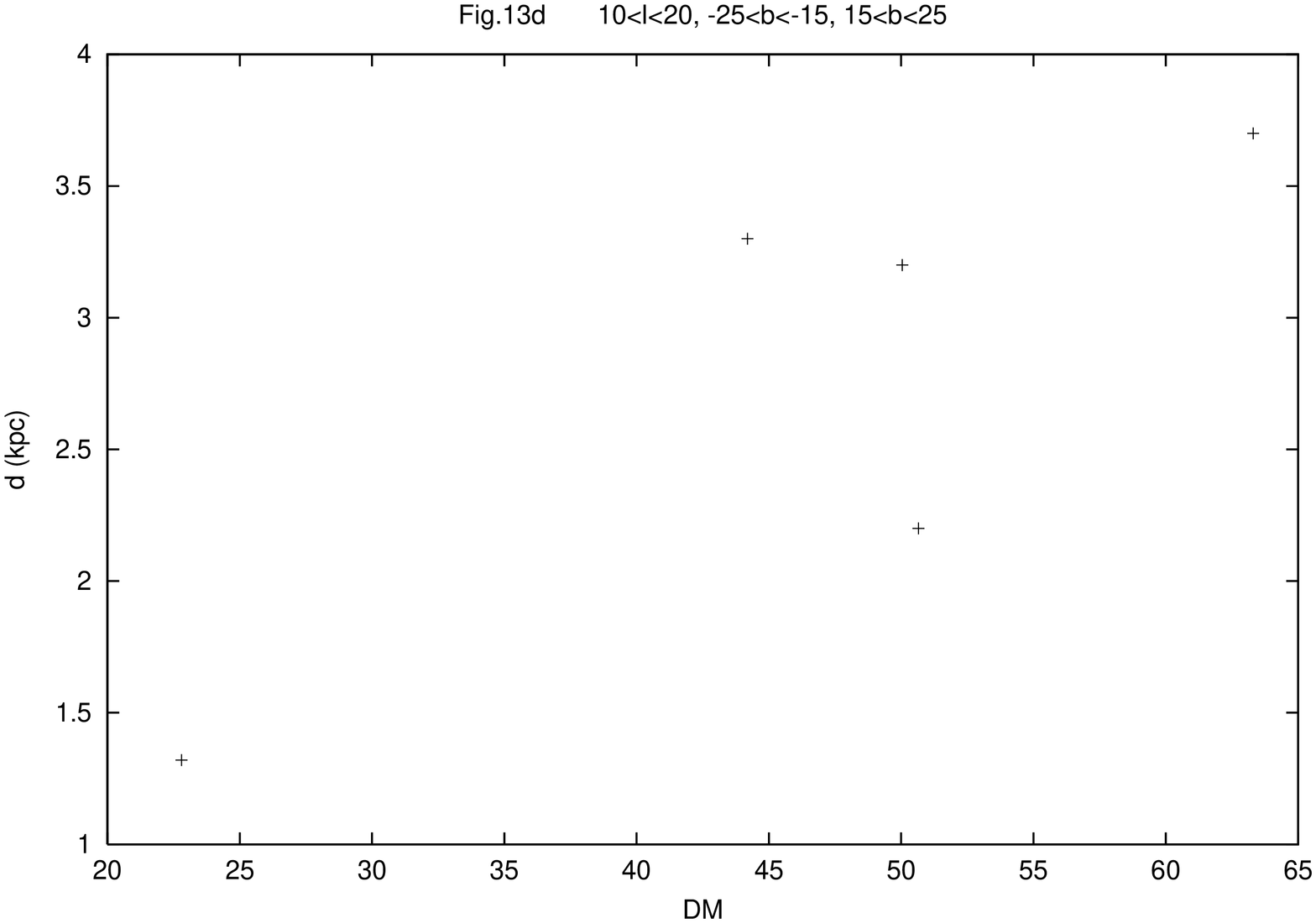,width=8cm,height=8cm,angle=-90} \\
\end{tabular}
\end{figure*}  
\begin{figure*}
\begin{tabular}{cc}
\psfig{file=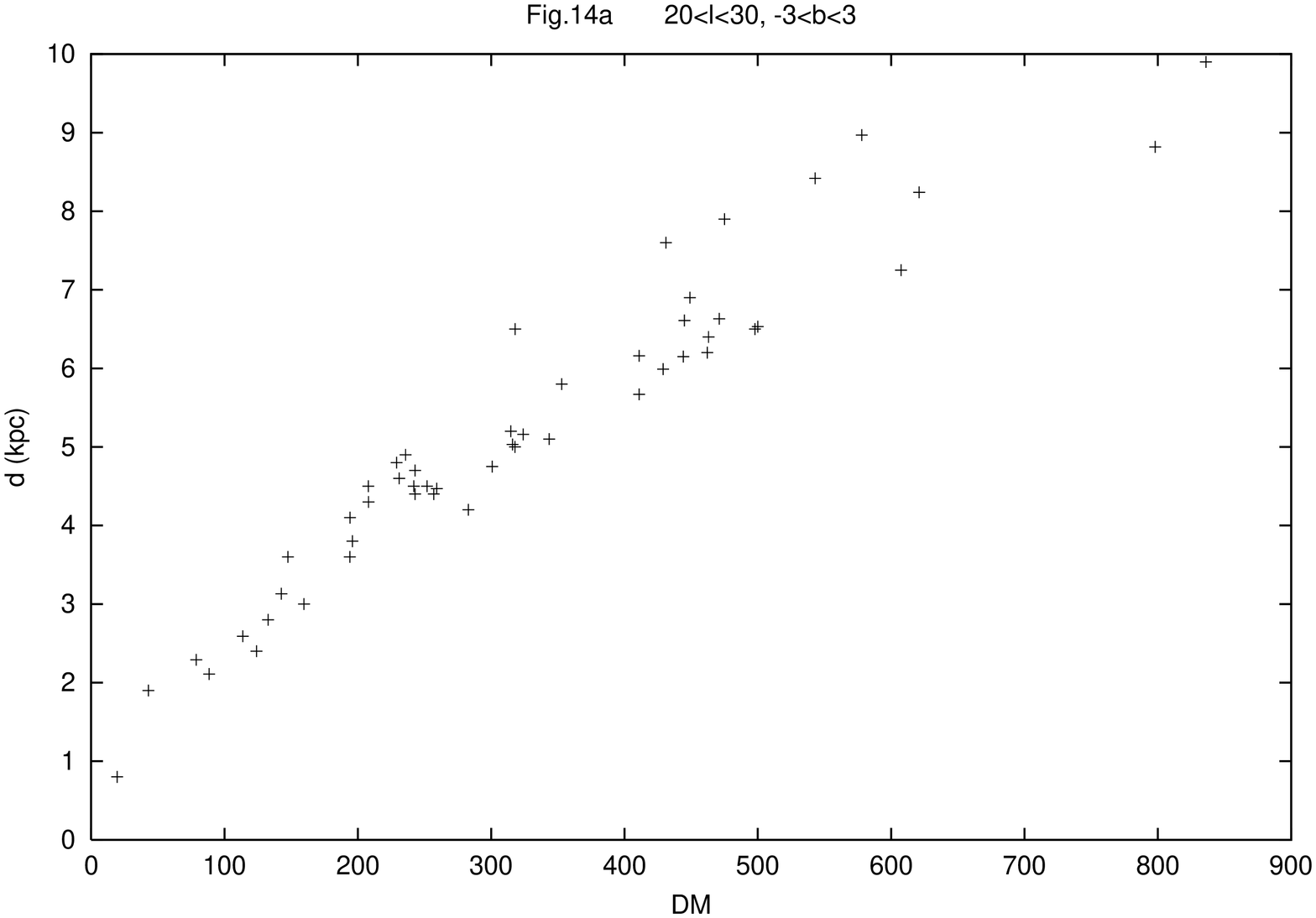,width=8cm,height=8cm,angle=-90} &
\psfig{file=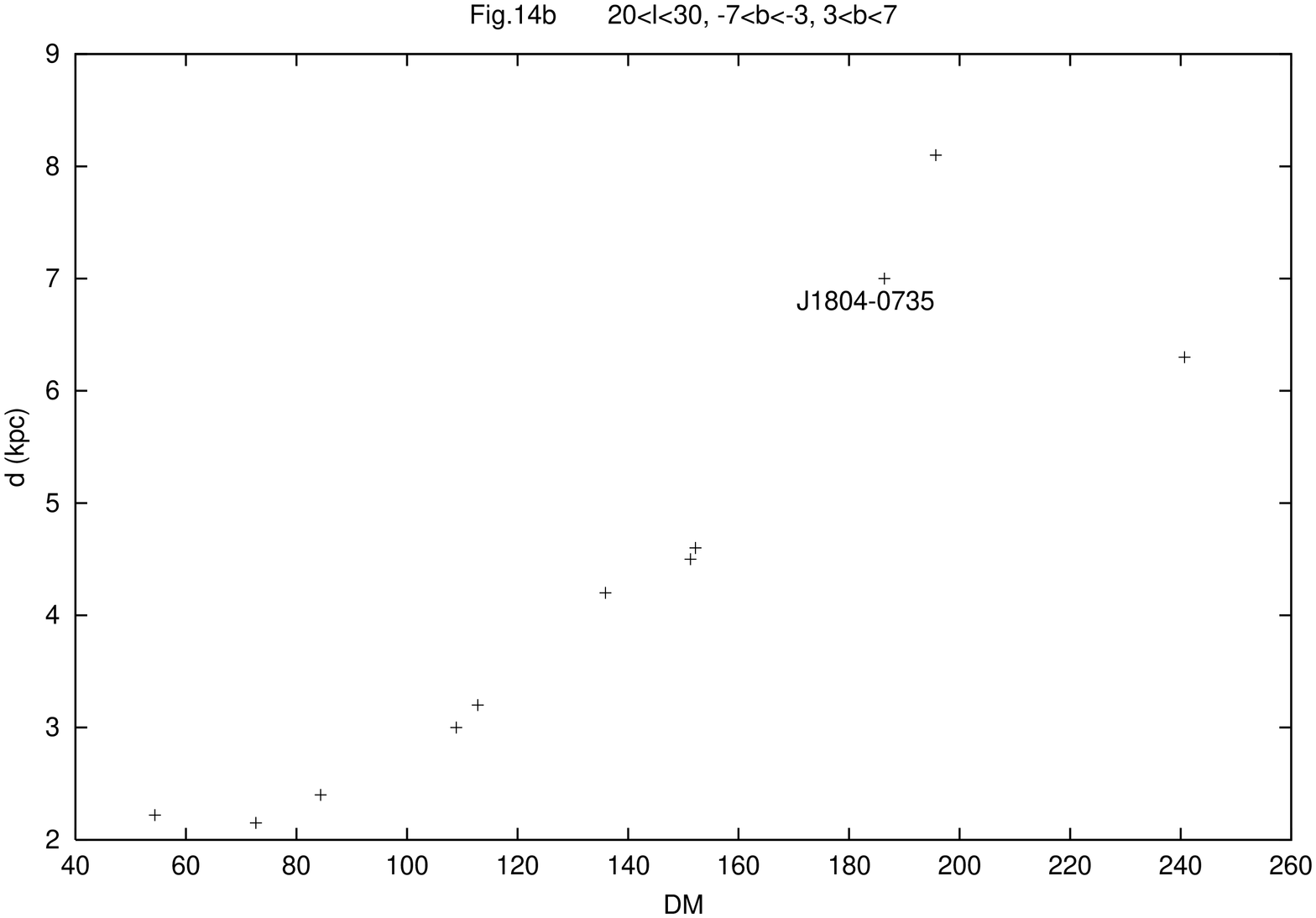,width=8cm,height=8cm,angle=-90} \\
\psfig{file=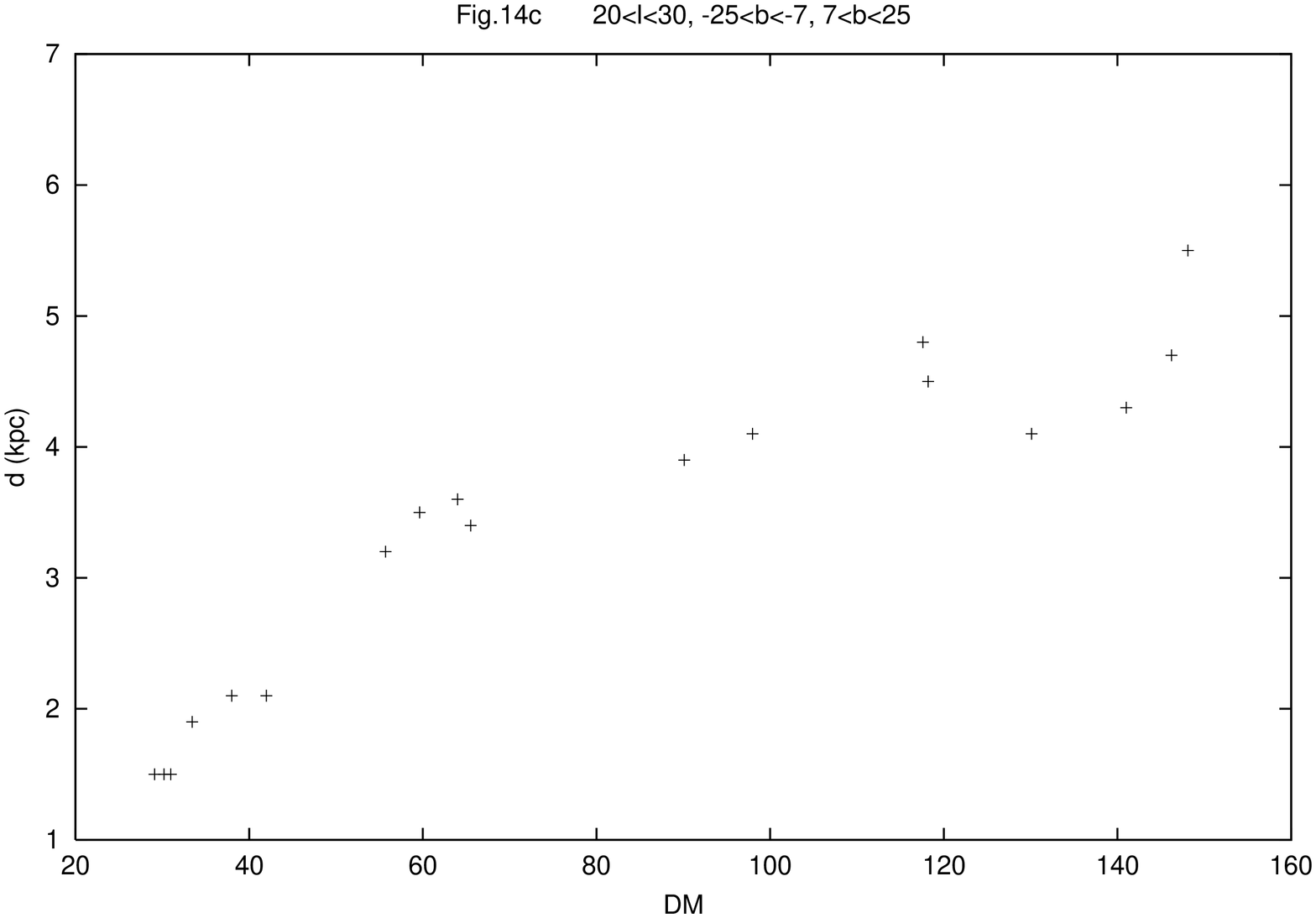,width=8cm,height=8cm,angle=-90} \\
\end{tabular}
\end{figure*}
\begin{figure*}
\begin{tabular}{cc}
\psfig{file=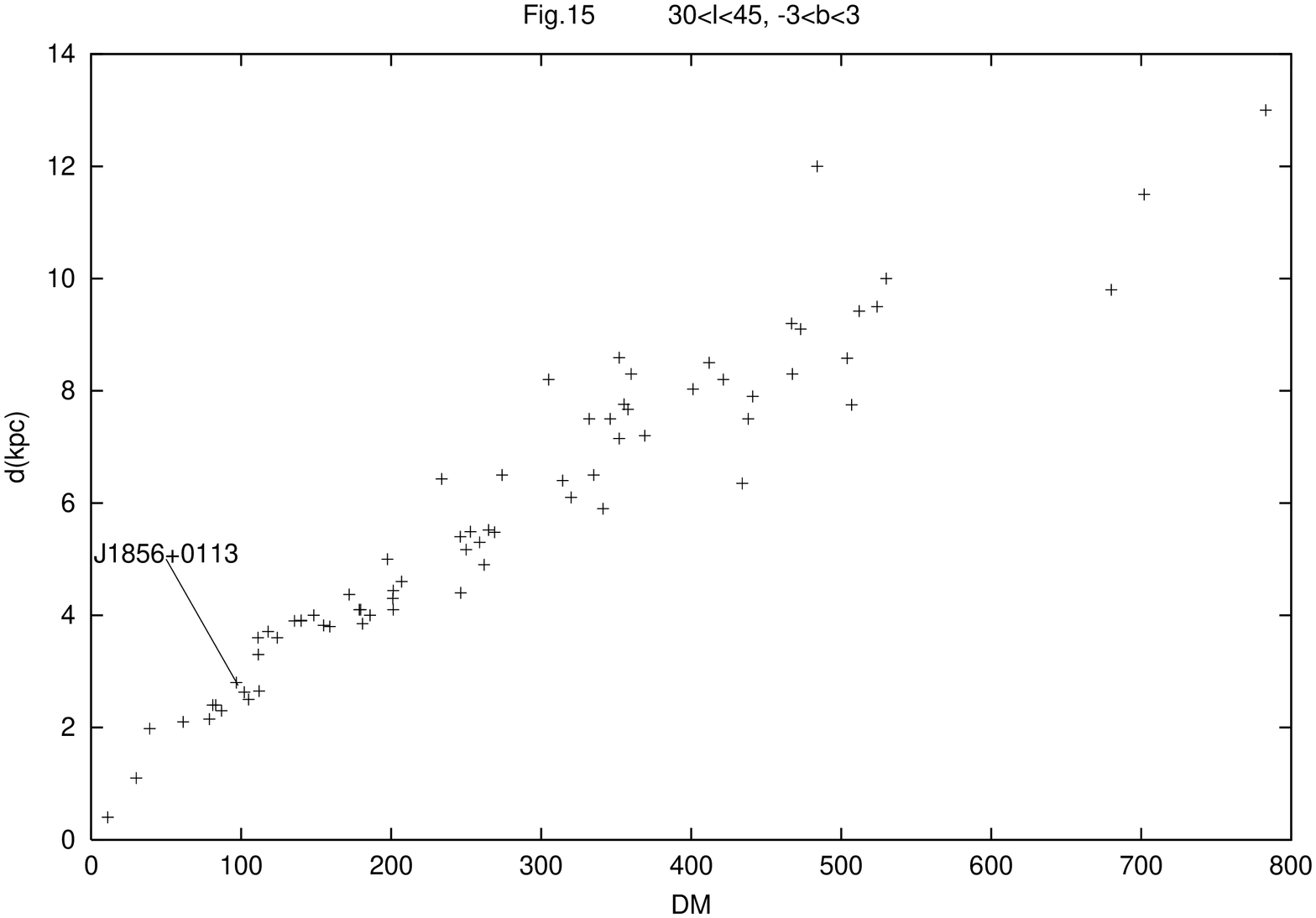,width=8cm,height=5cm,angle=-90} & 
\psfig{file=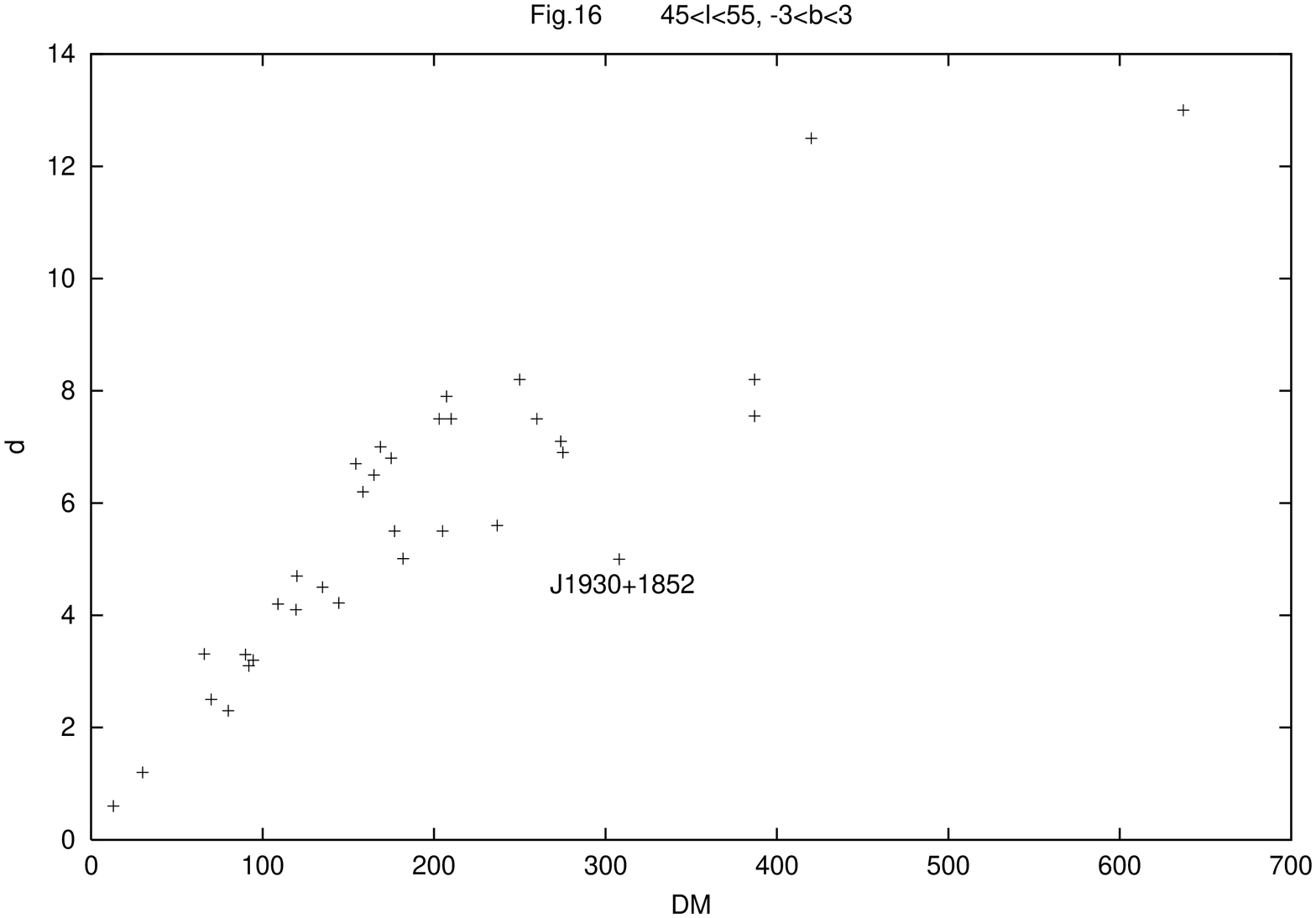,width=8cm,height=5cm,angle=-90} \\
\psfig{file=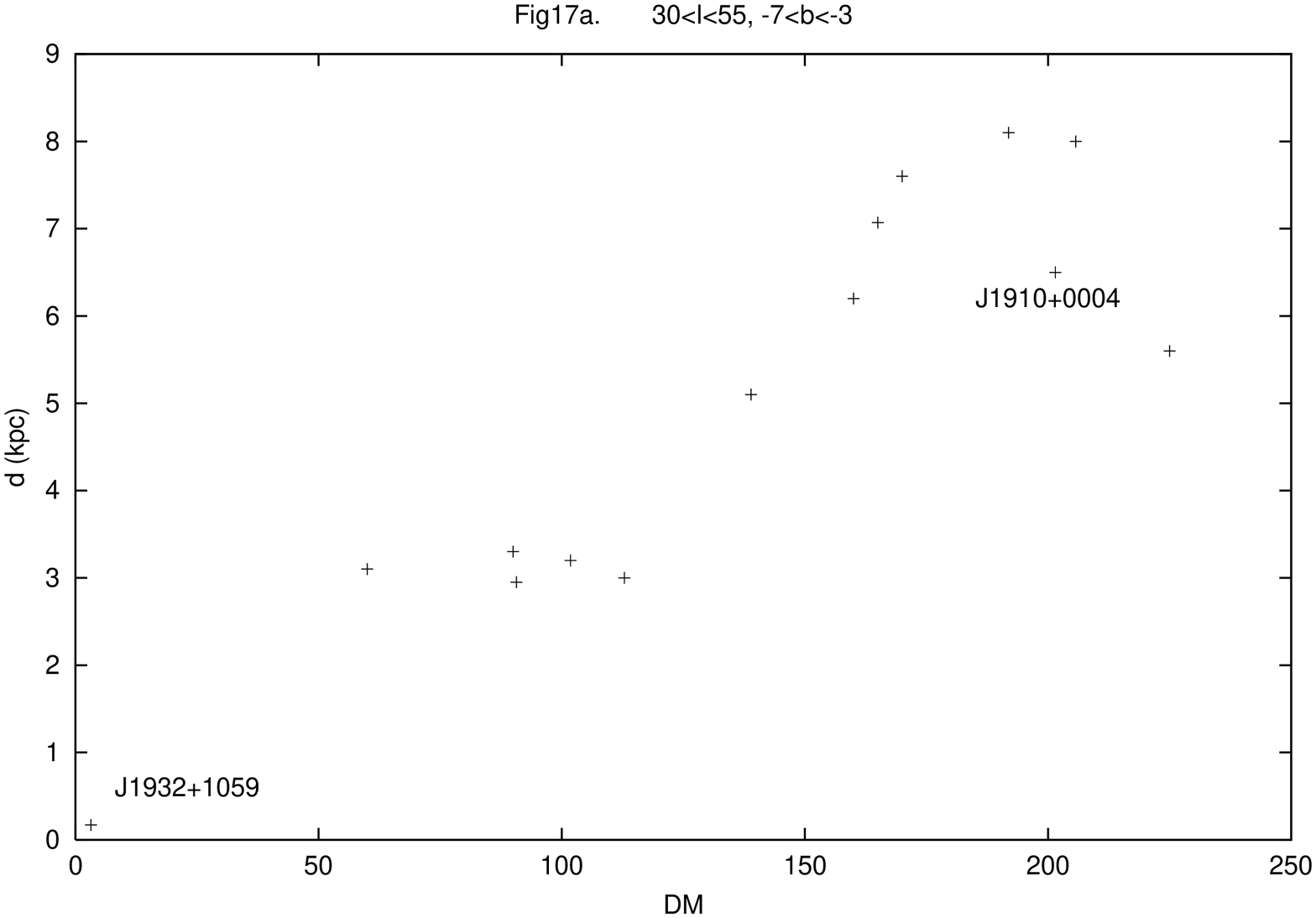,width=8cm,height=5cm,angle=-90} &
\psfig{file=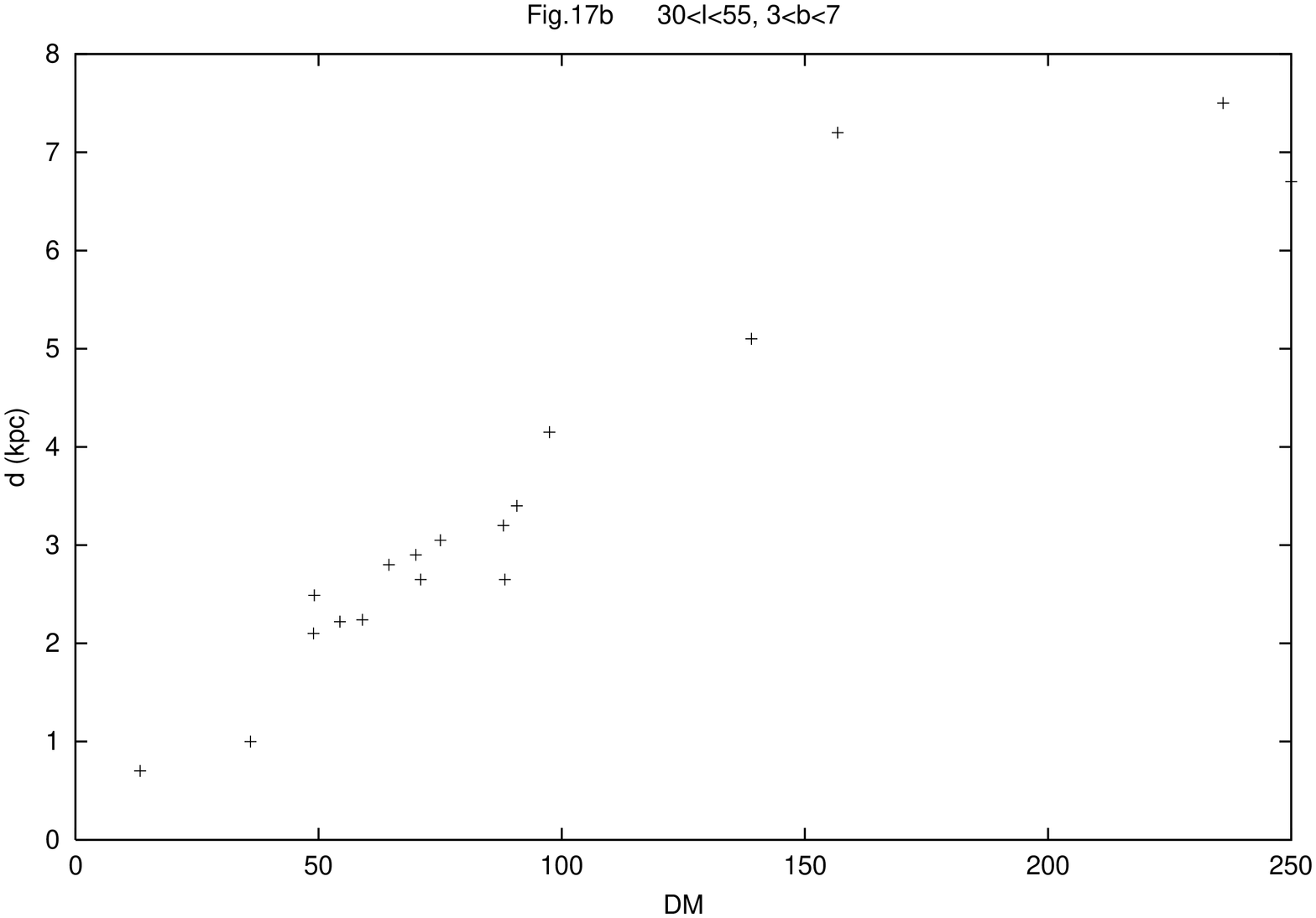,width=8cm,height=5cm,angle=-90} \\
\psfig{file=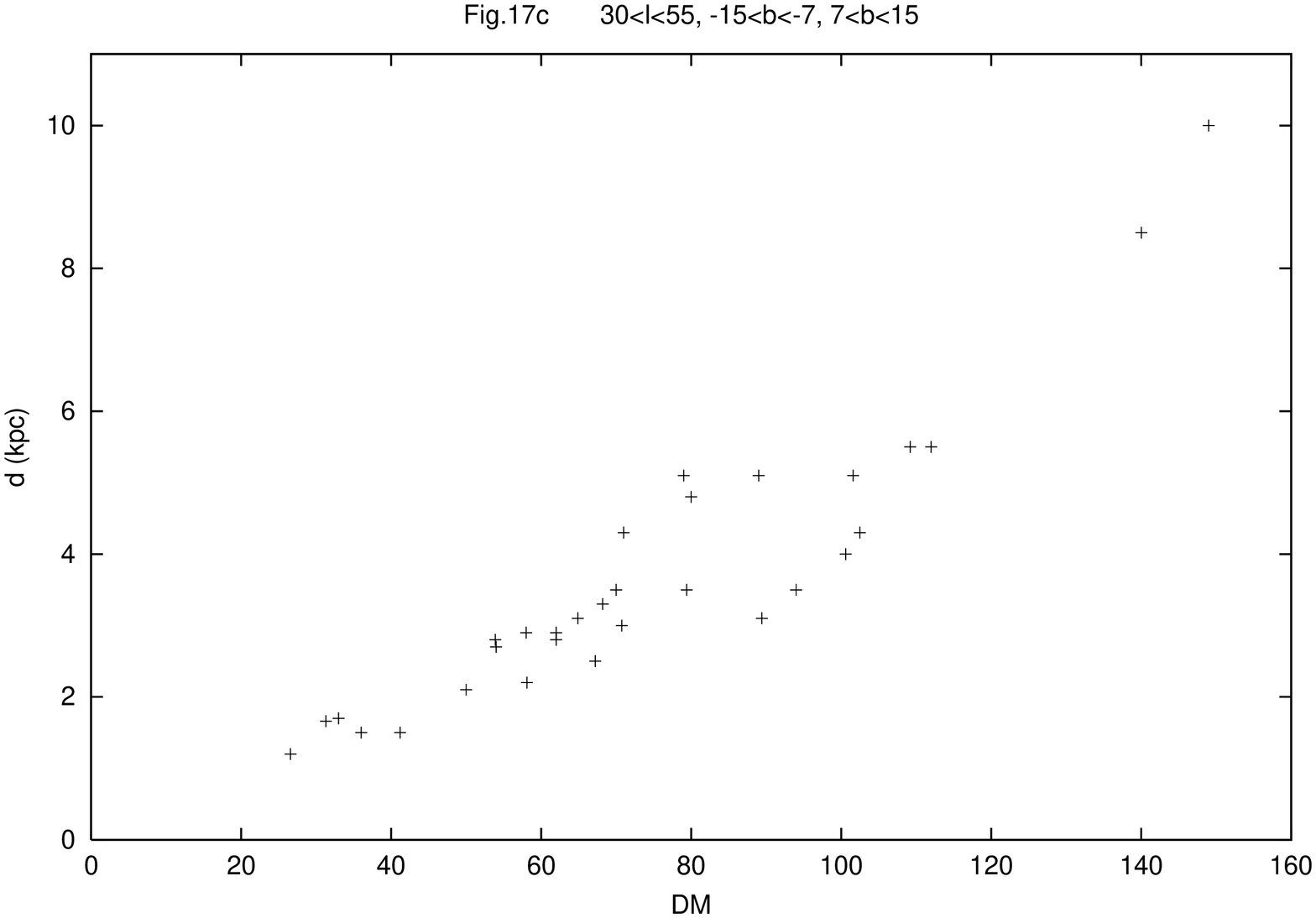,width=8cm,height=5cm,angle=-90} &
\psfig{file=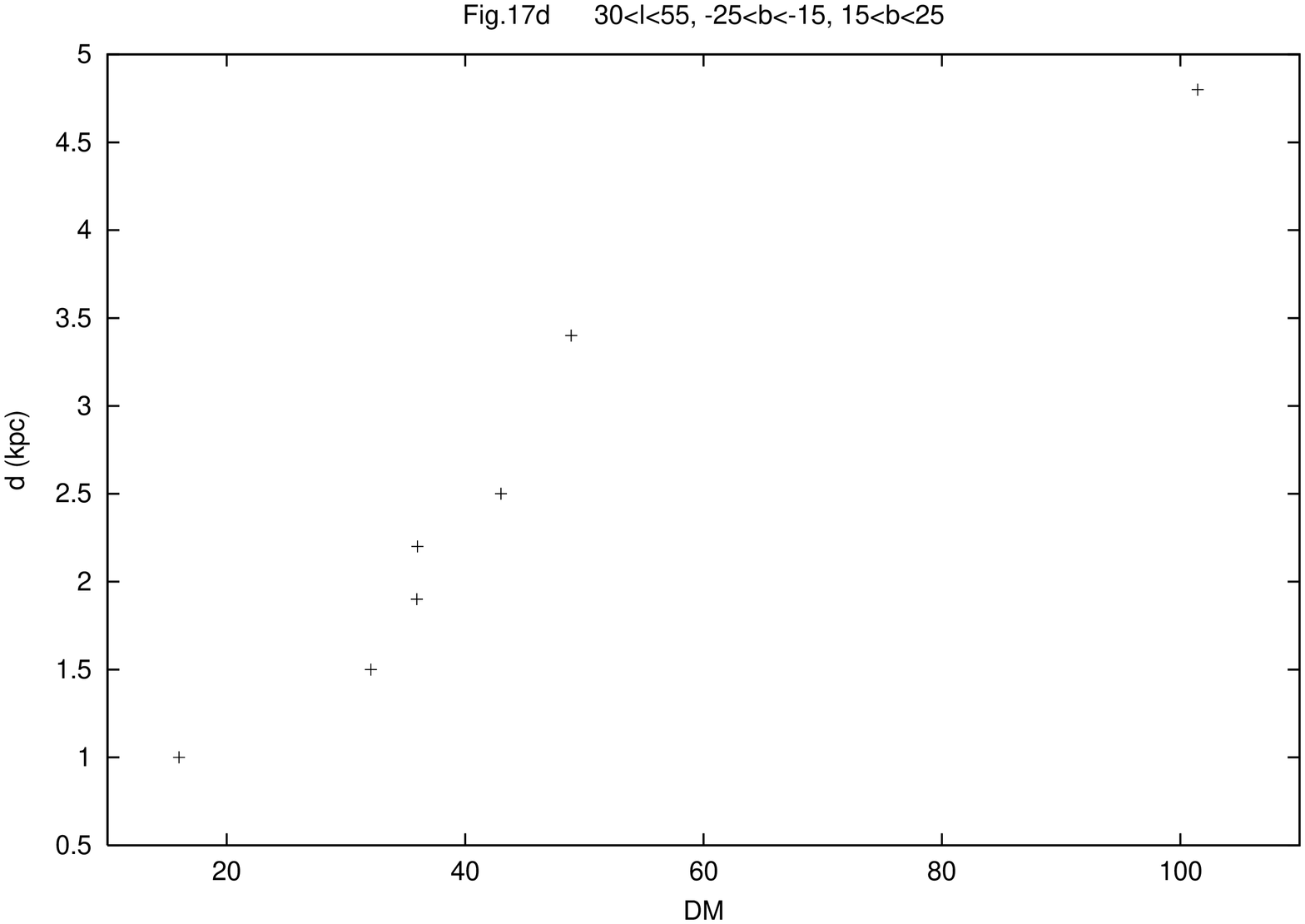,width=8cm,height=5cm,angle=-90} \\
\end{tabular}
\end{figure*}
\begin{figure*}
\begin{tabular}{cc}
\psfig{file=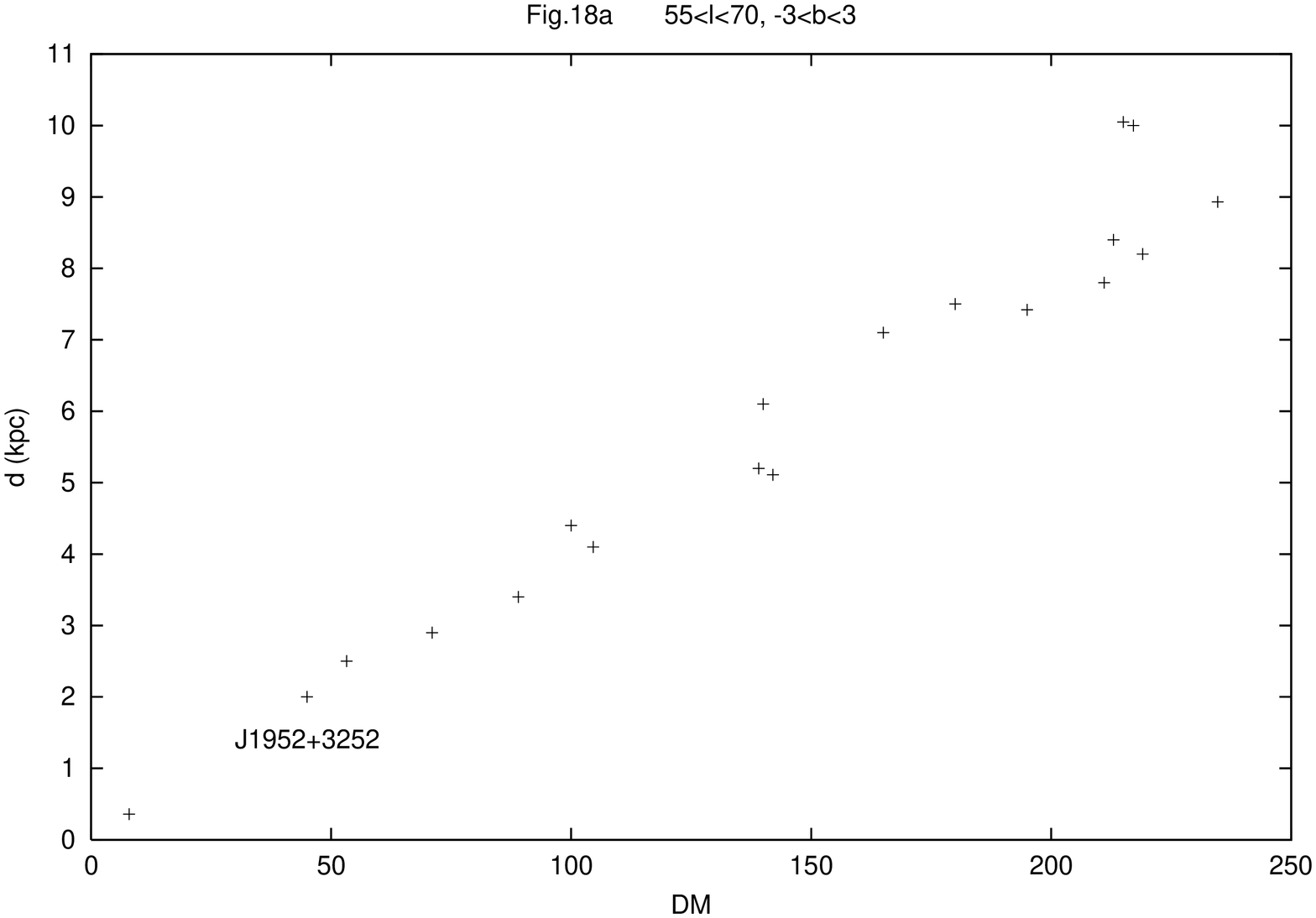,width=8cm,height=8cm,angle=-90} & 
\psfig{file=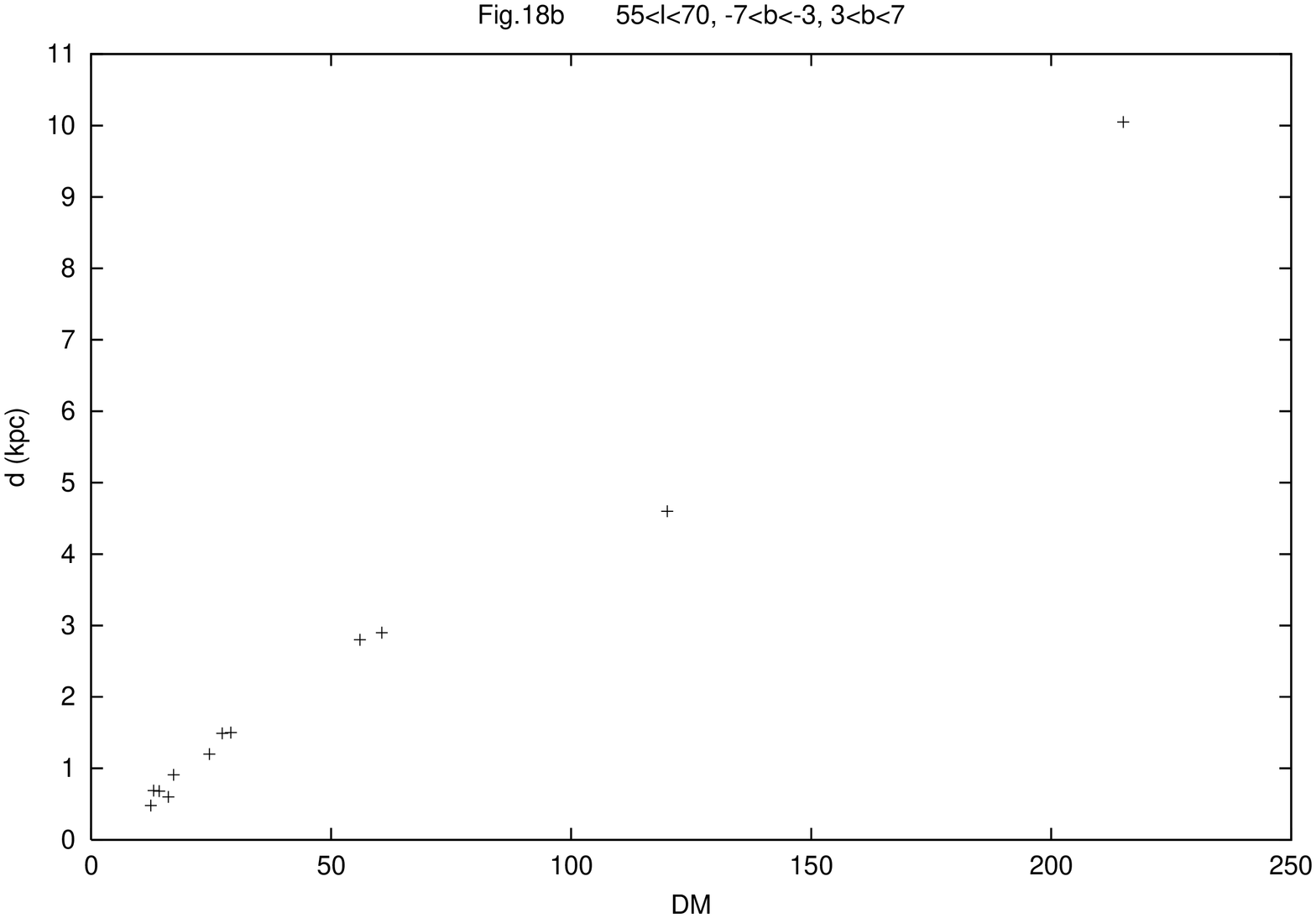,width=8cm,height=8cm,angle=-90} \\ 
\psfig{file=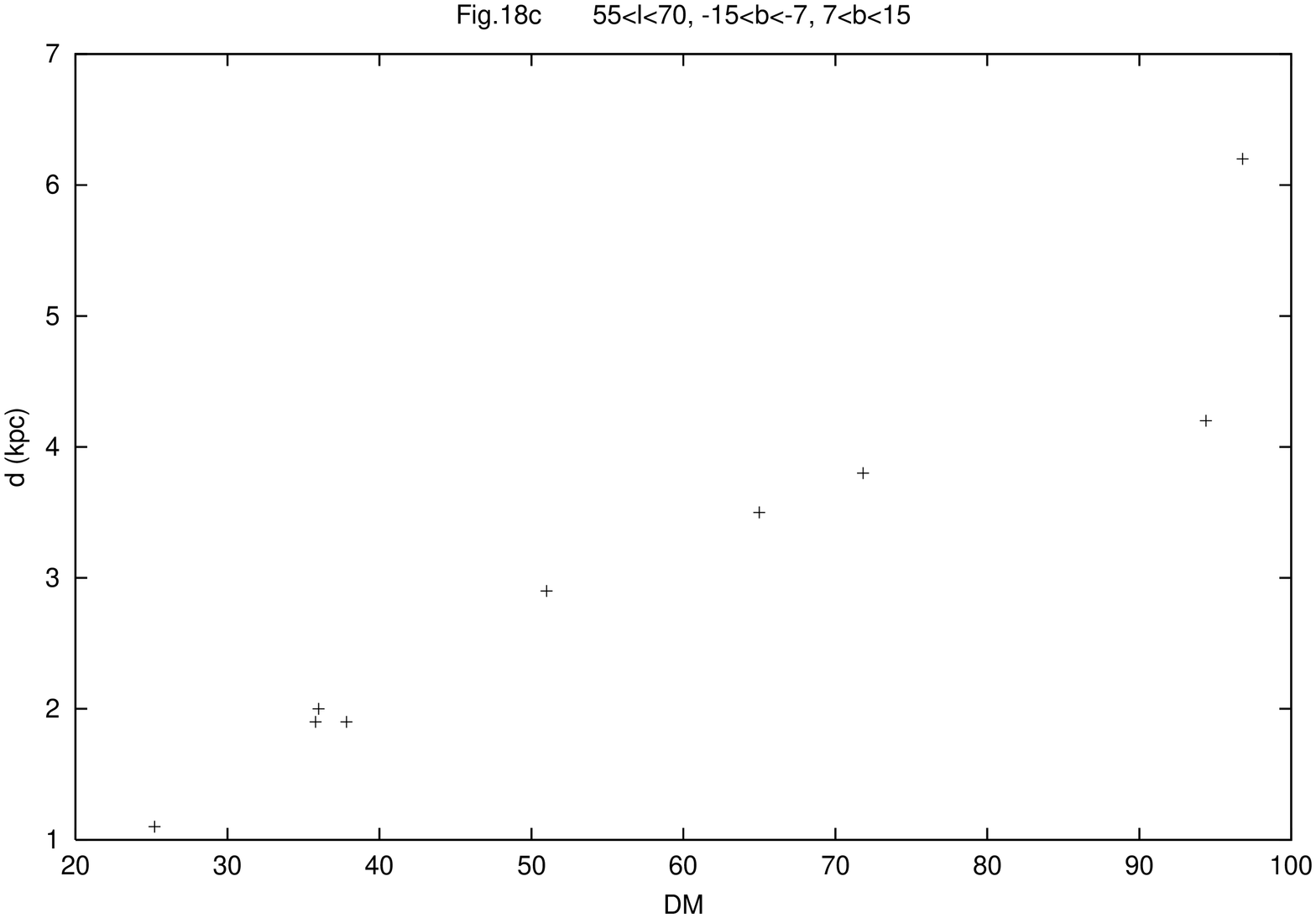,width=8cm,height=8cm,angle=-90} &
\psfig{file=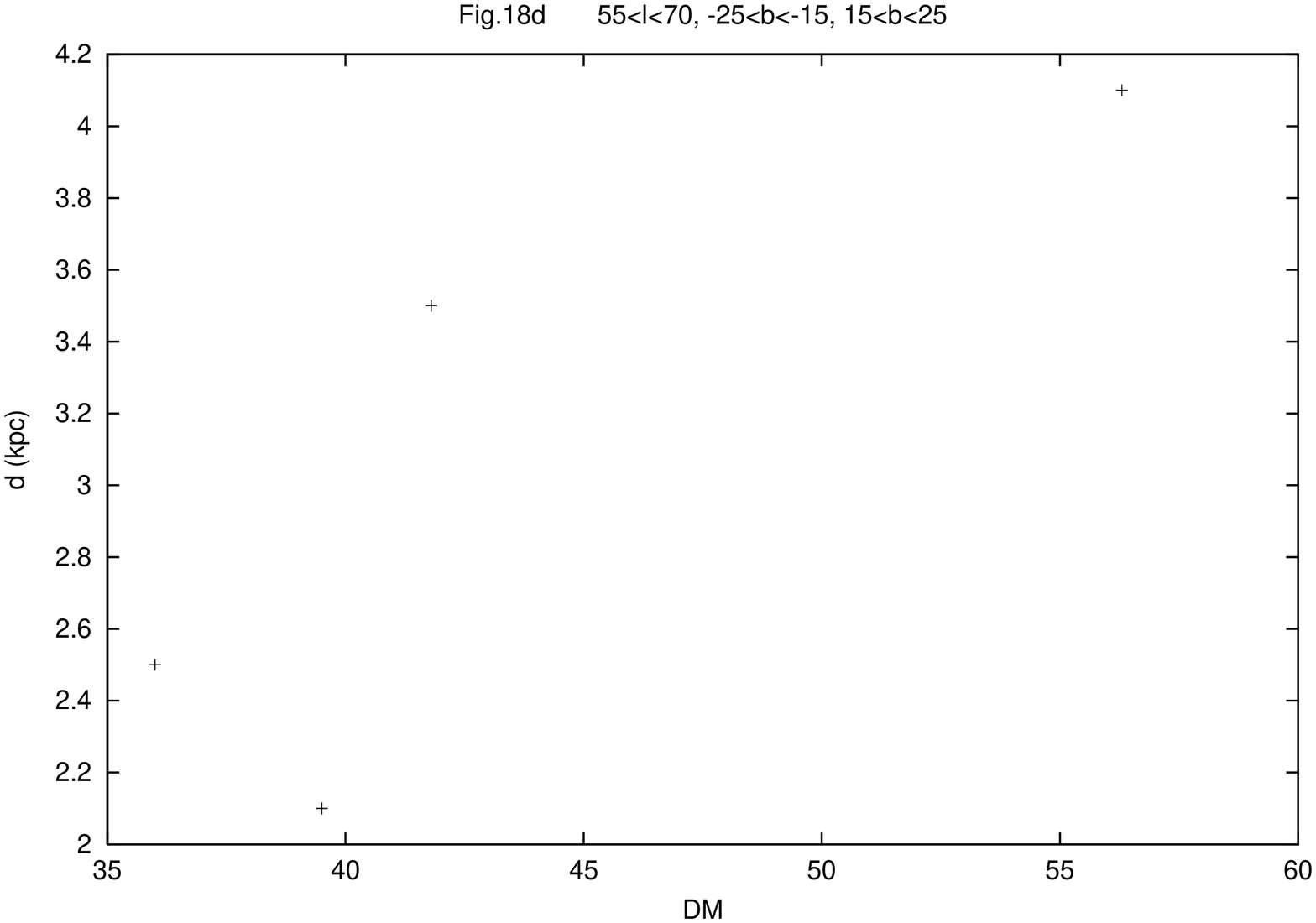,width=8cm,height=8cm,angle=-90} \\
\end{tabular}
\end{figure*}  
\begin{figure*}
\begin{tabular}{cc}
\psfig{file=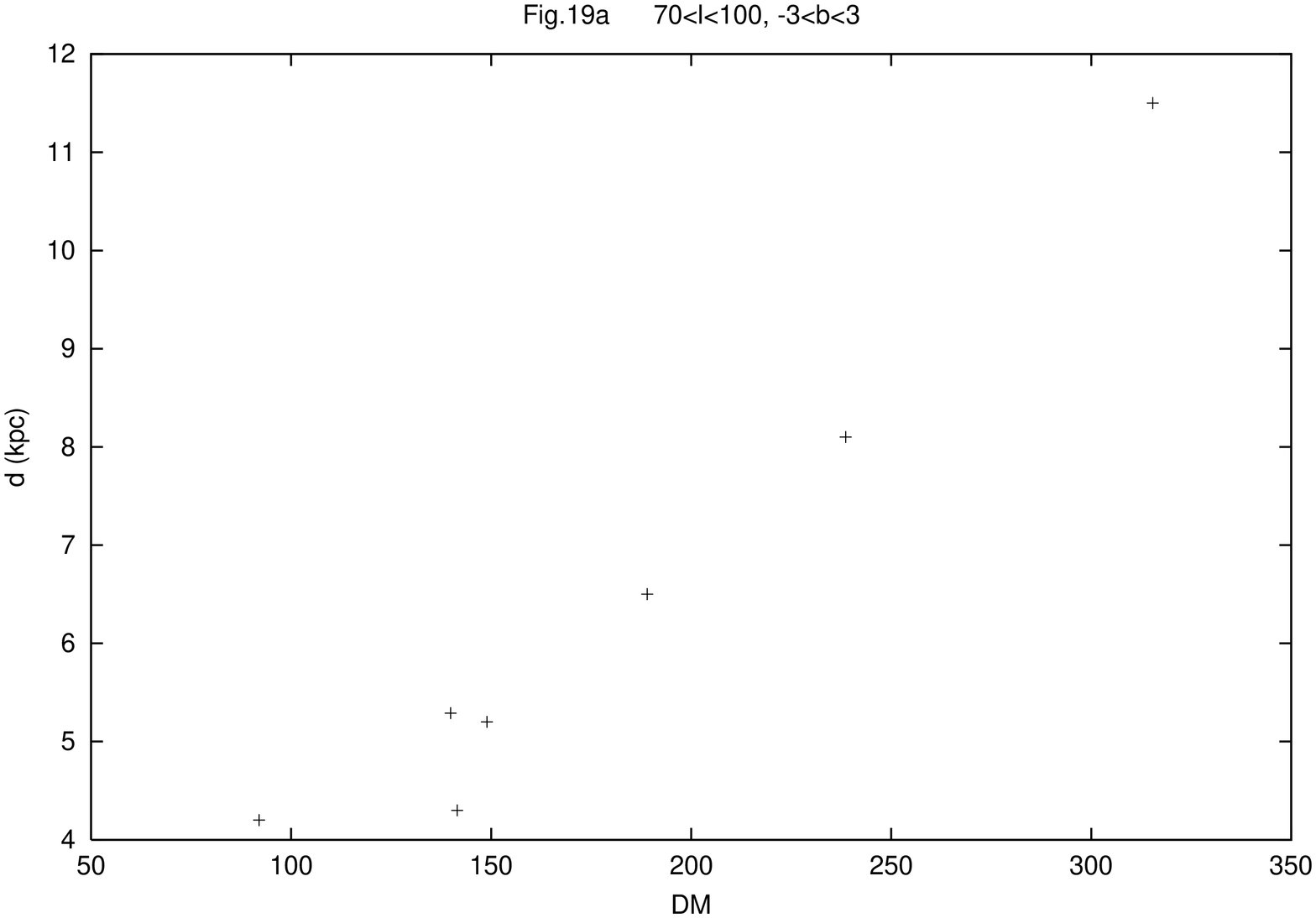,width=8cm,height=8cm,angle=-90} &
\psfig{file=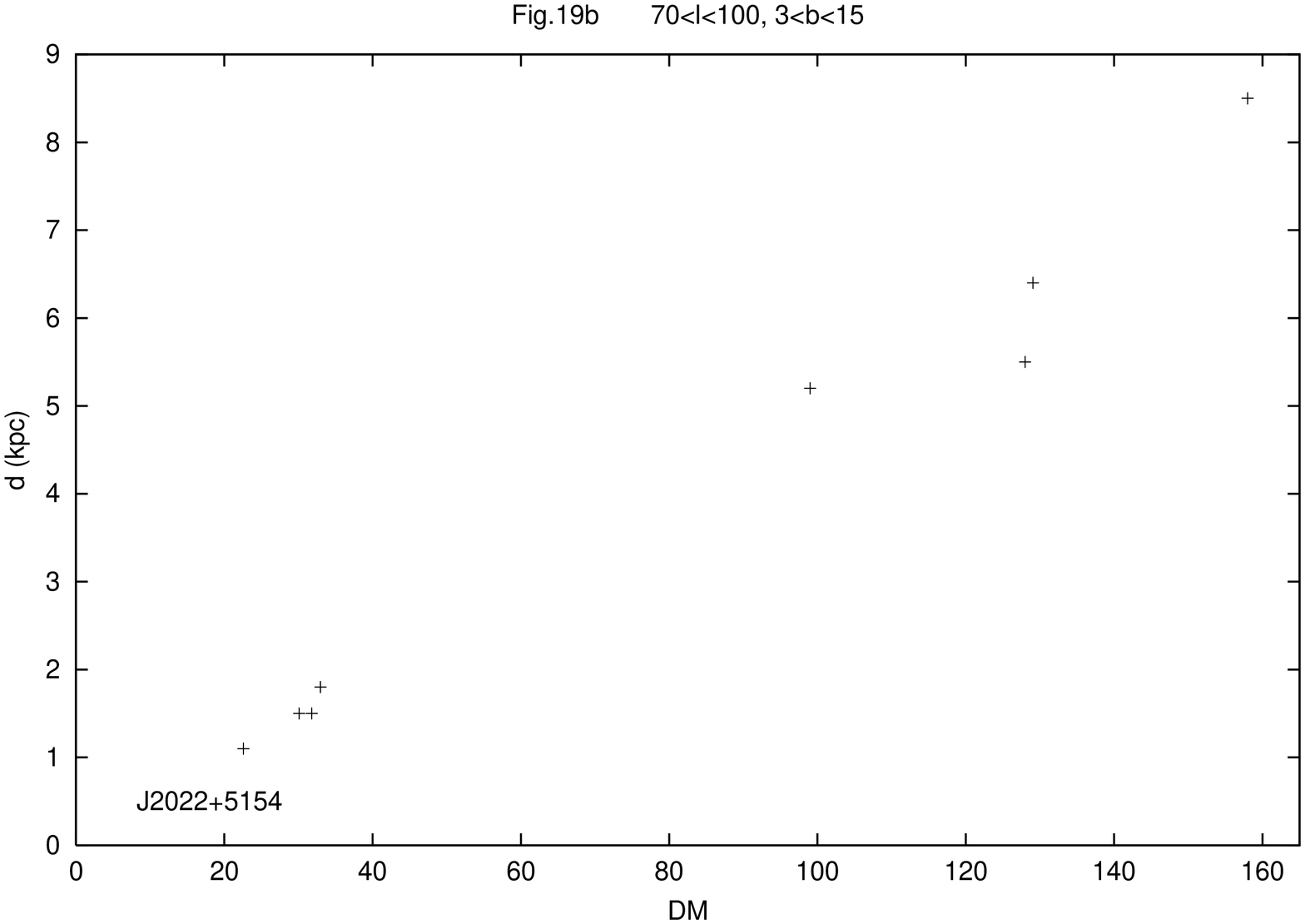,width=8cm,height=8cm,angle=-90} \\
\psfig{file=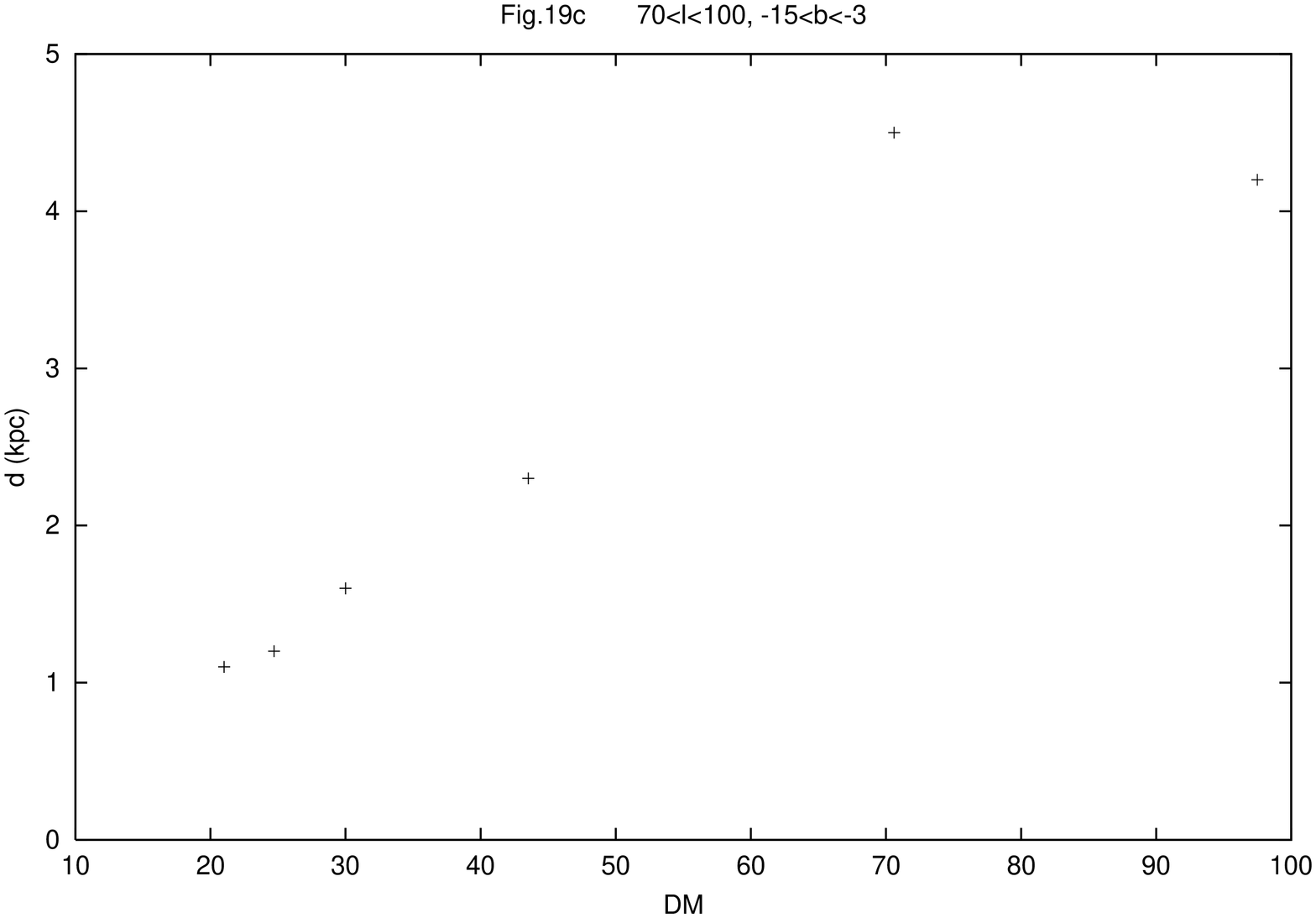,width=8cm,height=8cm,angle=-90} &
\psfig{file=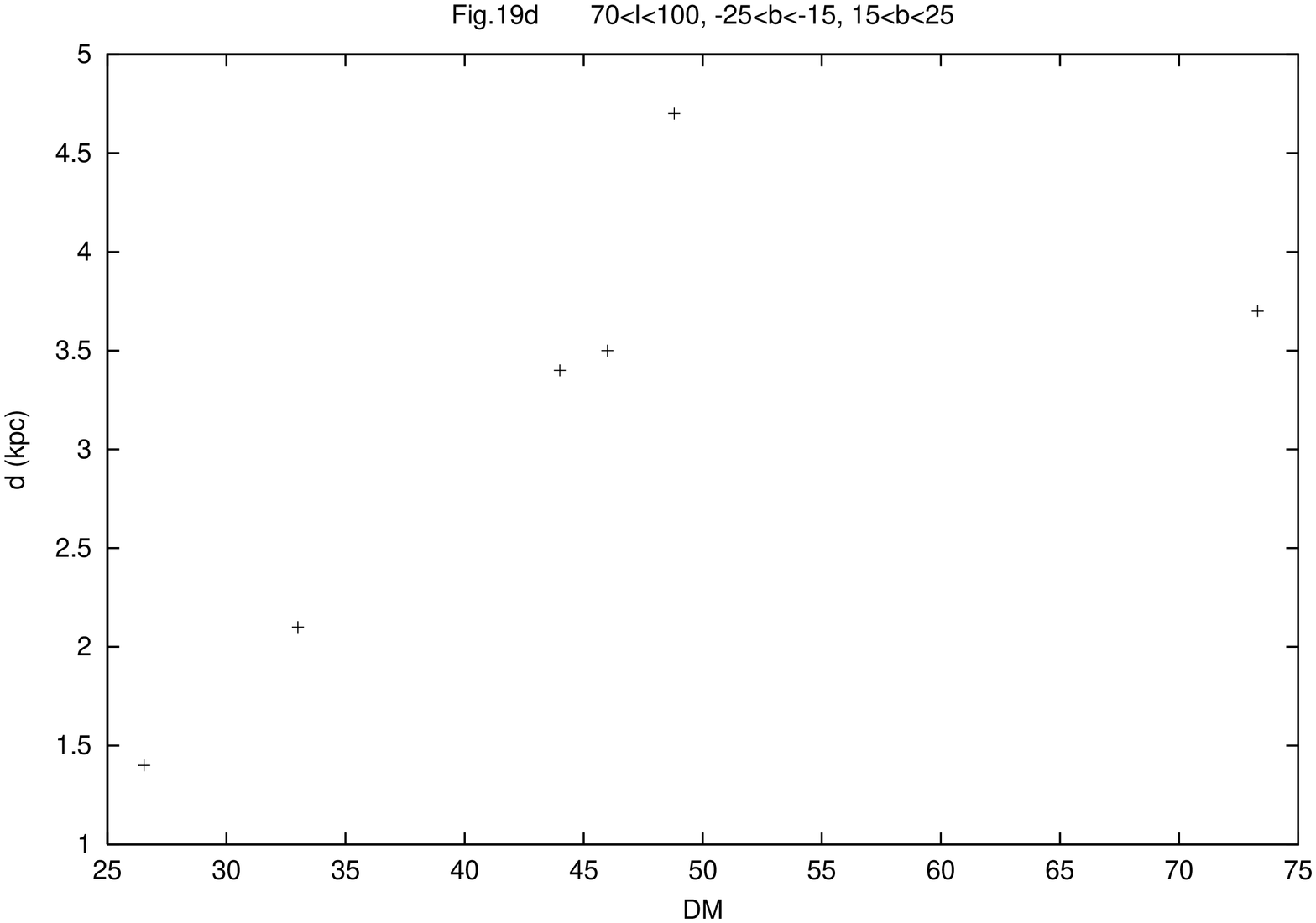,width=8cm,height=8cm,angle=-90} \\
\end{tabular}
\end{figure*}  

\newpage
\begin{figure}
\begin{tabular}{cc}
\psfig{file=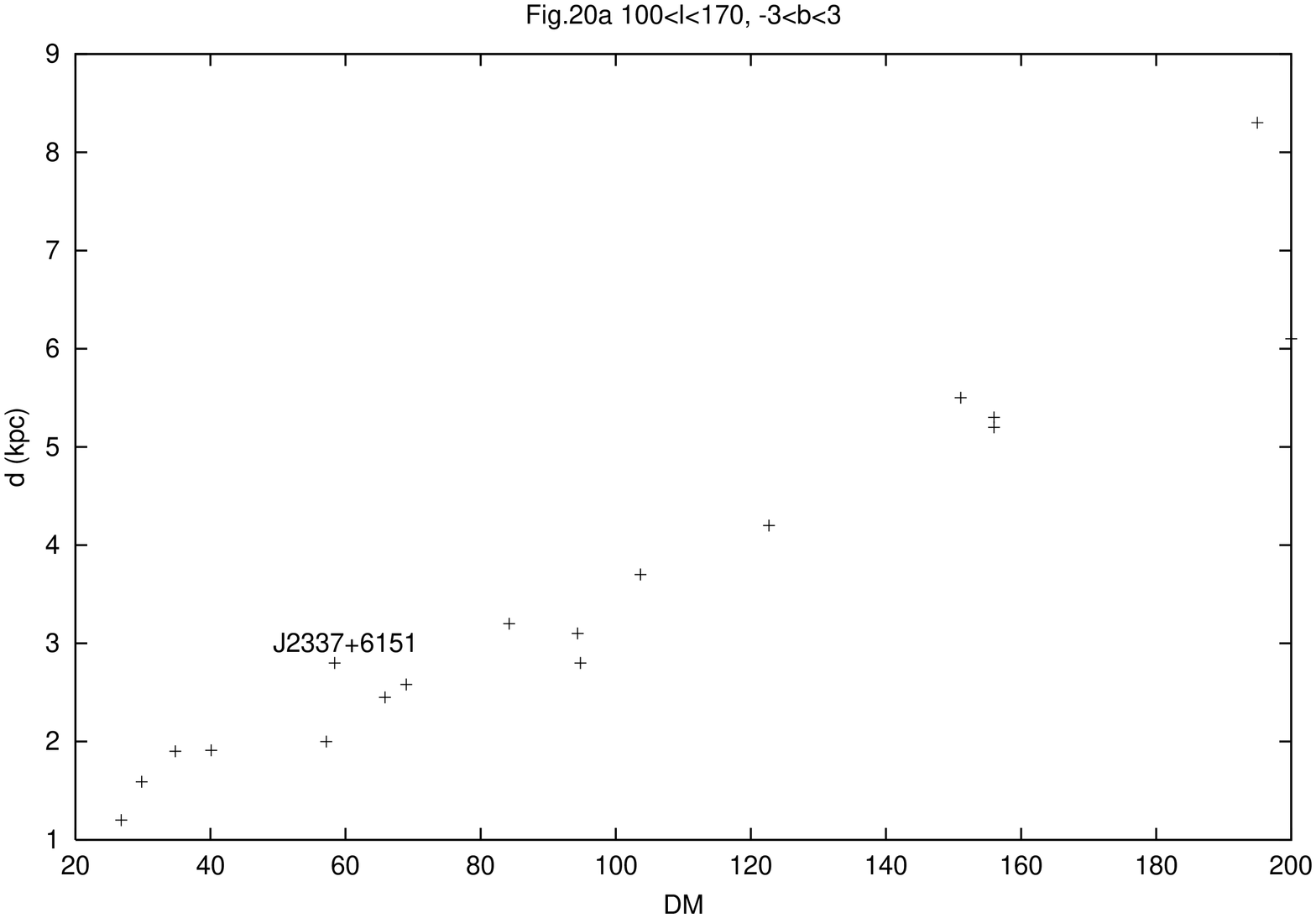,width=8cm,height=8cm,angle=-90} &
\psfig{file=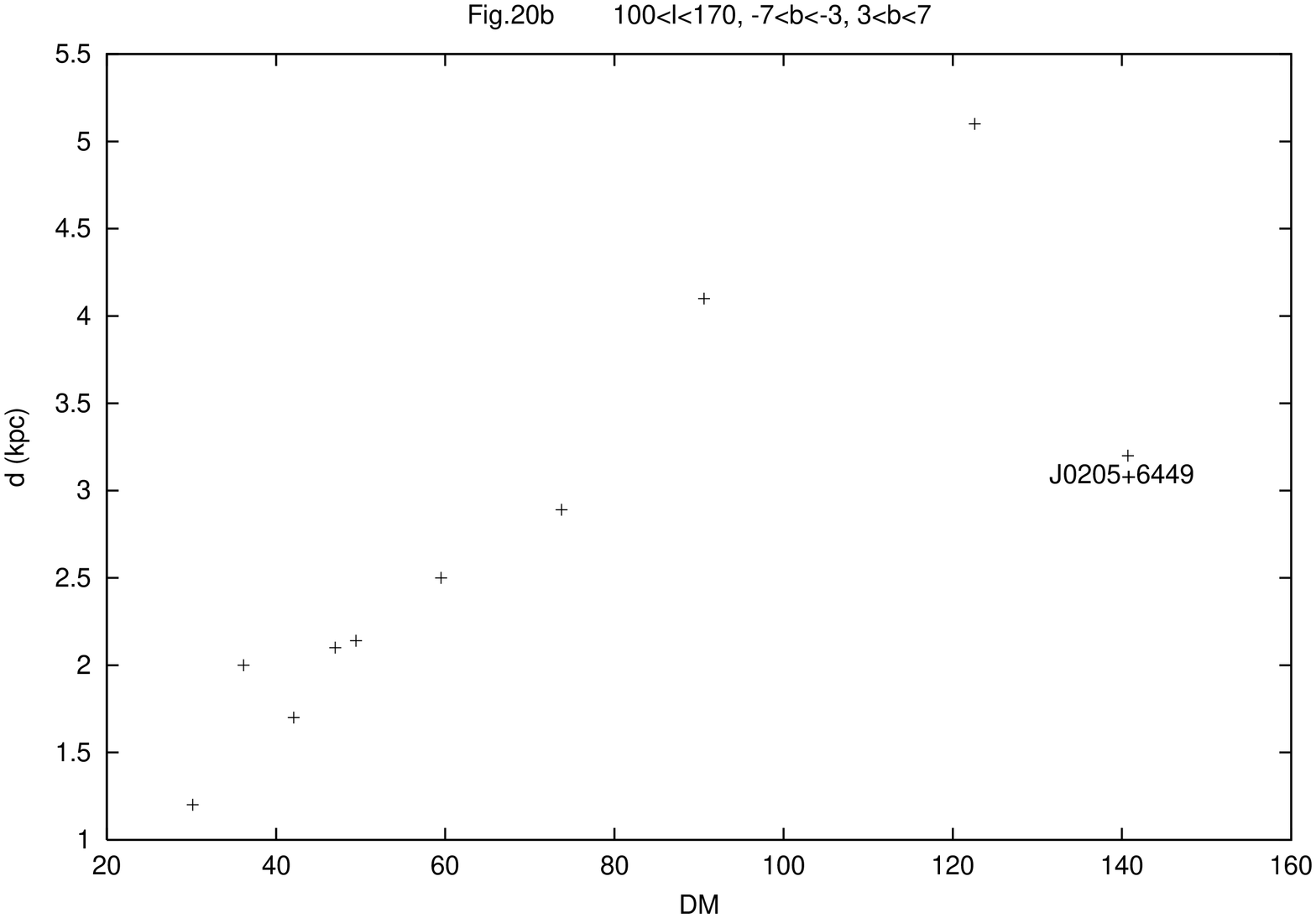,width=8cm,height=8cm,angle=-90} \\
\psfig{file=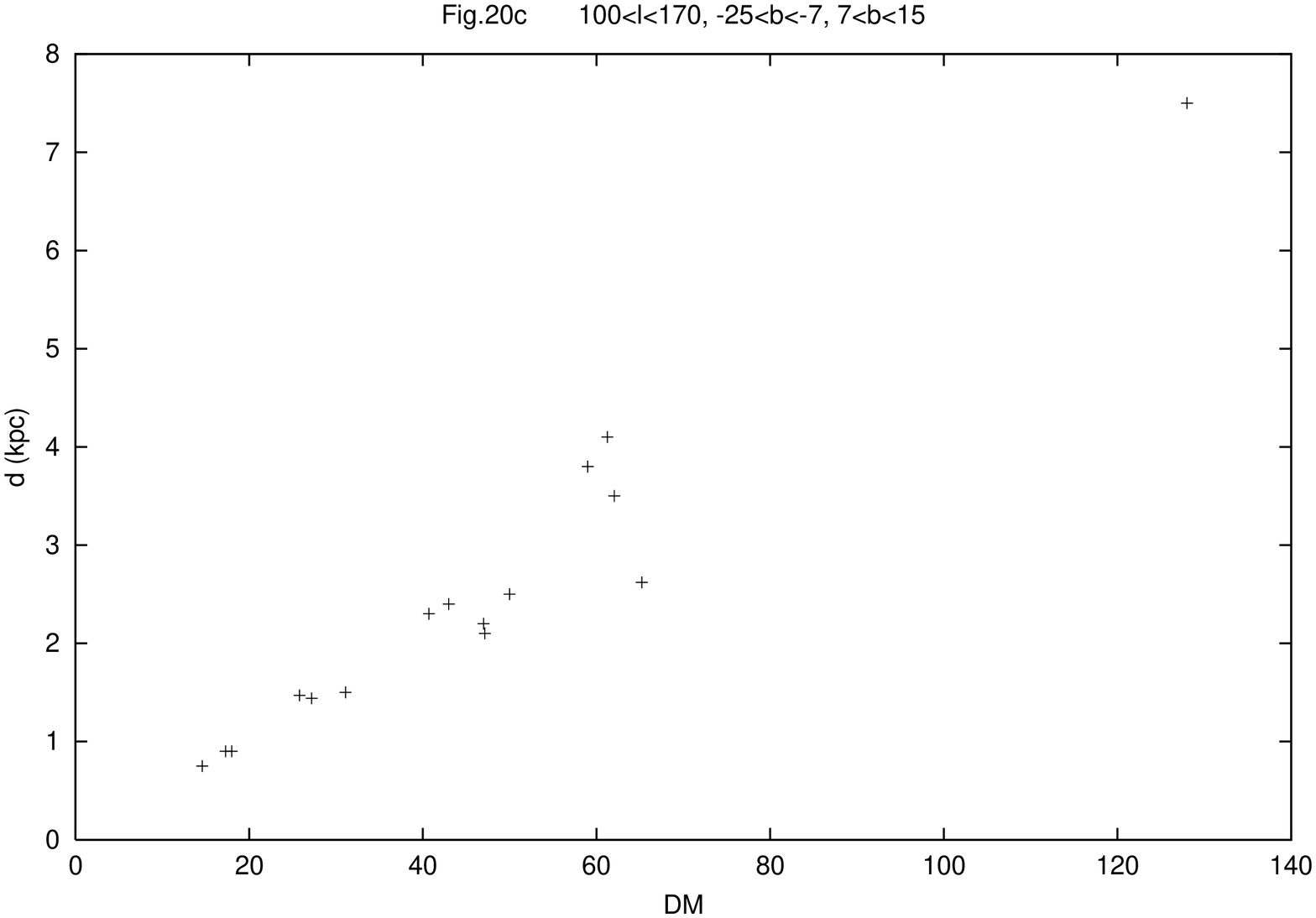,width=8cm,height=8cm,angle=-90} & \\
\end{tabular}
\end{figure}

\begin{figure}
\begin{tabular}{cc}
\psfig{file=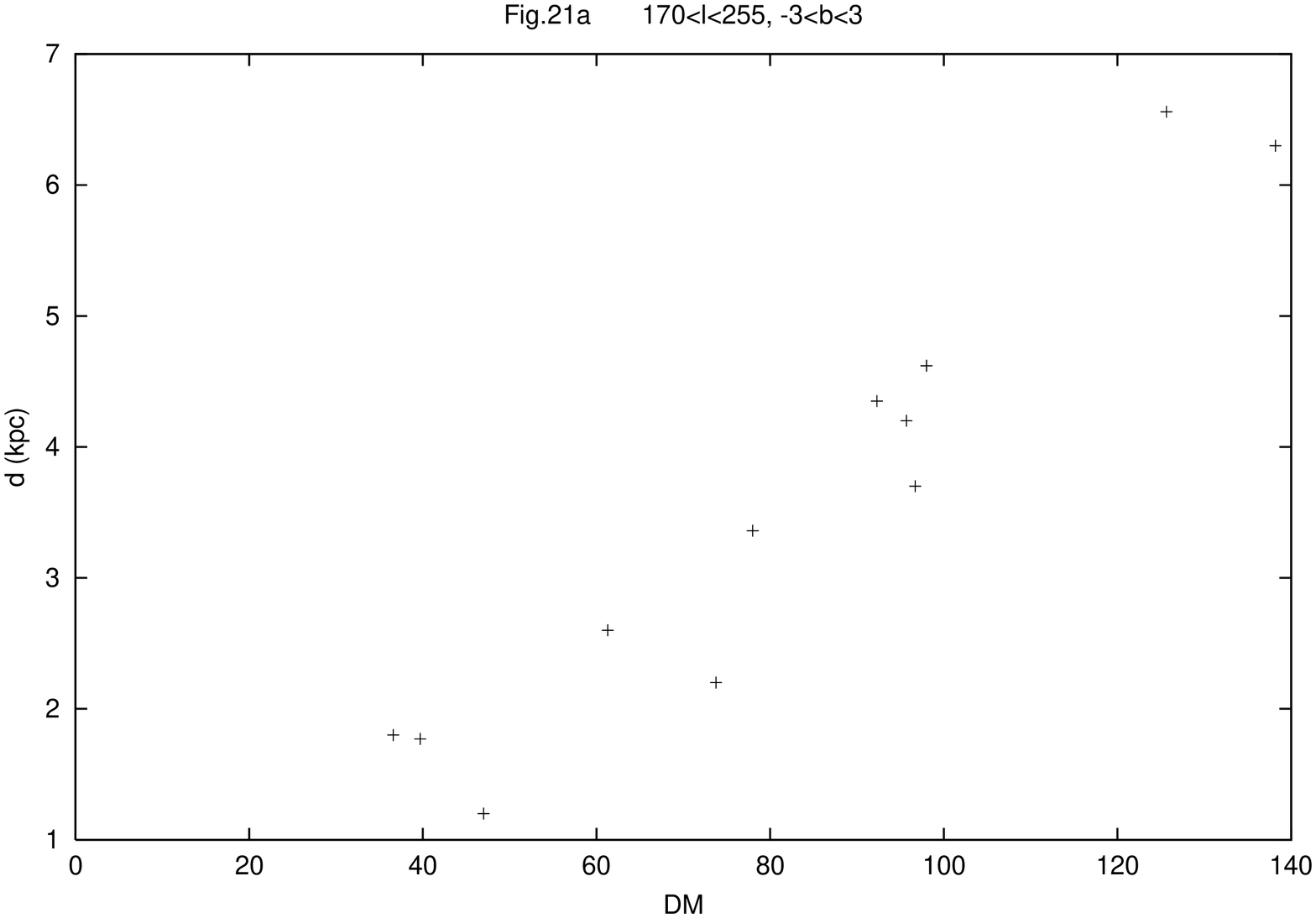,width=8cm,height=8cm,angle=-90} &
\psfig{file=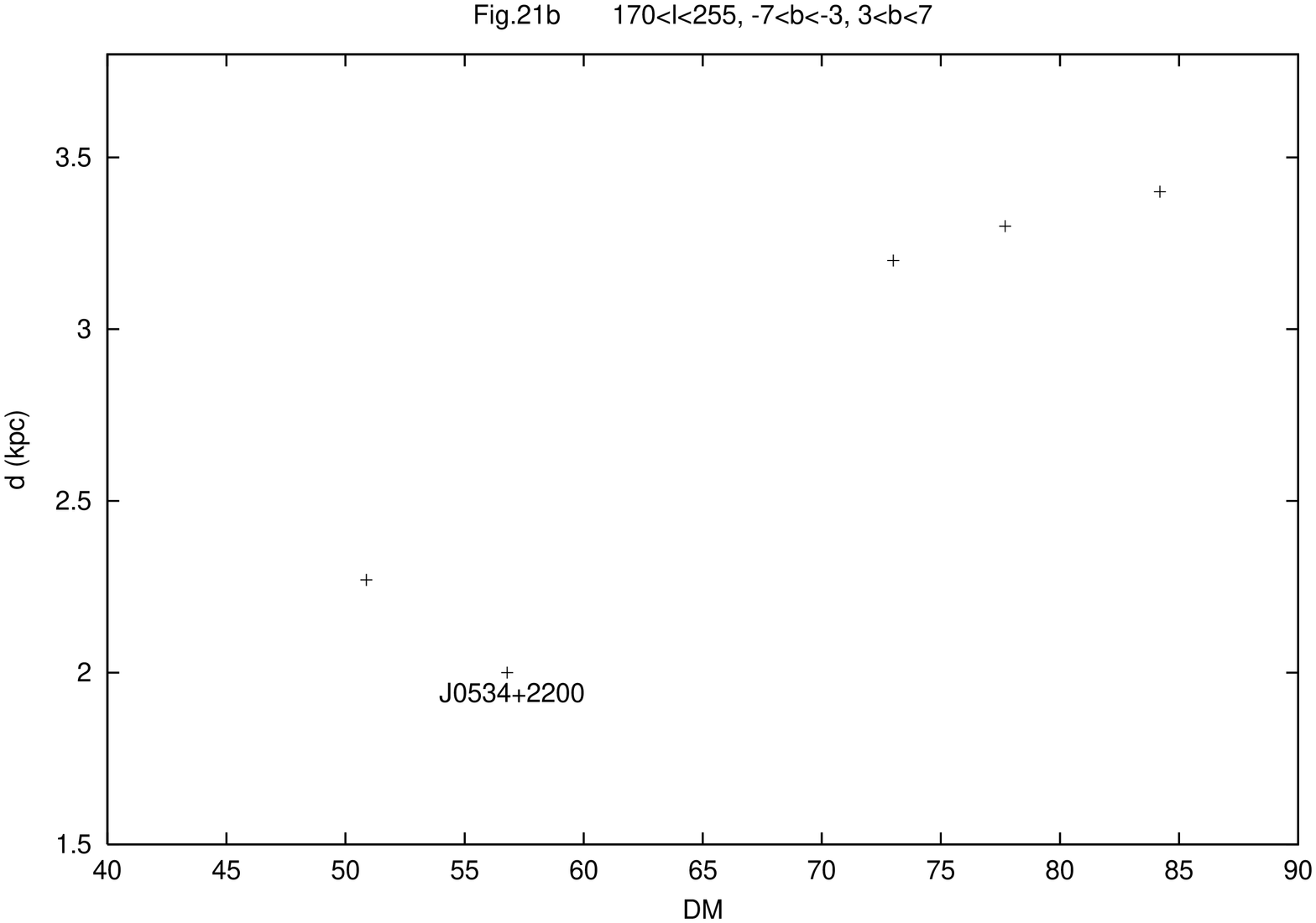,width=8cm,height=8cm,angle=-90} \\
\psfig{file=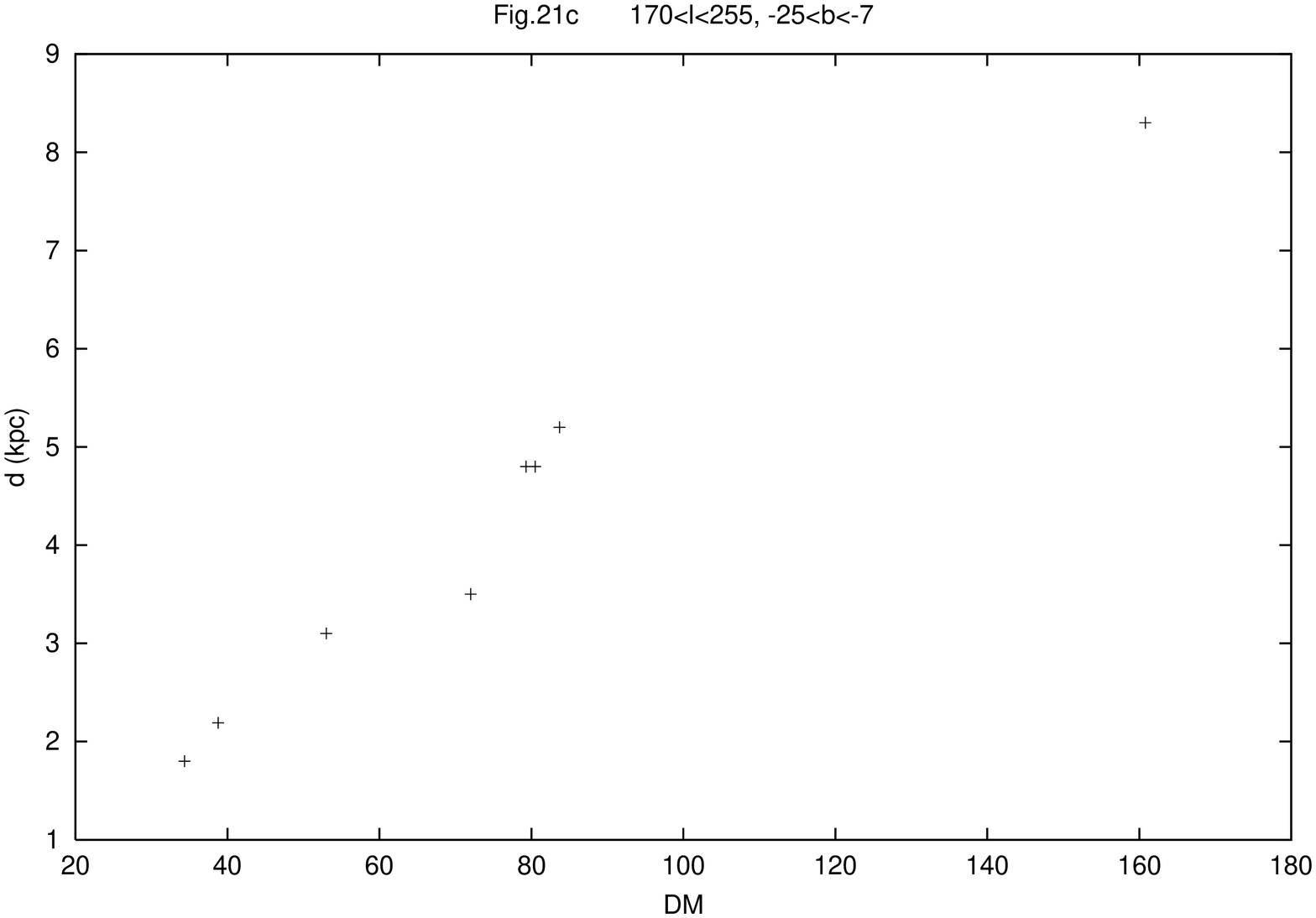,width=8cm,height=8cm,angle=-90} &
\psfig{file=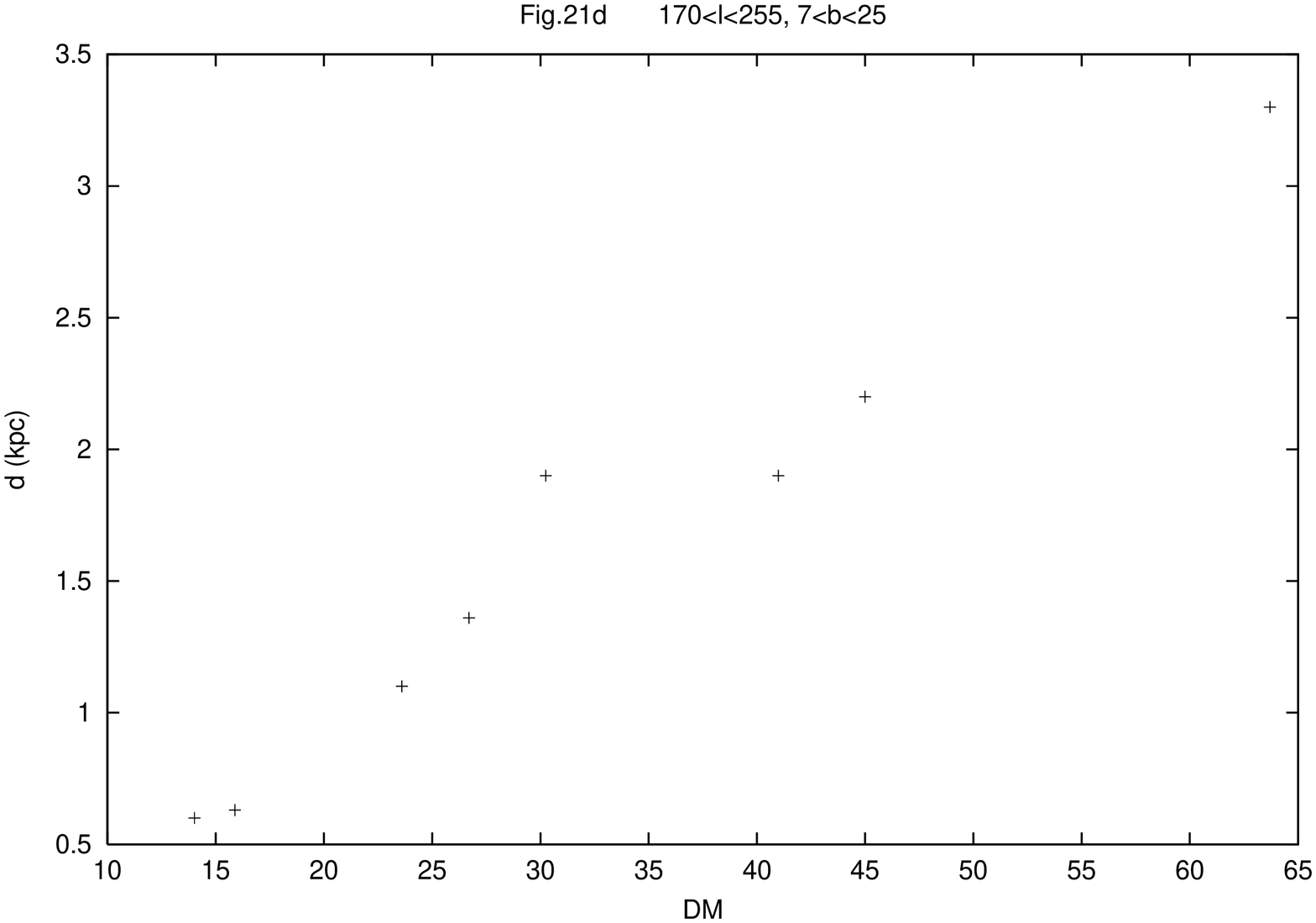,width=8cm,height=8cm,angle=-90} \\
\end{tabular}
\end{figure}  

\begin{figure}
\begin{tabular}{cc}
\psfig{file=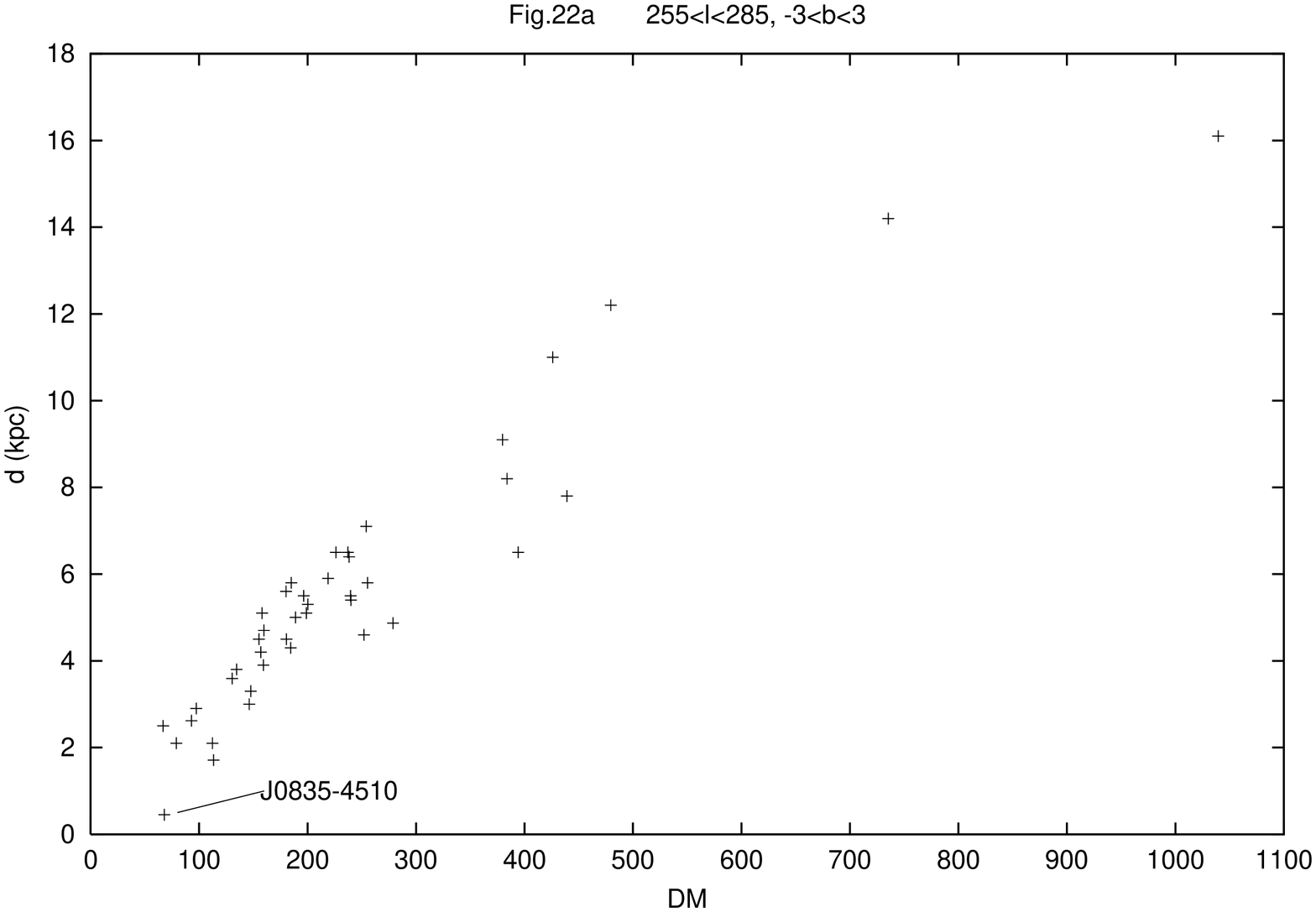,width=8cm,height=8cm,angle=-90} &
\psfig{file=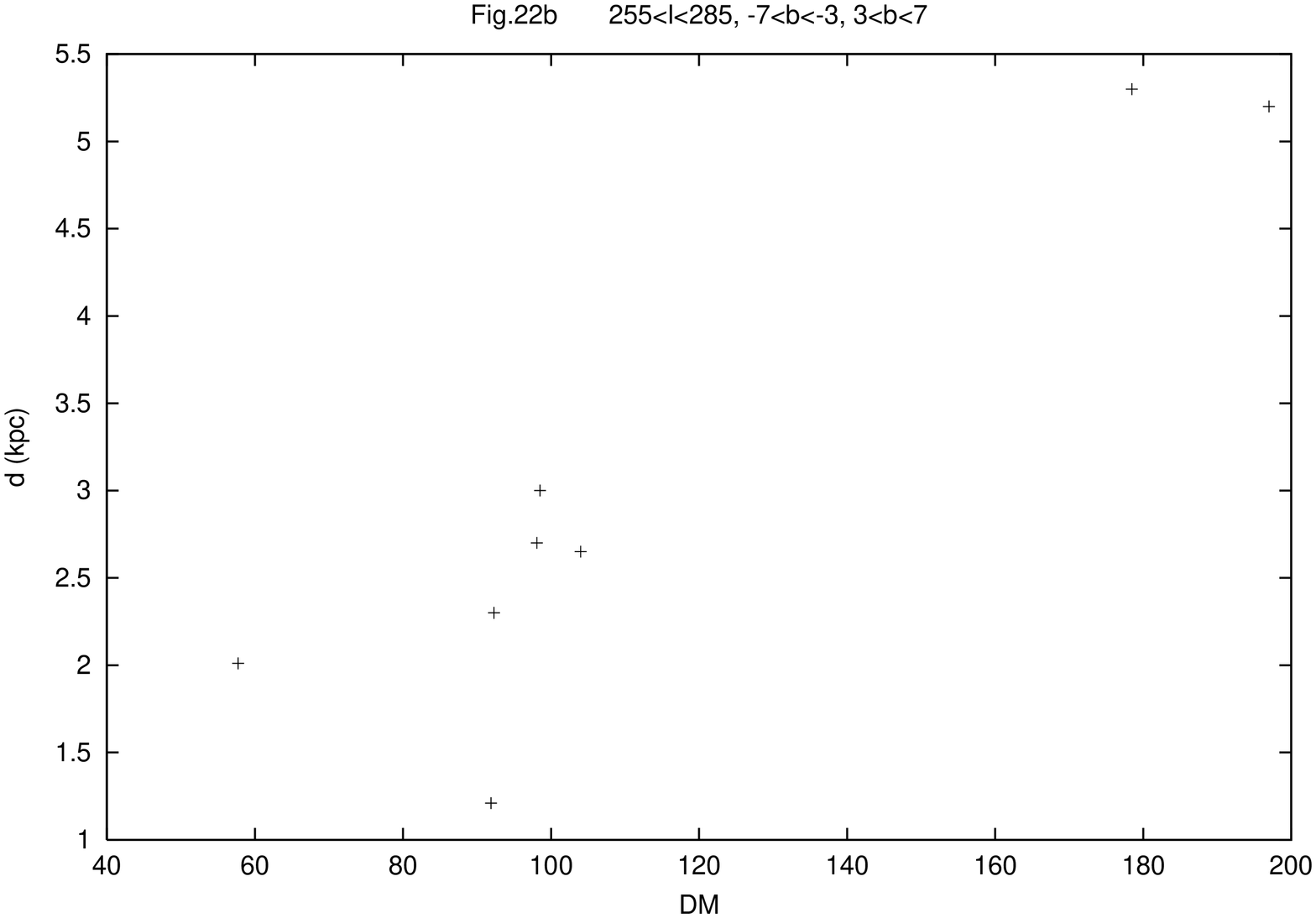,width=8cm,height=8cm,angle=-90} \\
\psfig{file=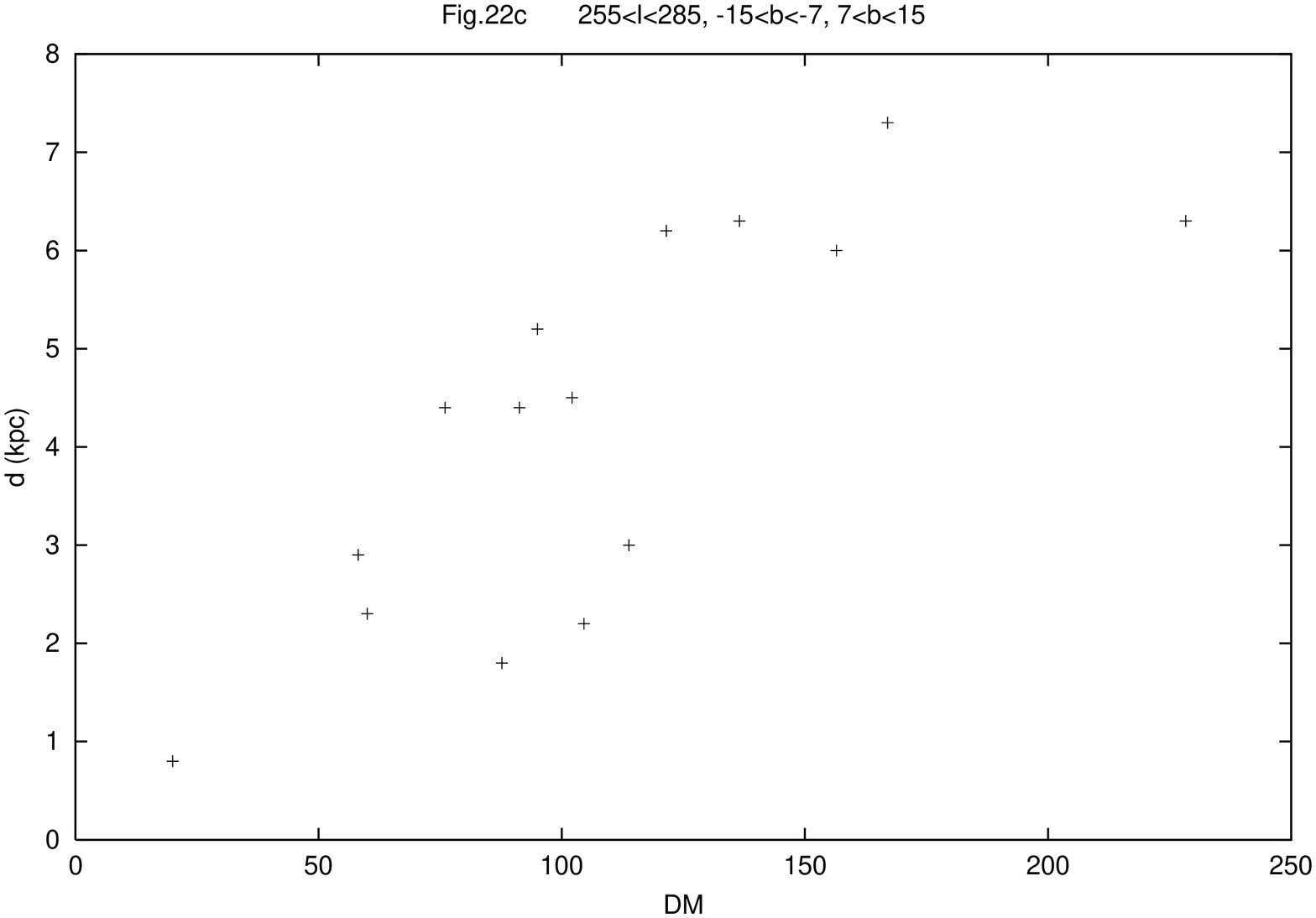,width=8cm,height=8cm,angle=-90} &
\psfig{file=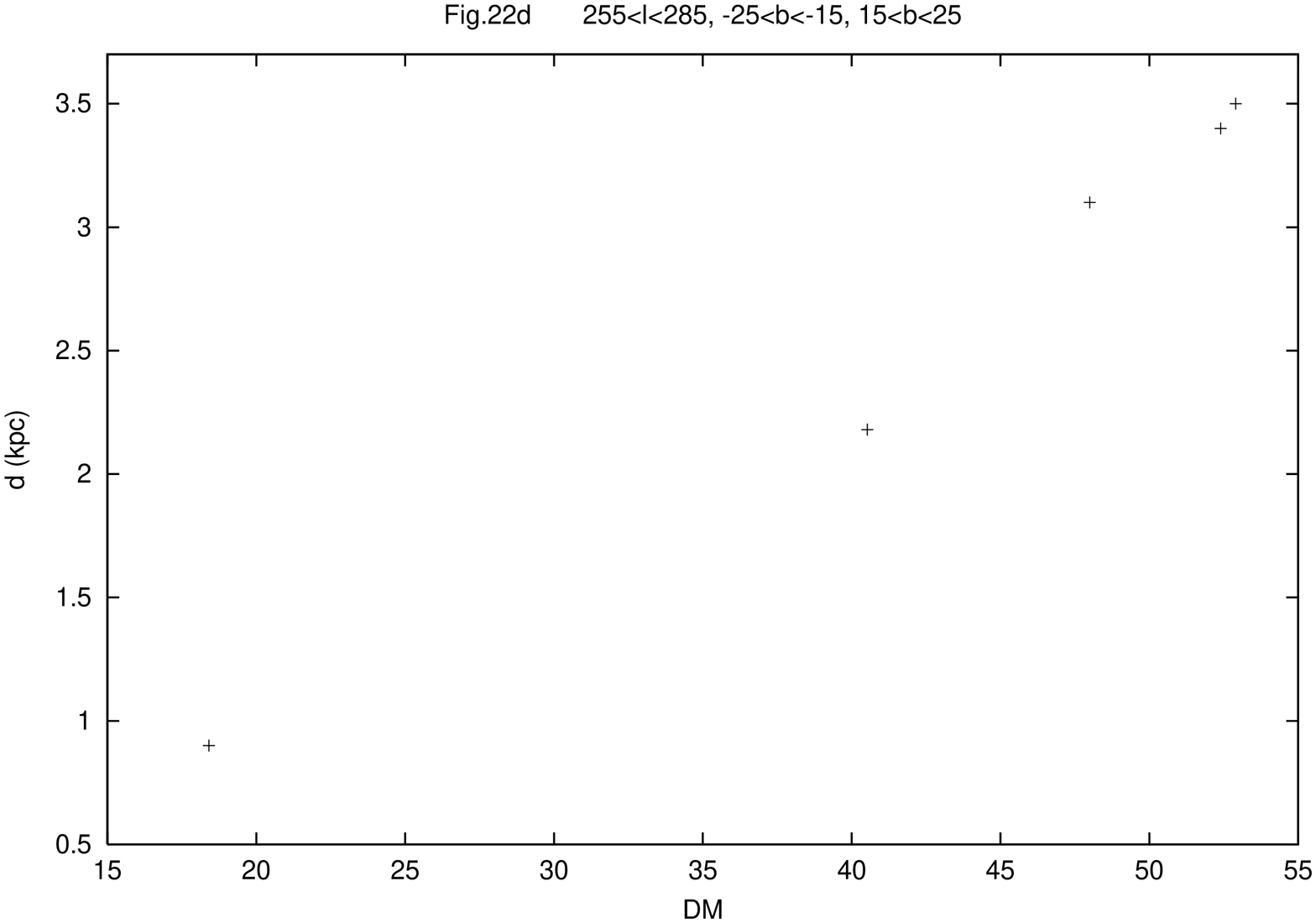,width=8cm,height=8cm,angle=-90} \\
\end{tabular}
\end{figure}  

\begin{figure}
\begin{tabular}{cc}
\psfig{file=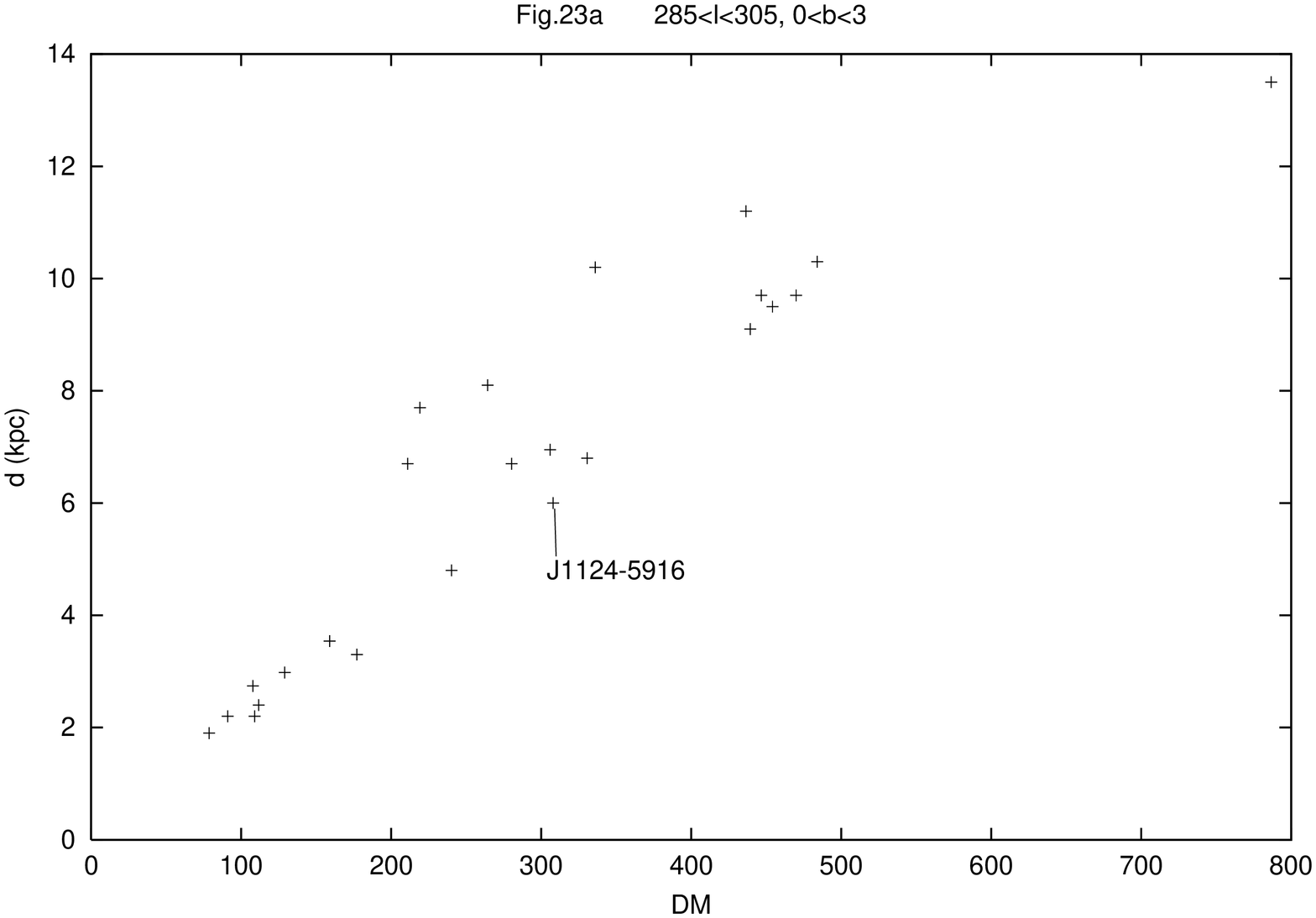,width=8cm,height=5cm,angle=-90} &
\psfig{file=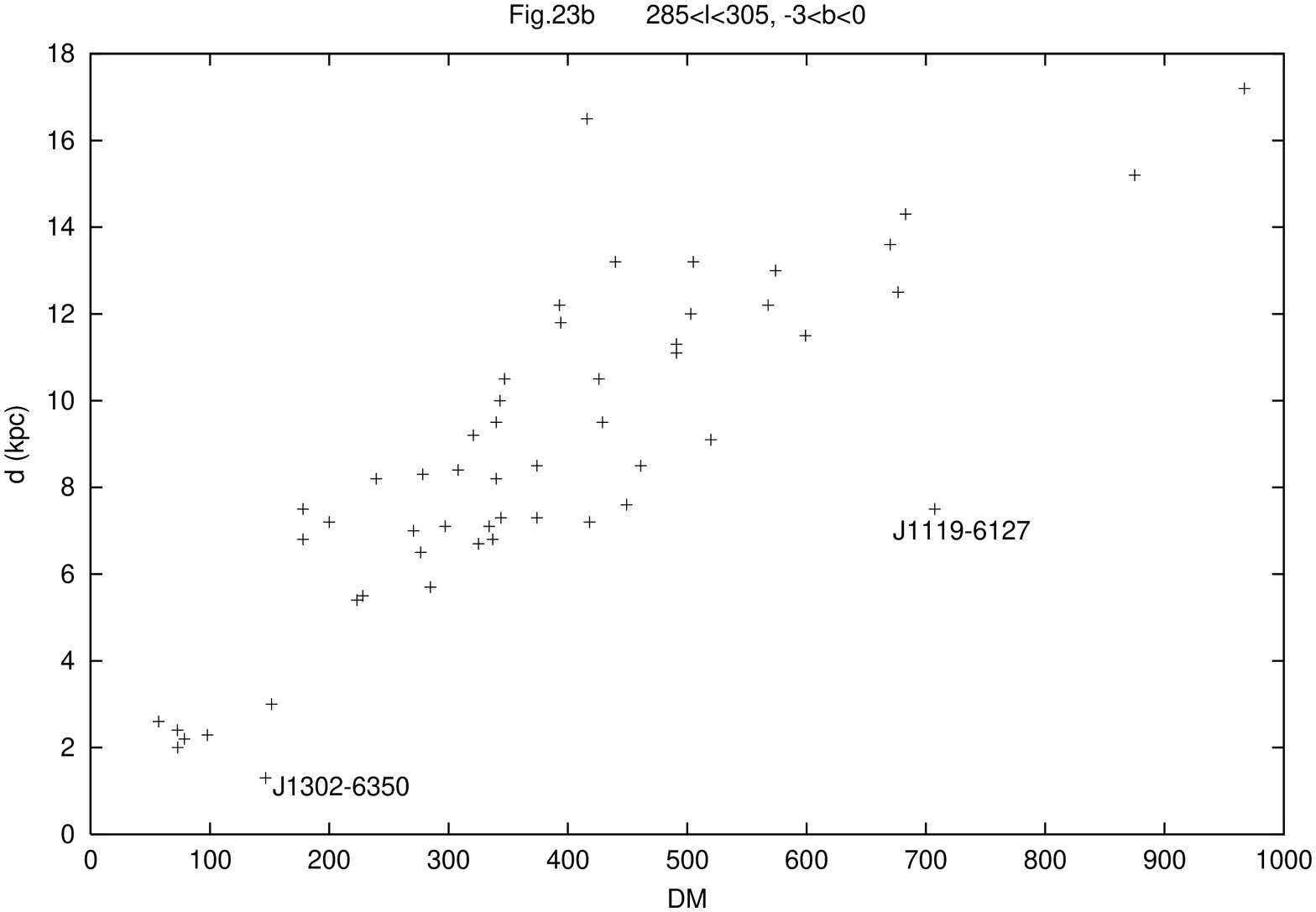,width=8cm,height=5cm,angle=-90} \\
\psfig{file=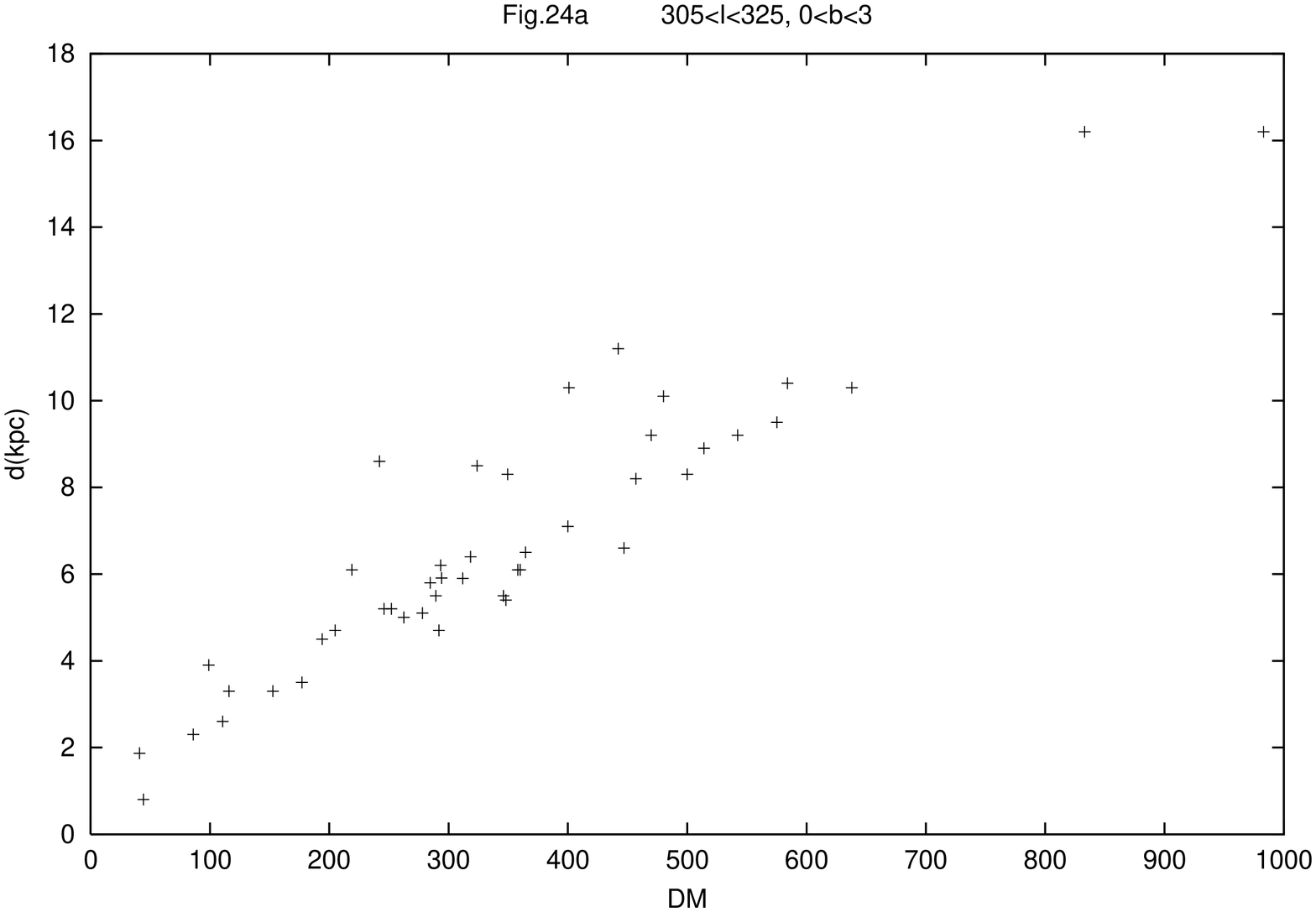,width=8cm,height=5cm,angle=-90} &
\psfig{file=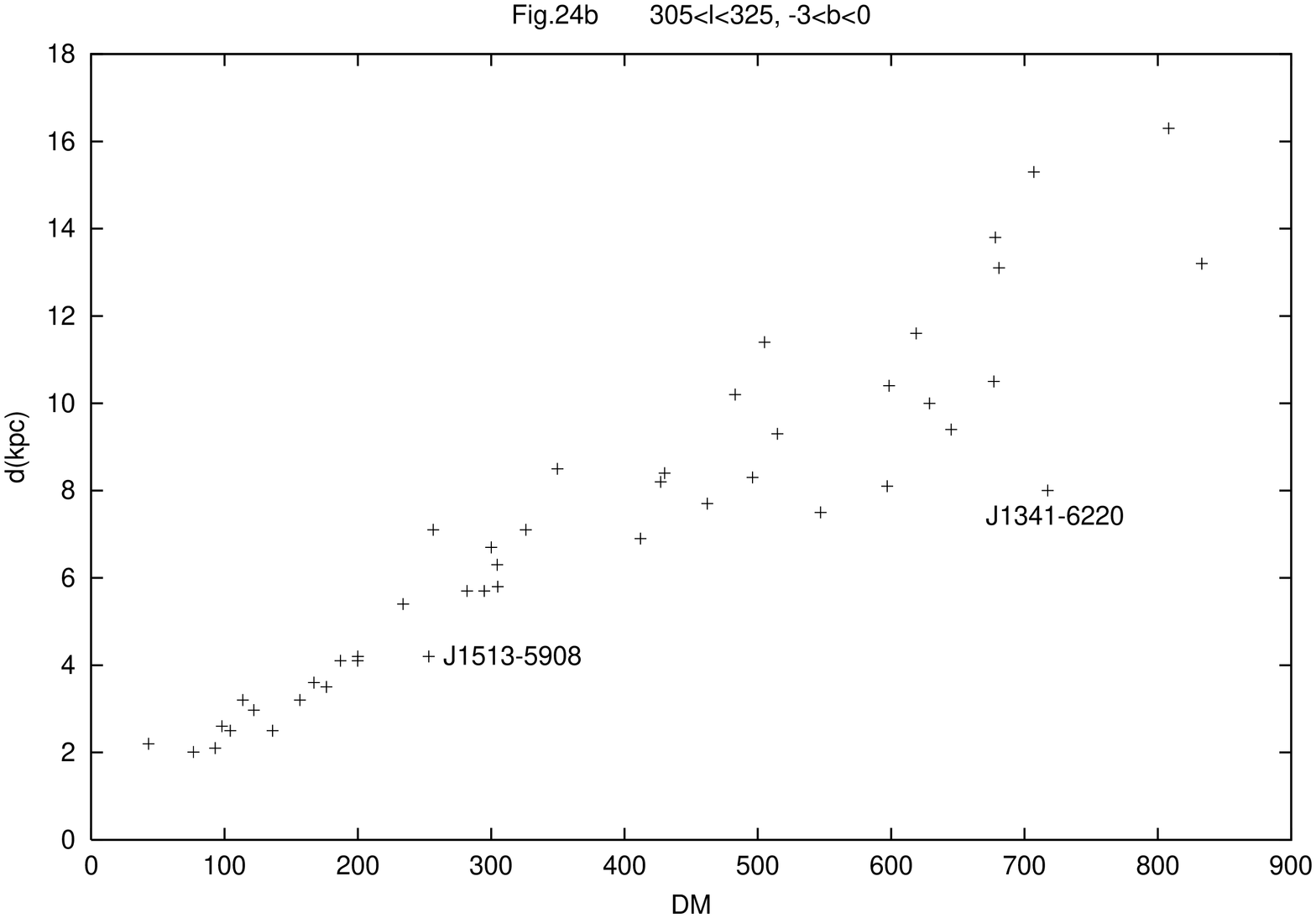,width=8cm,height=5cm,angle=-90} \\
\psfig{file=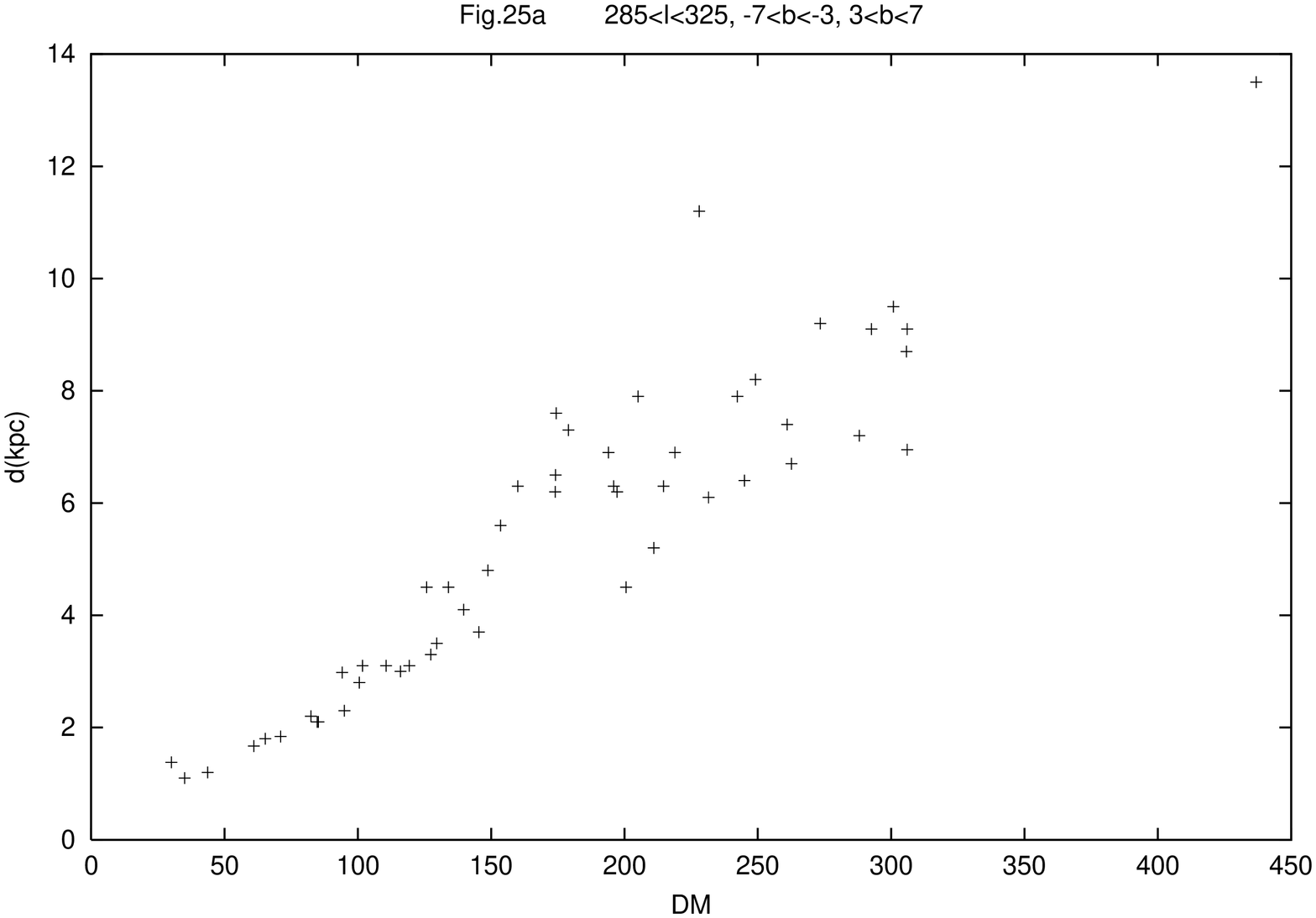,width=8cm,height=5cm,angle=-90} &
\psfig{file=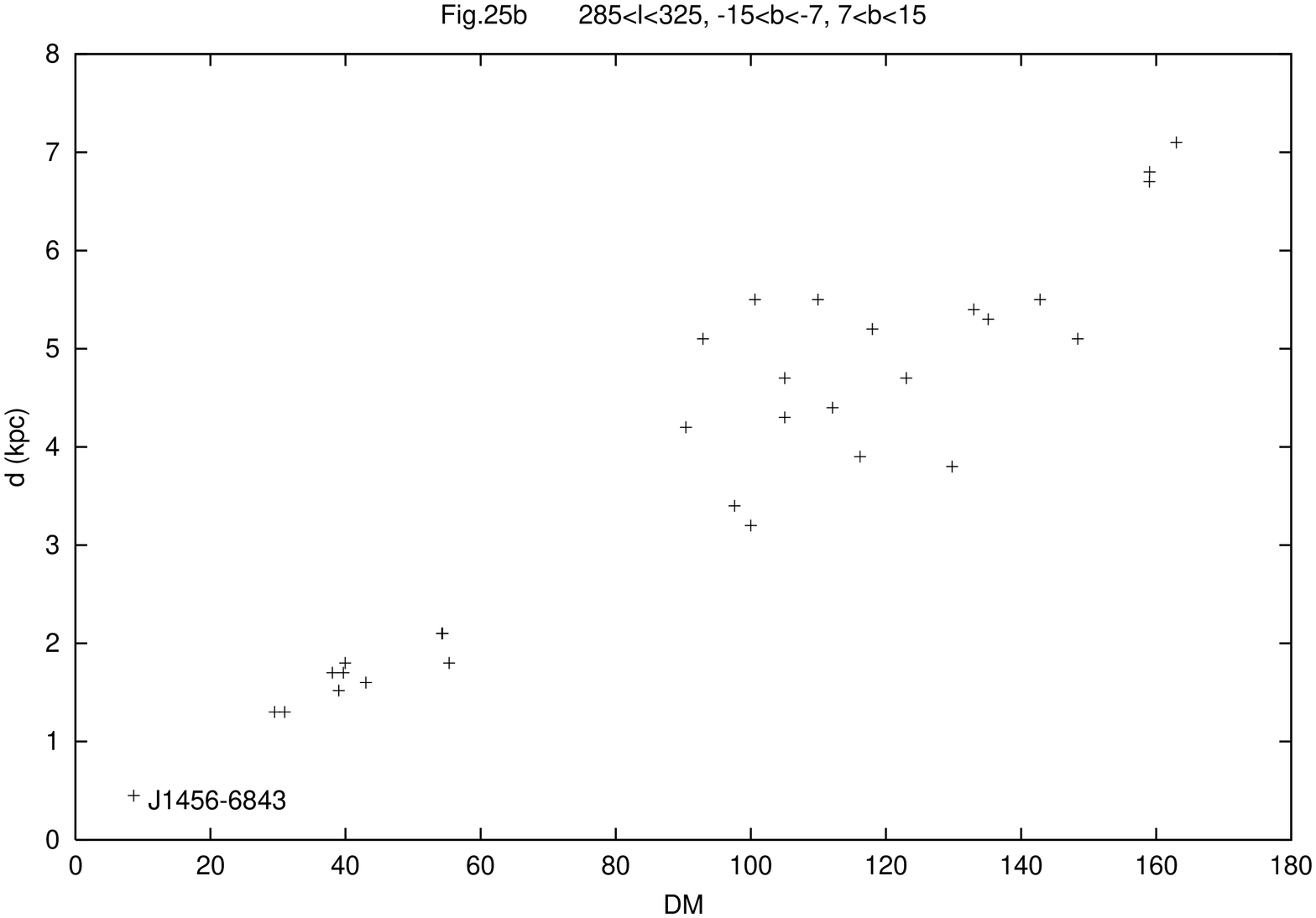,width=8cm,height=5cm,angle=-90} \\
\end{tabular}
\end{figure}

\begin{figure}
\begin{tabular}{cc}
\psfig{file=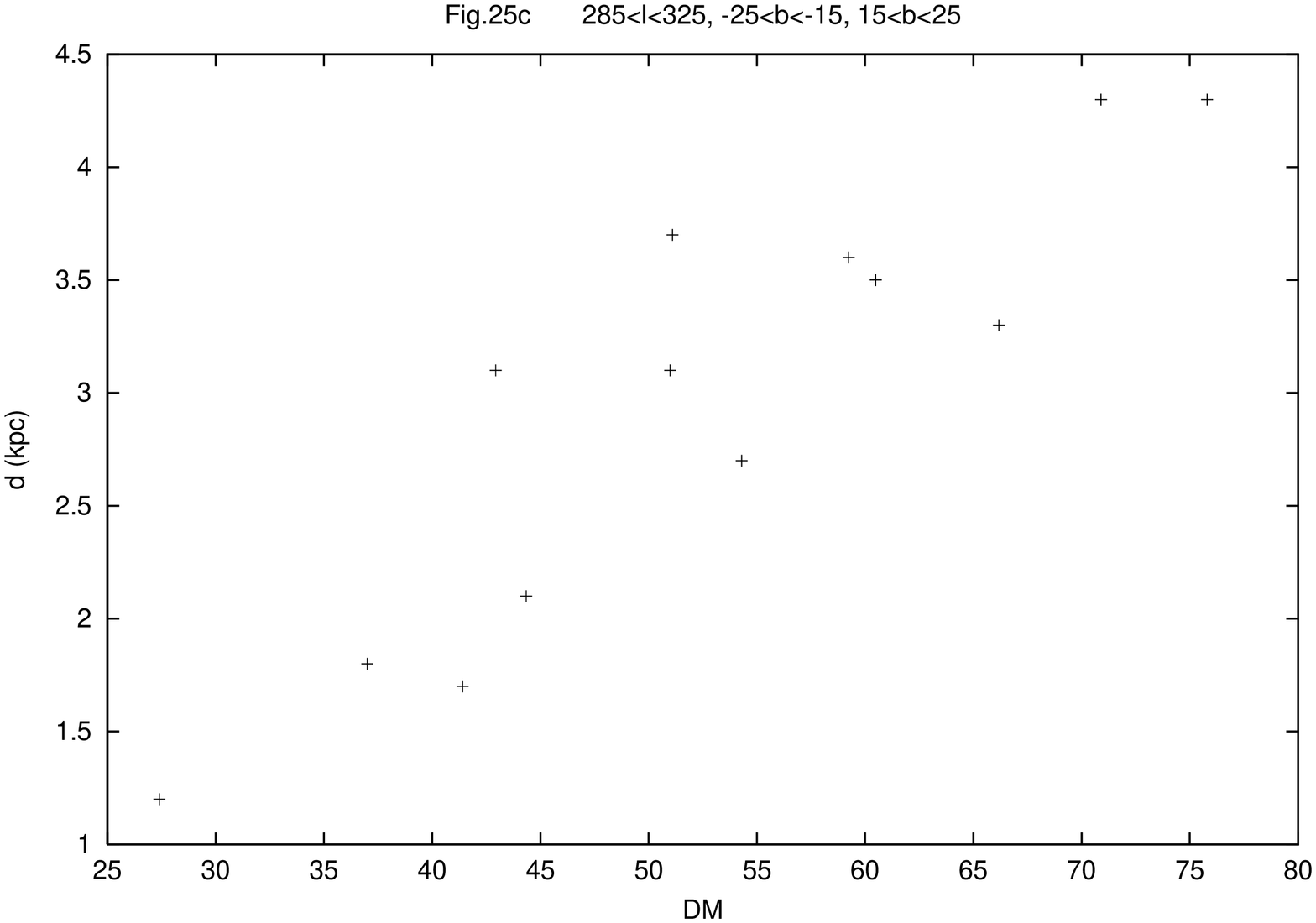,width=8cm,height=5cm,angle=-90} &
\psfig{file=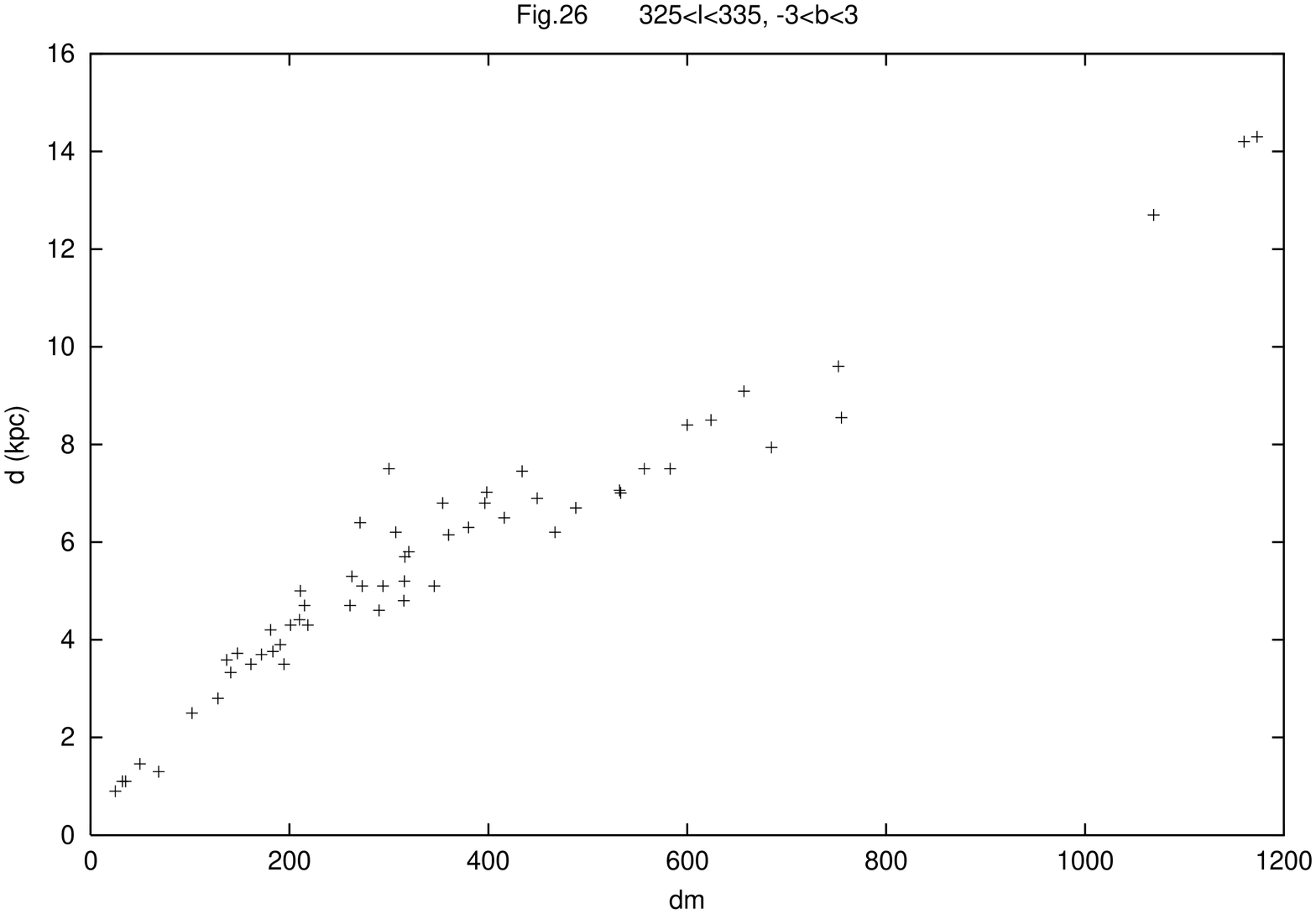,width=8cm,height=5cm,angle=-90} \\
\psfig{file=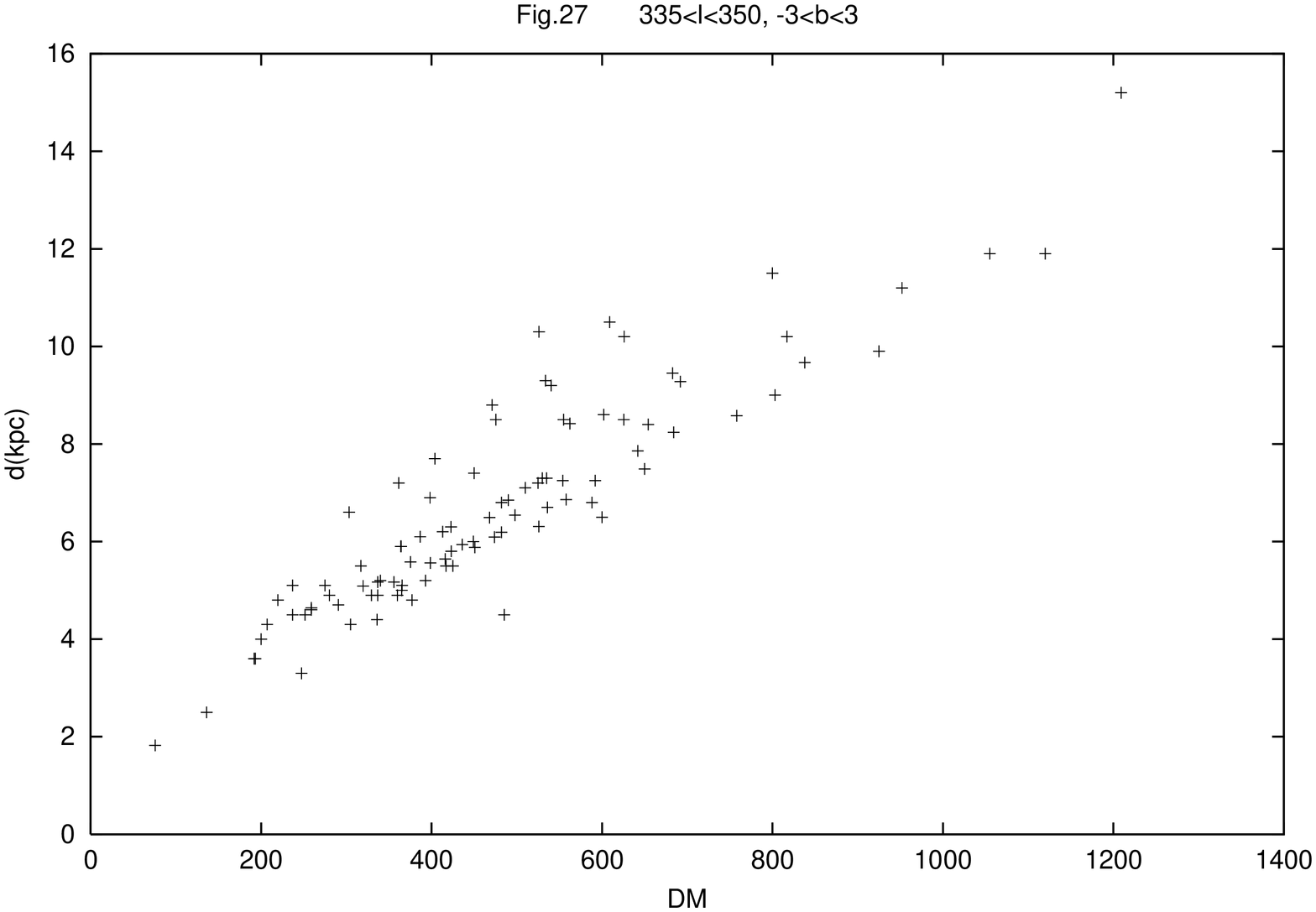,width=8cm,height=5cm,angle=-90} &
\psfig{file=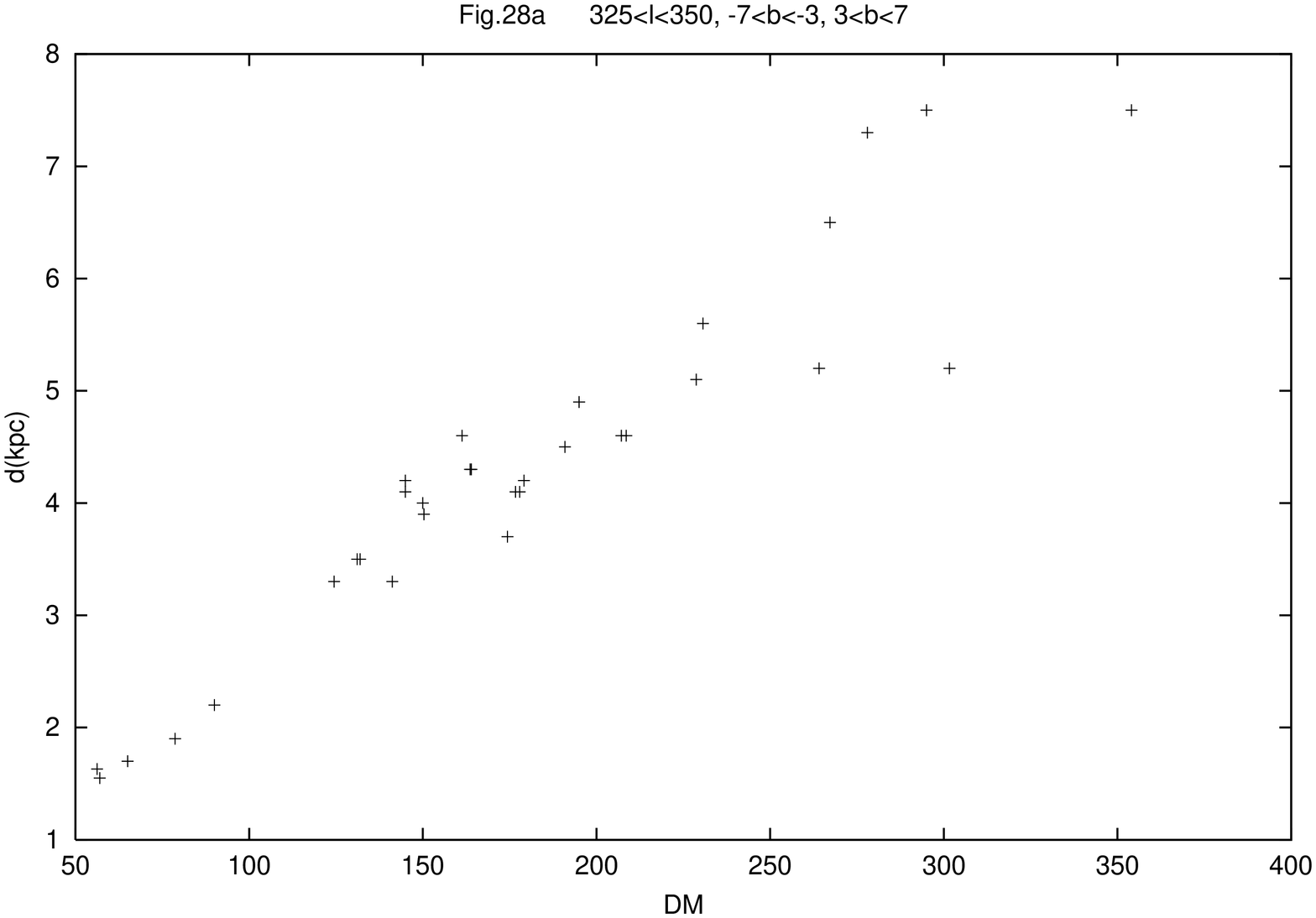,width=8cm,height=5cm,angle=-90} \\
\psfig{file=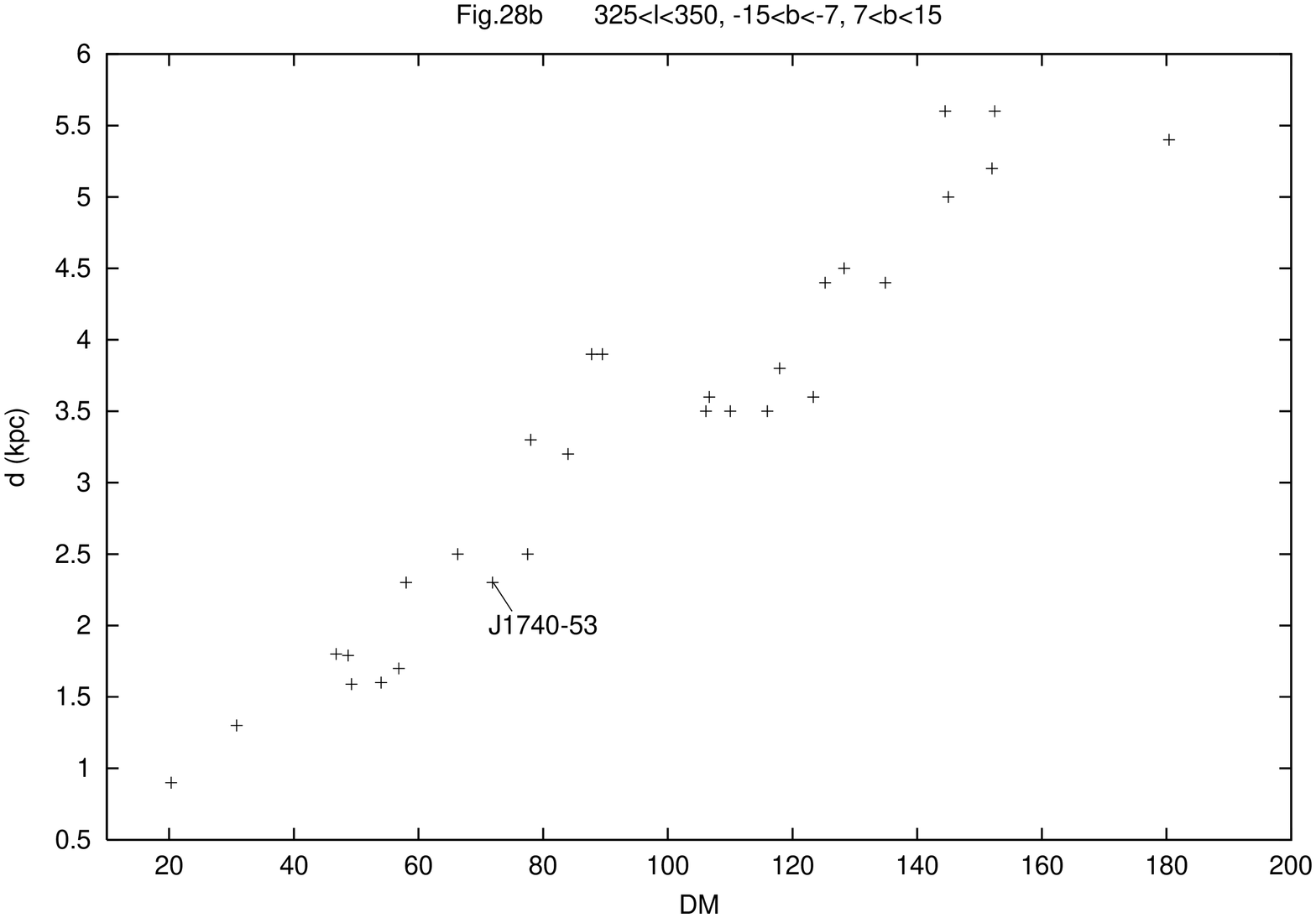,width=8cm,height=5cm,angle=-90} &
\psfig{file=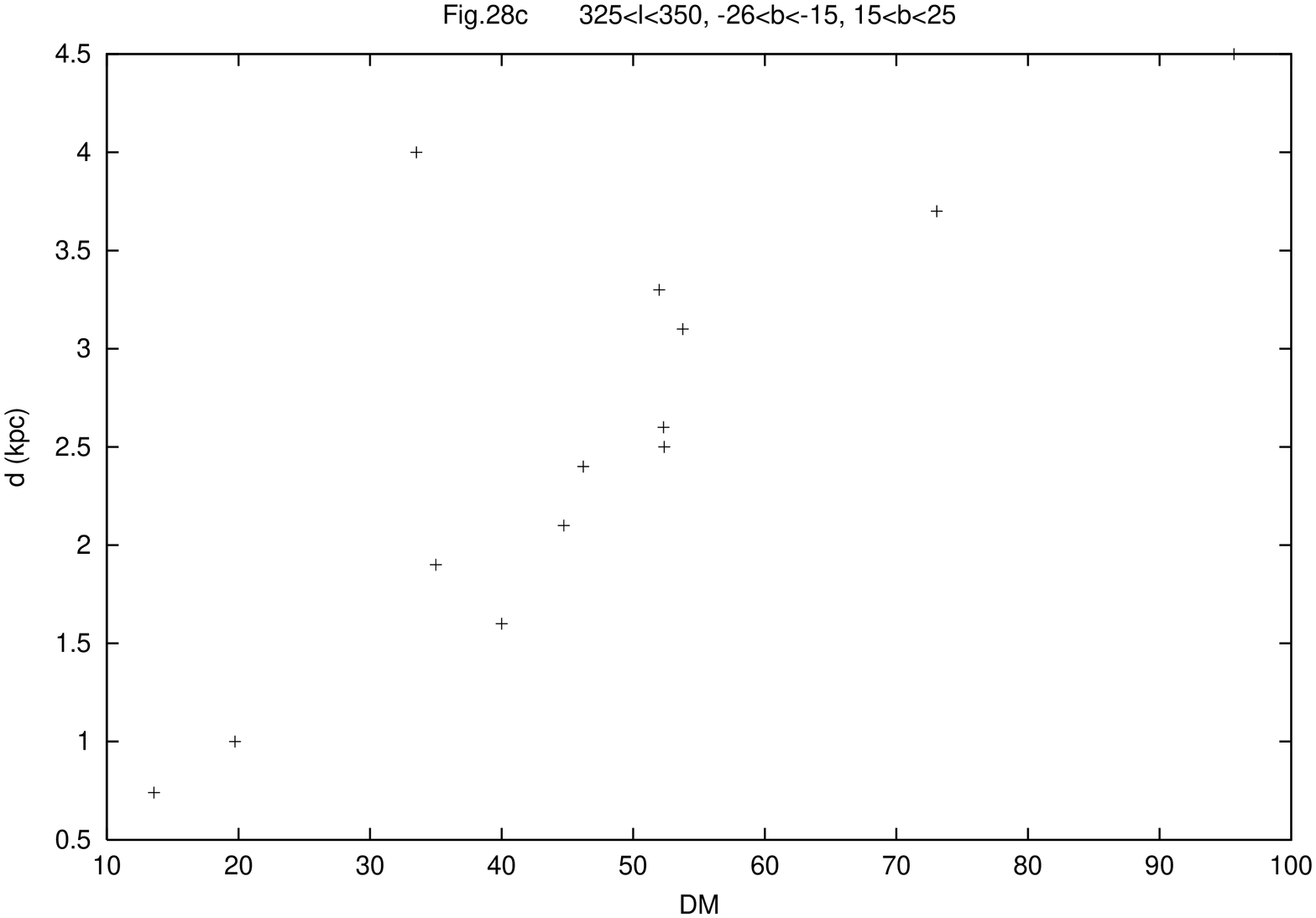,width=8cm,height=5cm,angle=-90} \\
\end{tabular}
\end{figure}

\end{document}